\documentclass[reprint,aps,prb,superscriptaddress]{revtex4-1}
\usepackage[applemac]{inputenc}
\usepackage[T1]{fontenc}
\usepackage[american]{babel}
\usepackage{graphicx,textcomp,booktabs,mathtools,amssymb,bm}
\usepackage{courier,framed,multirow}
\usepackage[scaled]{helvet}
\usepackage{bm}
\usepackage{subfigure}
\usepackage[colorlinks,pdfpagelabels,pdfstartview = FitH,bookmarksopen = true,bookmarksnumbered = true,linkcolor = black,plainpages = false,hypertexnames = false,citecolor = black] {hyperref}
\newcommand{\Dirac}[3]{\left\langle #1 \left| #2\right| #3\right\rangle}

\newcommand{\ket}[1]{\left|\left. #1 \right\rangle\right.}

\newcommand{\op}[2]{\left.\left| #1\right\rangle\left\langle #2\right|\right.}
\newcommand{\sket}[1]{\left|\left. #1 \right)\right.}
\newcommand{\sbra}[1]{\left(\left. #1 \right|\right.}
\newcommand{\sop}[2]{\left.\left| #1\right)\left( #2\right|\right.}
\newcommand{\figureshortname}{Fig.}
\newcommand{\equationshortname}{Eq.}
\newcommand{\tableshortname}{Tab.}
\graphicspath{{Pictures/}}

\begin{document}
\def\sectionautorefname{Sec.}

\title{Noise Analysis of Qubits Implemented in Triple Quantum Dot Systems in a Davies Master Equation Approach}
\author{Sebastian Mehl}
\email{s.mehl@fz-juelich.de}
\affiliation{Peter Grünberg Institute: Theoretical Nanoelectronics, Forschungszentrum Jülich, D-52425 Jülich, Germany}
\affiliation{Institute for Quantum Information, RWTH Aachen University, D-52056 Aachen, Germany}

\author{David P. DiVincenzo}
\affiliation{Peter Grünberg Institute: Theoretical Nanoelectronics, Forschungszentrum Jülich, D-52425 Jülich, Germany}
\affiliation{Institute for Quantum Information, RWTH Aachen University, D-52056 Aachen, Germany}
\affiliation{Jülich-Aachen Research Alliance (JARA), Fundamentals of Future Information Technologies, D-52425 Jülich, Germany}
\date{\today}

\begin{abstract}
We analyze the influence of noise for qubits implemented using a triple quantum dot spin system. We give a detailed description of the physical realization and develop error models for the dominant external noise sources. We use a Davies master equation approach to describe their influence on the qubit. The triple dot system contains two meaningful realizations of a qubit: We consider a subspace and a subsystem of the full Hilbert space to implement the qubit. The main goal of this paper is to test if one of these implementations is favorable when the qubit interacts with realistic environments. When performing the noise analysis, we extract the initial time evolution of the qubit using a Nakajima-Zwanzig approach. We find that the initial time evolution, which is essential for qubit applications, decouples from the long time dynamics of the system. We extract probabilities for the qubit errors of dephasing, relaxation and leakage. Using the Davies model to describe the environment simplifies the noise analysis. It allows us to construct simple toy models, which closely describe the error probabilities.
\end{abstract}

\maketitle

\section{\label{sec:Intro}
Introduction}

The spin eigenstates of the electron provide one of the most natural representations of a qubit - the building block of a logical unit in a quantum computer. In recent years great progress in the fabrication and the control of quantum dots containing only one electron has been reported.\cite{hanson2007} This progress is essential if a spin-based quantum computer is to be realized.

The first proposal of a spin-based quantum computer used the spin of a single electron in a quantum dot as a qubit.\cite{loss1998} In this proposal, single qubit rotations are performed by pulsed magnetic fields, and a two qubit gate is achieved by the Heisenberg coupling of two neighboring electrostatically tuned quantum dots. Since electrostatic control of a qubit is achievable on much faster time scales than control of pulsed external magnetic fields, single qubit rotations based on the exchange interaction have been proposed.\cite{levy2002,taylor2005} Here an encoded qubit in the Hilbert space of two singly occupied spin quantum dots is used. The singlet and the spinless triplet level on the two dots define the qubit. Manipulation of the singlet-triplet qubit has been achieved experimentally.\cite{petta2005,johnson2005} 

For universal quantum computation the singlet-triplet qubit requires, in addition to intradot exchange interaction,  a magnetic field gradient between the two quantum dots.  It was natural to ask if a different coding of the qubit would enable universal computation with the exchange interaction alone; this is realized if the qubit is embodied by the states of three singly occupied quantum dots.\cite{divincenzo2000} The exchange coupling of at least two of the three dot pairs should be controllable. Laird et al.\cite{laird2010} and Gaudreau et al.\cite{gaudreau2012} have now shown experimentally this universal exchange control of the three-electron states in a trio of quantum dots.

The objective of this paper is to explore in detail the robustness of this triple quantum dot qubit in contact with a realistic set of environments.  We have two major alternatives to assess, since
the spin Hilbert space of the triple quantum dot can accommodate a qubit in two fundamentally different ways\cite{viola2001,fong2011,west2012}.   Recall that three spin-$\frac{1}{2}$ degrees of freedom combine to form four ``doublets'' (total spin-$\frac{1}{2}$) and four ``quadruplets'' (total spin-$\frac{3}{2}$).  The first approach is to use two of the four doublet energy eigenstates of this system as the qubit levels. To manipulate the qubit, we need to control only the subspace spanned by these two states. Consequently this qubit is called \textit{subspace qubit}. However, there is a second alternative: working in the four-dimensional space of states with total spin quantum number 1/2, one considers the space as a tensor product of two two-dimensional systems.  One of these two-dimensional systems is taken as the coded qubit.  This is referred to as \textit{subsystem qubit}.\cite{viola2001}   Note that, although more abstract, this notion of subsystem is mathematically identical to that of an ordinary subsystem, e.g., the states of one quantum dot in a collection of many quantum dots.

A triple quantum dot offers both a subspace and a subsystem that are immune against various types of global noise.\cite{kempe2001,lidar2003} Defined in a subsystem a qubit is immune against \textit{strong collective decoherence}, and in a subspace it is protected from \textit{weak collective decoherence}.\cite{divincenzo2000,laird2010} Strong collective decoherence is any noise acting globally on the triple quantum dot. Weak collective decoherence involves just global phase noise.  Interaction with the real environment is not simply described by either of these limits so our coded qubits will be susceptible to decoherence.  However, the goal is to identify the encoded qubit that is as robust as possible against external influences, with the longest possible relaxation and dephasing time scales. 

This paper presents calculations of the robustness of the subspace and subsystem qubit coupled to realistic environments for semiconducting spin qubits in triple quantum dot systems. We give a detailed description of the qubit implementation and analyze the time evolution of the noisy qubit. We employ a specific Markov approximation, describing the limit of weak coupling of the quantum dot to its surroundings. This model was introduced by Davies\cite{davies1974} and is called the \textit{Davies model} (DM) in the following. We analyze the influence of noise and extract error probabilities for relaxation and dephasing phenomena, as well as for the leakage to other parts of the Hilbert space.

Numerical simulations show that the initial time evolution can behave quite differently from the long time evolution of the qubit. Since we are mainly interested in the errors of qubit manipulations that are achieved on short time scales, we focus on the description of the initial time evolution. We develop an effective master equation for the description of the qubit, while removing the influence of the environment, using a Nakajima-Zwanzig approach.\cite{fick1983,fick1990} The Nakajima-Zwanzig approach especially helps to develop a description for the initial time evolution. We analyze the initial dynamics in detail and describe how error probabilities can be extracted. The description in the DM allows us to controllably sort the generated dynamics of the quantum dot into groups of transition terms. This special structure strongly restricts the time evolution of the qubit. Additionally, it helps to analyze the generated dynamics. We describe the error probabilities using a few simple toy models.

While rather lengthy, we believe that this paper will be useful as handbook describing the many possible decoherence and relaxation scenarios that can arise for the triple dot qubit.  As more experiments are done to explore the various possible encodings of qubits in these systems, the results here should serve as a guide to help in arriving at the optimal design for making further progress towards functioning multi-qubit structures.

The organization of the paper is as follows. In \autoref{sec:Model} we introduce the model analyzed in this paper. We construct the triple dot Hamiltonian, describe the qubit implementation and introduce the noise model. In \autoref{sec:Systems} we describe the modeling of real spin qubits. Besides the description of the triple quantum dot system, we also introduce the noise parameters. In \autoref{sec:FullTime} we analyze the full time evolution of the qubit, while in \autoref{sec:Errors} error rates for the initial time evolution are extracted. We conclude with a summary and an outlook in \autoref{sec:Summary}. We include in the appendices a detailed description of techniques used in the main text of the paper. We discuss symmetry properties of the noise model and describe ways to analyze the short and long time evolutions. Finally, we discuss error models for the analysis of the noisy qubit evolution from a solid-state and quantum information perspective.

\section{\label{sec:Model}
Description of Model}

\subsection{\label{ssec:Hamiltonian}
Triple Dot Hamiltonian and Qubit Implementation}

\begin{figure}
\centering
\includegraphics[width=0.4\textwidth]{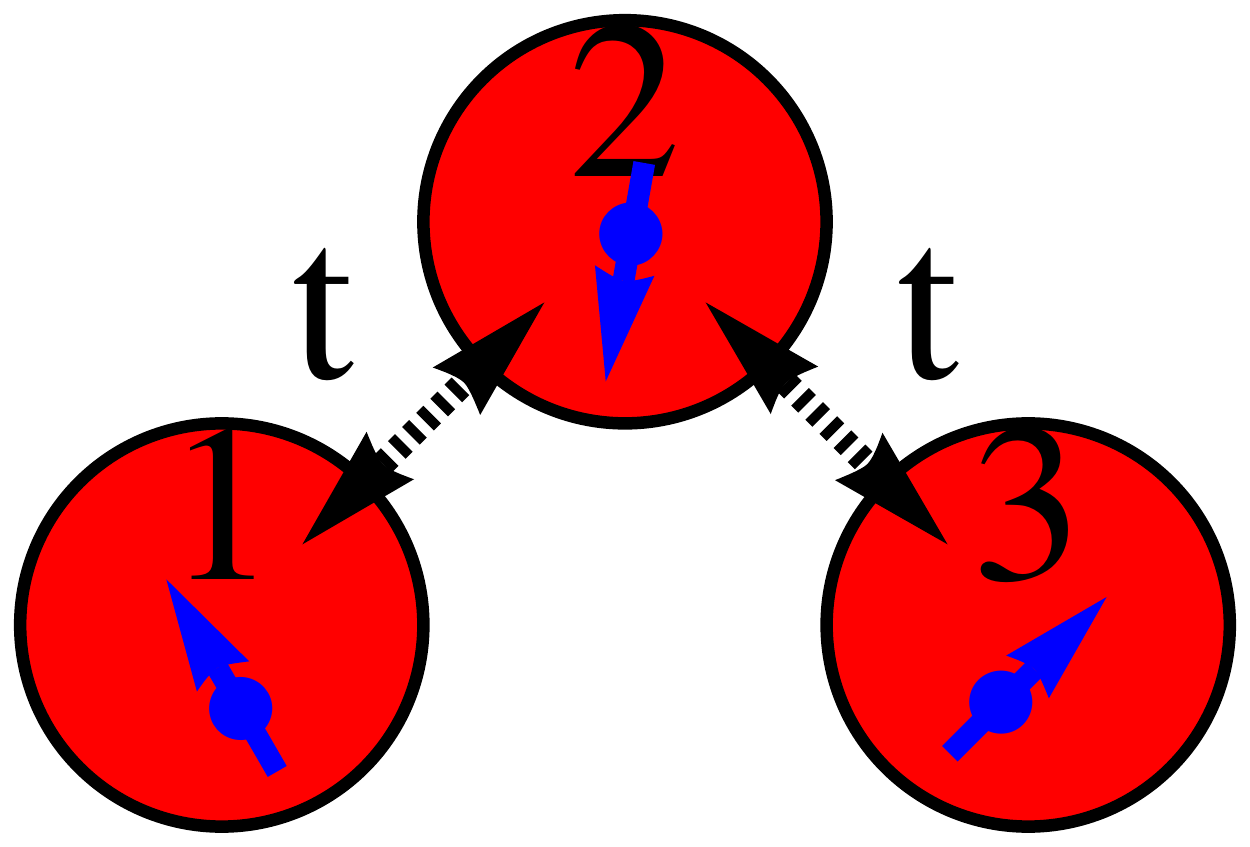}
\caption{\label{fig:Geometry}
Layout of the triple quantum dot setup. Each quantum dot is occupied with one electron. Neighboring dot pairs are tunnel coupled with the coupling strength $t$. An external electric bias is used to occupy either the left (dot 1) or the right (dot 3) quantum dot with two electrons.}
\end{figure}

The effective Hamiltonian $\mathcal{H}$ describing the triple quantum dot contains the exchange interaction between two neighboring quantum dot pairs. Additionally, an out-of-plane magnetic field is added. Our notations follow those of Laird et al.\cite{laird2010},

\begin{align}\nonumber
\mathcal{H}=&\frac{J_{12}}{4}\left(\bm{\sigma}^1\cdot\bm{\sigma}^2-\mathbf{1}\right)
+\frac{J_{23}}{4}\left(\bm{\sigma}^2\cdot\bm{\sigma}^3-\mathbf{1}\right)
\\&-\frac{E_z}{2}\sum_{i=1,\dots,3}\sigma ^i_z,
\label{eq:Hamiltonian}
\end{align}
where $E_z$ is the Zeeman energy of a magnetic field applied perpendicular to the quantum dots, $\sigma^i_{x,y,z}$ are the Pauli matrices at quantum dot $i$, and $J_{12}$ ($J_{23}$) represents the Heisenberg exchange interaction between two neighboring dots. It can be changed by applying an electric bias on the outer dots.

The coupling parameters $J_{12}$ and $J_{23}$ can be derived from a three-site Hubbard Hamiltonian describing the quantum dot layout in \figureshortname~\ref{fig:Geometry}:
\begin{align}\nonumber
	\mathcal{H}_{Hubbard}&=\sum_{\alpha,s} \epsilon_{\alpha}n_{\alpha,s}+
\sum_{\alpha}U_{\alpha}n_{\alpha,\uparrow}n_{\alpha,\downarrow}\\&+
t\sum_{\left\langle \alpha,\beta \right\rangle,s} \left(a_{\alpha,s}^\dagger a_{\beta,s}+h.c.\right),\label{eq:Hubbard}
\end{align}
where $\epsilon_\alpha$ is the single particle energy, $U_\alpha$ the coulomb repulsion, and $t$ the tunnel coupling between neighboring quantum dots. For simplicity we take only tunneling into account for the left (right) quantum dot with the center dot ($\left\langle1,2\right\rangle$ and $\left\langle2,3\right\rangle$). Additionally, we assume these tunnel couplings to be equal. When going from $\mathcal{H}_{Hubbard}$ to $\mathcal{H}$, we take into account single occupation of all three qubits [$\left(1,1,1\right)$ configuration] and use an electric bias to go to a double occupied left (right) dot [$\left(2,0,1\right)$ and $\left(1,0,2\right)$ configurations]. For the double occupied states, we consider only the orbital ground state.

In analogy to the case of double dots,\cite{coish2005,taylor2007} we describe all three charge regimes in a common basis. We eliminate the higher energetic states by adiabatic elimination\cite{taylor2005} and work only with the low-energy subspace of all possible charge distributions. This approach is mainly adopted from adiabatic manipulation protocols, where the manipulation velocity is slower than transition rates to higher excited states. It therefore allows computation within the low-energy subspace. The tunnel coupling causes transitions between the singlet states of all charge distributions. We eliminate the excited states separately for the charge transition to a double occupied left and right dot. One arrives at the exchange parameters:
\begin{align}
	J_{12}&=\frac{\epsilon_{-}-\epsilon}{2}+\sqrt{\left(\frac{\epsilon_{-}-\epsilon}{2}\right)^2+ 2 t^2},\label{eq:ExInt1}
\end{align}
\begin{align}
	J_{23}&=\frac{\epsilon-\epsilon_{+}}{2}+\sqrt{\left(\frac{\epsilon-\epsilon_{+}}{2}\right)^2+ 2 t^2}.\label{eq:ExInt2}
\end{align}
The bias parameter $\epsilon$ lowers the energy of the left quantum dot for $\epsilon<0$, while the right quantum dot is favored for $\epsilon>0$. $\epsilon_{-}$ is the bias at which the $\left(1,1,1\right)$ and $\left(2,0,1\right)$ configurations have the same energy in the absence of tunnel coupling (and similarly for $\epsilon_+$).\footnote{These parameters are related to the ones from the Hubbard Hamiltonian in \equationshortname~\eqref{eq:Hubbard} by $\epsilon_{-}\equiv-\left(\epsilon_{1}-\epsilon_{2}+U_{1}\right)$ and $\epsilon_{+}\equiv\epsilon_{3}-\epsilon_{2}+U_{3}$} The eigenstates of the Hamiltonian $\mathcal{H}$ are
\begin{widetext}
\begin{align}
Q_{\frac{3}{2}}&=\left|\left. \uparrow\uparrow\uparrow\right\rangle\right.,\\
\label{eq:state2}
Q_{\frac{1}{2}}&=\frac{1}{\sqrt{3}}\left(
\left|\left. \uparrow\uparrow\downarrow\right\rangle\right.
+\left|\left. \uparrow\downarrow\uparrow\right\rangle\right.
+\left|\left. \downarrow\uparrow\uparrow\right\rangle\right.
\right),\\
\label{eq:state3}
\Delta_{\frac{1}{2}}&=\frac{
\left(J_{12}-J_{23}+\Omega\right)\left|\uparrow\uparrow\downarrow\right.\rangle+
\left(J_{23}-\Omega\right)\left|\uparrow\downarrow\uparrow\right.\rangle-
J_{12}\left|\downarrow\uparrow\uparrow\right.\rangle
}
{\sqrt{4\Omega^2+2\Omega\left(J_{12}-2J_{23}\right)}},\\
\label{eq:state4}
\Delta^{\prime}_{\frac{1}{2}}&=\frac{
\left(-J_{12}+J_{23}+\Omega\right)\left|\uparrow\uparrow\downarrow\right.\rangle-
\left(J_{23}+\Omega\right)\left|\uparrow\downarrow\uparrow\right.\rangle+
J_{12}\left|\downarrow\uparrow\uparrow\right.\rangle
}
{\sqrt{4\Omega^2+2\Omega\left(2J_{23}-J_{12}\right)}}.
\end{align}
\end{widetext}
where $\Omega= \sqrt{J_{12}^2+J_{23}^2-J_{12}J_{23}}$. For later analysis we introduce the notation $W_{\frac{1}{2}}$, indicating the $s_z=\frac{1}{2}$ subspace ($W\in\left\{Q,\Delta,\Delta^\prime\right\}$). All remaining eigenstates can be obtained by flipping all three spins to obtain from the states $Q_{\frac{3}{2}}$ and $W_{\frac{1}{2}}$ the corresponding states $Q_{-\frac{3}{2}}$ and $W_{-\frac{1}{2}}$. The eigenenergies of $\mathcal{H}$ are
\begin{align}
E_{Q_k}&=-k\cdot E_z\label{eq:Energy1},\ k\in\left\{\pm\frac{3}{2},\pm\frac{1}{2}\right\},
\end{align}

\begin{align}
E_{\Delta_{\pm\frac{1}{2}}}&=-\frac{1}{2}\left(J_{12}+J_{23}-\Omega\right)\mp \frac{E_z}{2}\label{eq:Energy2},
\end{align}

\begin{align}
E_{\Delta^\prime_{\pm\frac{1}{2}}}&=-\frac{1}{2}\left(J_{12}+J_{23}+\Omega\right)\mp \frac{E_z}{2}\label{eq:Energy3}.
\end{align}

\begin{figure}
\centering
\subfigure{\label{fig:EnergyDiag1}\includegraphics[width=0.49\textwidth]{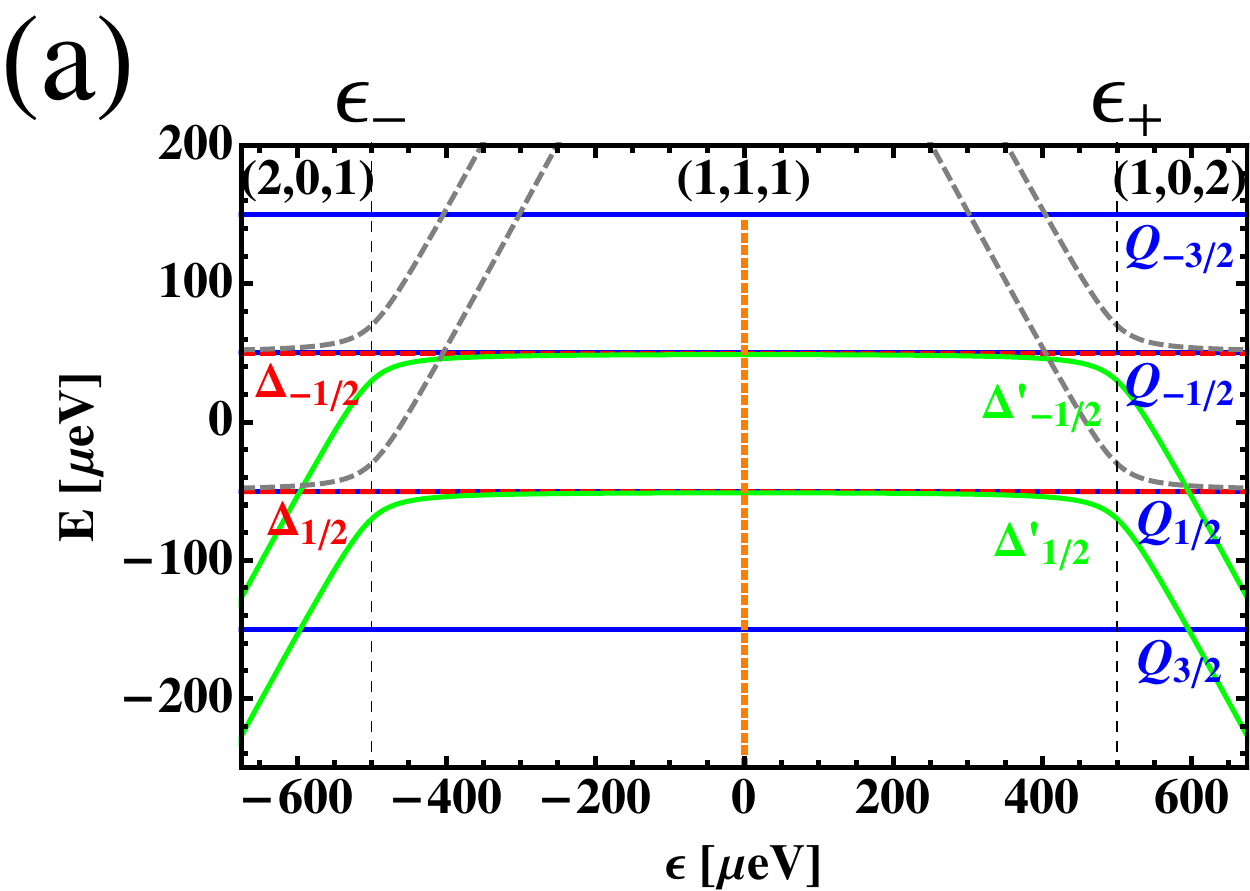}}
\subfigure{\label{fig:EnergyDiag2}\includegraphics[width=0.49\textwidth]{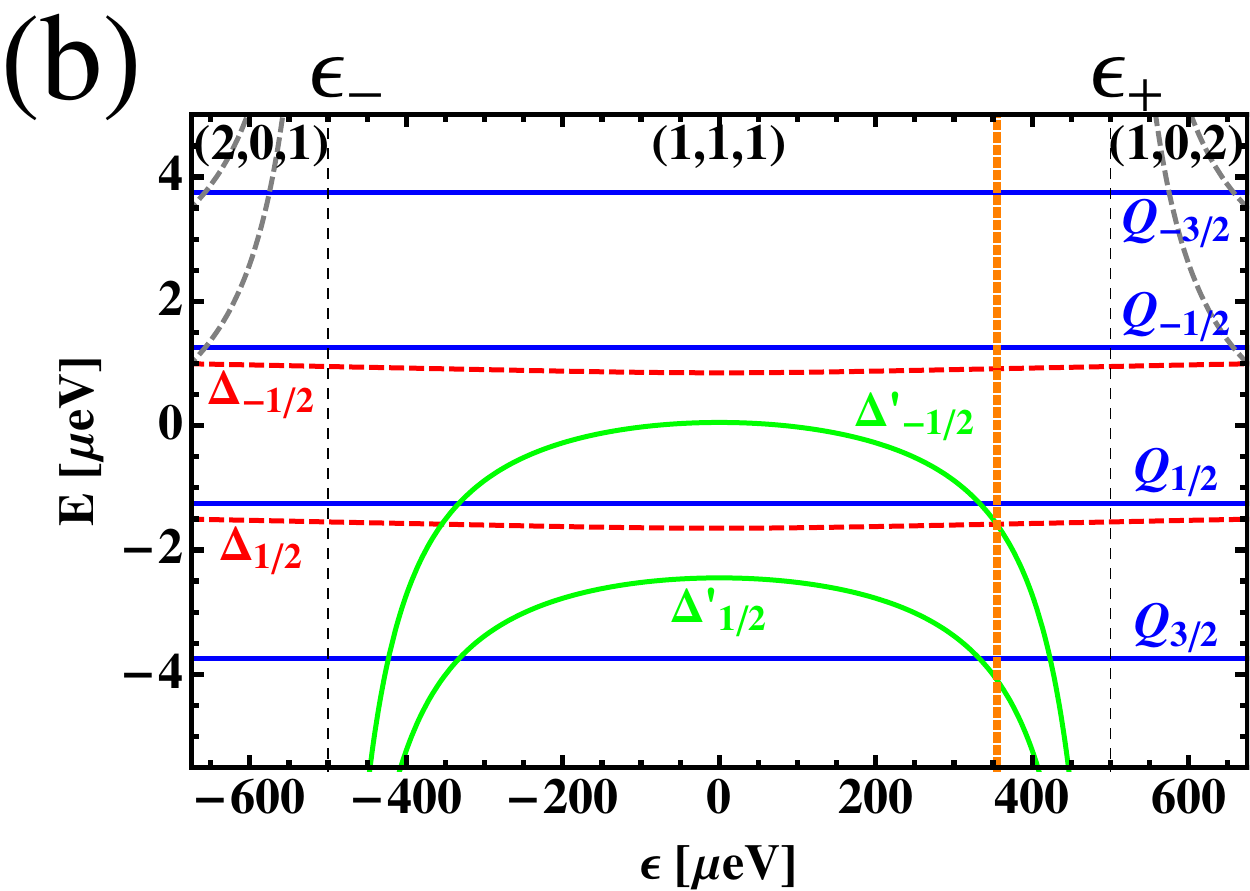}}
\caption{\label{fig:EnergyDiagram}
Diagram of the eigenenergies of the exchange Hamiltonian of \equationshortname~\eqref{eq:Hamiltonian} as a function of the bias parameter $\epsilon$ with $E_z>0$. The dashed gray lines are the higher energy states that are not treated by the $\mathcal{H}$ in \equationshortname~\eqref{eq:Hamiltonian}. They are removed from the $\mathcal{H}_{Hubbard}$ by adiabatic elimination. (a) The case of large external magnetic fields: $E_z=100\ \mu\text{eV}$ ($\approx7\ \text{T}$ in GaAs). In (b) the external magnetic fields are small ($E_z=2.5\ \mu\text{eV}$, corresponding to $200\ \text{mT}$ in GaAs). The dashed orange lines mark regimes analyzed in the analysis of \autoref{sec:FullTime}.}
\end{figure}

The energy diagram is sketched in \figureshortname~\ref{fig:EnergyDiagram}. For the upcoming analysis we introduce three quantum numbers $\left(l,S,s_z\right)$, which fully characterize the eigenstates. $S$ describes the total spin of the eigenstates. It has the value $3/2$ for all $Q_k$ states and $1/2$ for the remaining ones. The $s_z$-quantum number labels the spin projection in the $z$ direction. It has values $\pm3/2$ and $\pm1/2$. Furthermore, we introduce a third formal quantum number $l$. It distinguishes the $\Delta_k$ ($l=1$) and $\Delta_k^\prime$ states ($l=0$).

\subsection{\label{ssec:SsubAndSsys}
Subspace and Subsystem Qubits}

For our later use we construct a subsystem and a subspace qubit inside the eight-dimensional Hilbert space spanned by the eigenstates of the triple dot Hamiltonian in \equationshortname~\eqref{eq:Hamiltonian}.\cite{fong2011,west2012} The basis states of the subspace qubit define the computational subspace: $span\left\{\Delta_{\frac{1}{2}},\Delta^\prime_{\frac{1}{2}}\right\}$. We identify the state $\Delta_{\frac{1}{2}}$ as logical ``1'', $\Delta^\prime_{\frac{1}{2}}$  as logical ``0''. For the subsystem qubit we use the larger subspace $span\left\{\Delta_{\frac{1}{2}},\Delta^\prime_{\frac{1}{2}},\Delta_{-\frac{1}{2}},\Delta^\prime_{-\frac{1}{2}}\right\}$ as the computational subspace. We identify both states $\Delta_{\frac{1}{2}}$ and $\Delta_{-\frac{1}{2}}$ as logical ``1'' (represented by the quantum number $l=1$). The $l=0$ states are identified as logical ``0''. For the subsystem qubit the $s_z$ population does not matter. In the Nakajima-Zwanzig approach (see Appendix \ref{ssec:InitSubsystem}), we fix this population to a constant value. The thermal distribution over the Zeeman-split eigenstates will be a reasonable choice:

\begin{equation}
\rho_0^{s_z}=e^{- \frac{-\frac{E_z}{2}\sigma_z}{T_K}}/Tr\left(e^{- \frac{-\frac{E_z}{2}\sigma_z}{T_K}}\right).
\label{eq:subsysbath}
\end{equation}

These two possible ways of defining a qubit are motivated by the experimental possibilities for initializing the qubit into a controlled state. For the subspace qubit, initialization into the $\Delta^\prime_{\frac{1}{2}}$ state seems experimentally achievable. When doing all experiments at high external magnetic fields [cf. \figureshortname~\ref{fig:EnergyDiag1}], the $s_z=-\frac{1}{2}$ states may be avoided; they are far up in energy compared to the  $s_z=\frac{1}{2}$ states [cf. \equationshortname~\eqref{eq:Energy1}-\eqref{eq:Energy3}]. Initialization into $\Delta^\prime_{\frac{1}{2}}$ can be achieved by coupling two of the three dots strongly, creating effectively a strongly coupled double quantum dot and an uncoupled single dot. $\Delta^\prime_{\frac{1}{2}}$ is described by a singlet eigenstate on the double dot, while the $\Delta_{\frac{1}{2}}$ state involves triplet eigenstates. Initialization is now identical to the initialization of the singlet-triplet qubit in double dots.\cite{taylor2007} The uncoupled single dot needs to be in its ground state $\left(\ket{\uparrow}\right)$.

For the subsystem qubit the initialization works in the same way. Here, however, it does not matter if we initialize into the $\Delta^\prime_{\frac{1}{2}}$ state or the corresponding $\Delta^\prime_{-\frac{1}{2}}$ state; both are labeled by $l=0$. These states differ only by the population of the weakly coupled single dot. For the subsystem qubit it is satisfactory to produce a thermal distribution between the spin-up and spin-down states on the weakly coupled dot. This property is described by the density matrix \eqref{eq:subsysbath}. The strongly coupled quantum dot should again be initialized into the singlet state. We can accomplish the initialization of the subsystem qubit for small and large magnetic fields.

\subsection{\label{ssec:Noise}
Noise Description}

To study the influence of noise in the triple dot setup we first discuss a Lindblad master equation on the eight-dimensional Hilbert space:

\begin{equation}
	\dot\rho\left(t\right)=\left(\mathcal{L}_0+\mathcal{L}_D\right)\rho\left(t\right)=-i\left[\mathcal{H},\rho\left(t\right)\right]+\mathcal{L}_D\left(\rho\left(t\right)\right).
	\label{eq:LindblMasEq}
\end{equation}

We set $\hbar=1$ and $k_B=1$. We add to the coherent evolution, given by the Hamiltonian from \equationshortname~\eqref{eq:Hamiltonian}, a dissipative Lindblad term $\mathcal{L}_D\left(\rho\left(t\right)\right)=\sum_{\mathcal{A}}\Upsilon_{\mathcal{A}}\mathcal{D}\left[\mathcal{A}\right]\left(\rho\left(t\right)\right)$, where $\mathcal{D}\left[A\right]\left(B\right)\equiv A B A^\dagger-\frac{1}{2}\left(A^\dagger AB+BA^\dagger A\right)$. An external bath couples through the operators $\mathcal{A}$ to the system. $\Upsilon_{\mathcal{A}}\in\mathbb{R}$ determines the coupling strength. We analyze the effects of dephasing and relaxation from external baths. Dephasing of spin qubits is generated by fluctuating magnetic fields parallel to the external magnetic field. Relaxation is generated by fluctuating perpendicular magnetic fields (compare also the description in \autoref{ssec:Rates}). The coupling operators $\mathcal{A}$ act either globally on the triple dot system or individually on each of the dots:

\begin{align}\label{eq:Lindbl1}
\mathcal{L}_{glob}\left(\rho\right)&=
\Upsilon^z\mathcal{D}\left[Z\right]\left(\rho\right)+
\Upsilon^x\mathcal{D}\left[X\right]\left(\rho\right),
\end{align}

\begin{align}
\label{eq:Lindbl2}
\mathcal{L}_{loc}\left(\rho\right)&=\sum_{i=1,2,3}\left(
\Upsilon^z_i\mathcal{D}\left[\sigma_z^i\right]\left(\rho\right)+
\Upsilon^x_i\mathcal{D}\left[\sigma_x^i\right]\left(\rho\right)
\right),
\end{align}
with $Z=\sum_{i=1,2,3}\sigma_z^i$ and $X=\sum_{i=1,2,3}\sigma_x^i$.

This model represents a specific Markov approximation to describe for the time evolution of an open quantum system.\cite{davies1974} The procedure for making a Markov approximations is mathematically not strict.\cite{celio1989} For our analysis we modify \equationshortname~\eqref{eq:Lindbl1} and \eqref{eq:Lindbl2} by employing a different Markov approximation. Our goal is to make sure that the system equilibrates in the long time limit. We directly adopt the description from the paper of Bravyi and Haah.\cite{bravyi2011} They describe a specific Markov approximation, originally introduced by Davies for the weak coupling limit of a system and a bath.\cite{davies1974} The modified Lindbladian in the DM is
\begin{equation}
 \widetilde{\mathcal{L}}_D\left(\rho\right)=\sum_{\mathcal{A},\omega}h\left(\mathcal{A},\omega\right)\mathcal{D}\left[\mathcal{A}_{\omega}\right]\left(\rho\right).
 \label{eq:DaviesModel}
\end{equation}
All coupling operators $\mathcal{A}$ are decomposed into transition terms between equidistant energy eigenstates of the free Hamiltonian: 
\begin{equation}
	\mathcal{A}=\sum_{\omega}\mathcal{A}_{\omega}.
	\label{eq:DaviesSumExpansion}
\end{equation}
$\mathcal{A}$ is grouped into terms $\mathcal{A}_{\omega}$, defined by
\begin{equation}
	\Dirac{A}{\mathcal{A}_{\omega}}{B}=\left\{ 
	\begin{array}{l}
	\Dirac{A}{\mathcal{A}_{\omega}}{B}\ \text{if}\ E_{A}-E_{B}=\omega,\\
	0\ \text{otherwise}.
	\end{array}
	\right.\label{eq:overlap}
\end{equation}
The rate of quantum jumps $h\left(\mathcal{A},\omega\right)\in\mathbb{R}$ is set by the transition frequencies $\omega$ between energy eigenstates of the system induced from the bath. It needs to fulfill the detailed balance condition, connecting positive and negative energy differences:
\begin{equation}
	h\left(\mathcal{A},-\omega\right)=e^{-\frac{\omega}{T_K}}h\left(\mathcal{A},\omega\right).
\end{equation}
As shown by Spohn,\cite{spohn1977} the Gibbs state is a fixed point of the dynamics in the DM:
\begin{equation}
	\widetilde{\mathcal{L}}_D\left(e^{-\frac{\mathcal{H}}{T_K}}\right)=0.
\end{equation}

For simplicity we neglect the tilde on the redefined Lindbladian \eqref{eq:DaviesModel} for the remainder of the paper. When inspecting the energy diagrams in \figureshortname~\ref{fig:EnergyDiagram}, one sometimes finds that sets of energy levels become equidistant at specific exchange interactions, which are not equidistant for each $\epsilon$. We do not add these ``accidental'' degeneracies to the construction \eqref{eq:DaviesModel} of the DM.

\section{\label{sec:Systems}
Approach to Model Real Systems}

\subsection{\label{ssec:Parameter}
System Parameters}

All system parameters to define the Hamiltonian of \equationshortname~\eqref{eq:Hamiltonian} are matching the triple-dot experiments by Gaudreau et al. (compare especially the Supplemental Material\cite{gaudreau2012}). The parameters are summarized in \tableshortname~\ref{tab:characteristicvalues}. We use a typical experimental time scale $\delta t$ of $10$ ns.\footnote{We refer especially to typical qubit manipulation times. In the mentioned publication by Gaudreau et al. pulse times well below $10$ ns are used\cite{gaudreau2012}} The temperature $T_K$ is set to $125$ mK ($\sim 10\ \mu\text{eV}$). A high magnetic field accounts for the case $E_z\gg T_K$, while the low magnetic field case describes the opposite limit ($E_z\lesssim T_K$). When deriving predictions for the initial time evolution, we never go far into the regime of double occupied quantum dots. We limit the bias $\epsilon$ on the interval $\left[\epsilon_-,\epsilon_+\right]$ of the exchange interaction parameter. The tunnel coupling parameter $t$ is also extracted from the paper.\footnote{In the paper by Gaudreau et al. the tunnel couplings $T$ are defined differently than the parameters $t$ in \equationshortname~\eqref{eq:ExInt1} and \eqref{eq:ExInt2}. The two constants are however connected by $T=t/\sqrt{2}$}

\begin{table}
\centering
\begin{tabular}{ c c }
\hline
\hline
Value & Size\\
\hline
$\delta t / ns$ & 10\\
$T_K$ & $10\ \mu\text{eV}$ ($\approx 125\ \text{mK}$)\\ \\
\multirow{2}{*}{Magnetic field strength}  &High: $7\ \text{T}$ ($E_z\approx 100\ \mu\text{eV}$)\\
& Low: $200\ \text{mT}$ ($E_z\approx 2.5\ \mu\text{eV}$)\\ \\
\multirow{2}{*}{Exchange interaction parameter} & $\epsilon_+=\left|\epsilon_-\right|=500\ \mu\text{eV}$\\
& $t=14\ \mu\text{eV}$\\ \\
Analyzed interval of applied $\epsilon$ & $\left[-500\ \mu\text{eV},500\ \mu\text{eV}\right]$\\
\hline
\hline
\end{tabular}
\caption{\label{tab:characteristicvalues}
Characteristic values for the analysis of GaAs triple quantum dots. The values are used according to the publication by Gaudrau et al.\cite{gaudreau2012}}
\end{table}

\subsection{\label{ssec:Rates}
Transition Rates for Noise Description}

Our model does not contain a microscopic description of interactions with the environment. In the DM the influence of the surroundings is modeled only by the generated transition rates between quantum states (see description of \autoref{ssec:Noise}). We especially focus in our analysis on experiments in GaAs. This material was used in all previous experiments on triple quantum dots.\cite{laird2010,gaudreau2012}

\subsubsection{\label{sssec:RateHyperfine}
Hyperfine Interaction}

As in the experiments on single and double quantum dots, the nuclear magnetic fields are one major source of noise for triple dots. The magnetic moments of the nuclei in GaAs couple through the hyperfine interactions to the spin of the electron. Extensive studies of the generated dynamics have been carried out in single quantum dots\cite{coish2004,coish2010,cywinski2009,cywinski2009-2} and double dots.\cite{coish2005} Also very recently a study on triple quantum dots appeared.\cite{ladd2012} We do not follow the arguments of these papers in detail, but extract transition rates for our later analysis.

In a semiclassical picture the nuclear magnetic moments add up to a macroscopic magnetic field.\cite{merkulov2002,hanson2007} Fluctuations in the nuclear magnetic field are slow compared to the precession time of the electron spin in this magnetic field. For the initial time evolution of the electron, the nuclear magnetic field can therefore be described as static. Due to the large number of spins interacting with the electron, one approximates the hyperfine fields by a Gaussian distribution with zero mean and a root mean square (rms) $\delta E_{nuc}$ [represented by the distribution function $f\left(B\right)=\frac{1}{\sqrt{2\pi}\delta E_{nuc}}e^{-\frac{B^2}{2\delta E_{nuc}^2}}$]. A typical values in GaAs quantum dots is $\delta E_{nuc}\approx 0.3\ \mu\text{eV}$.

We argue that fluctuating hyperfine fields generate locally spin dephasing and relaxation. Its influence is reflected by two distinct transition rates, which exponentially decrease for increasing transition energies $\omega$: 

\begin{equation}
	h\left(\sigma^i_{z}/ \sigma^i_{x},\omega\right)=
	\left\{\begin{array}{l}
		\Upsilon_i^{z,x} e^{-\frac{\omega^2}{2 \delta E^2_{nuc}}}\ \text{for}\ \omega\ge0,\\
		\Upsilon_i^{z,x} e^{-\left(\frac{\omega^2}{2 \delta E^2_{nuc}}+\frac{\omega}{T_K}\right)}\ \text{for}\ \omega<0,
	\end{array}\right.
	\label{eq:ratehyperfine}
\end{equation}
where $h\left(\sigma^i_{z},\omega\right)$ arises from local magnetic field fluctuations at quantum dot $i$ ($i=1,2,3$) in the direction of the external magnetic field. These operations lead to dephasing of individual spins. $h\left(\sigma^i_{x},\omega\right)$ describes local spin relaxation through in-plane magnetic field fluctuations (coupling through raising and lowering operators $\sigma_{x}^i=\sigma_{+}^i+\sigma_{-}^i$\footnote{In single spin experiments fluctuating in-plane magnetic fields, both in x- and y-direction, cause spin flips. The generated dynamics is very similar. We just consider the coupling operators $\sigma_x^i$, since they directly relate to single spin flips $\sigma_{\pm}$}). Quantum jumps between energy levels are possible at energy differences smaller than or in the range of the hyperfine interaction $\delta E_{nuc}$. We argue that both transitions generated from fluctuating out of plane magnetic fields and in plane magnetic fields have been observed in previous experiments.  We refer especially to experiments on single and double quantum dots to support the modeling through \equationshortname~\eqref{eq:ratehyperfine}.

\begin{figure}
\centering
\includegraphics[width=0.4\textwidth]{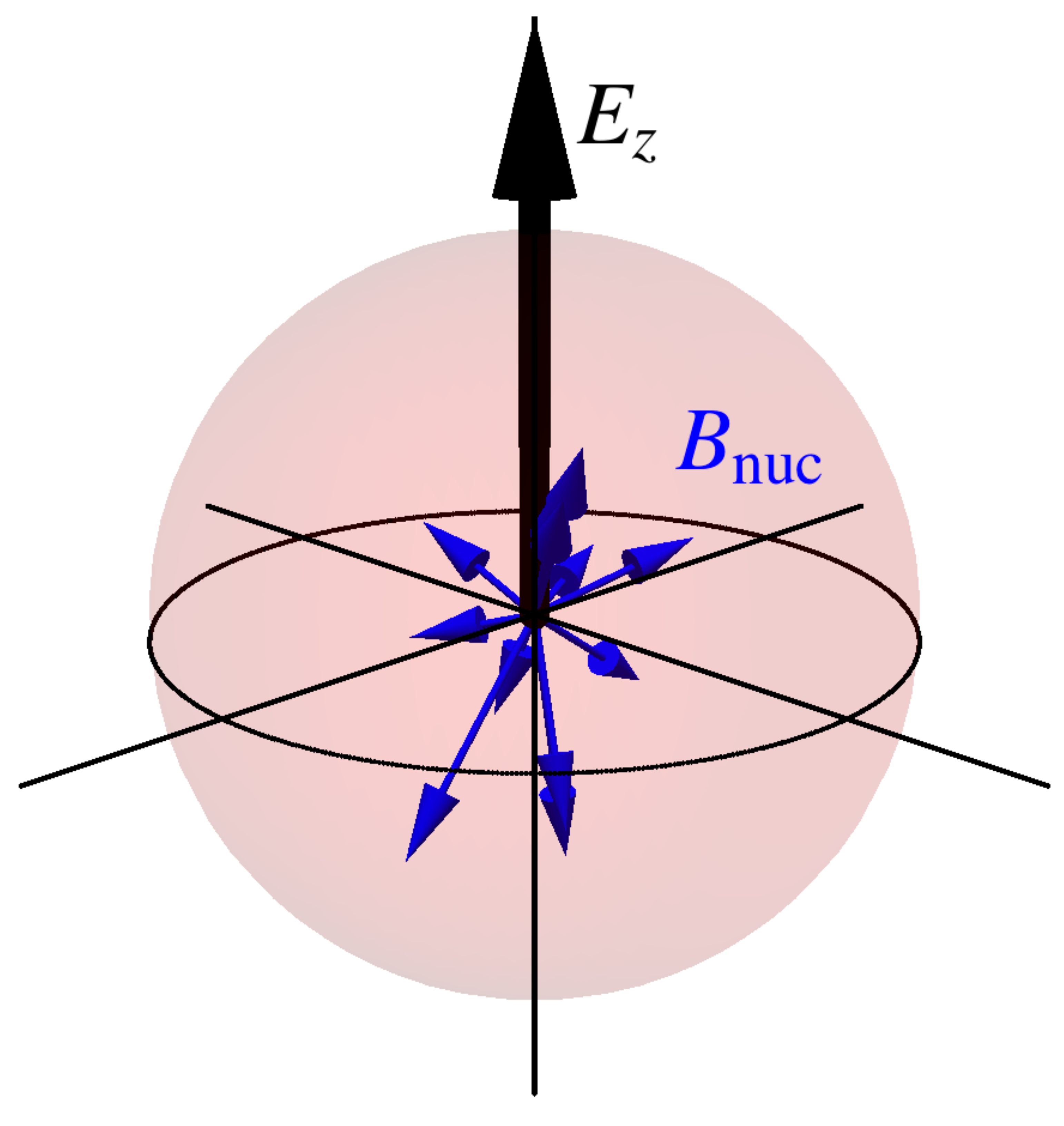}
\caption{\label{fig:semiclpicturenoise}
Semiclassical picture used for the time evolution of a single electron spin in a distribution of hyperfine configurations. For every experiment the hyperfine field acts like a static magnetic field, which is randomly varying between every run of the experiment. A constant external magnetic field $E_z$ is applied in the $z$ direction.}
\end{figure}

For single quantum dots a fluctuating magnetic field parallel to the static external field leads to dephasing, while a fluctuating perpendicular magnetic field causes spin flips. In experiments, the time evolution of a single spin [$\mathbf{S}_{B_{nuc}}\left(t\right)$ $= Tr\left(\bm{\sigma}\rho_{B_{nuc}}\left(t\right)\right) = Tr\left( \bm{\sigma}e^{i\mathcal{H}_{B_{nuc}}t}\rho\left(0\right)e^{-i\mathcal{H}_{B_{nuc}}t}\right)$] is measured as the average result of many runs of the experiment. For each measurement a Hamiltonian $\mathcal{H}_{B_{nuc}}$, corresponding to a specific hyperfine field $B_{nuc}$, determines the time evolution (cf. \figureshortname~\ref{fig:semiclpicturenoise}). The measured result reflects the ensemble average over the hyperfine field distribution $\left\langle \mathbf{S}\left(t\right)\right\rangle\equiv\int \mathbf{S}_B\left(t\right) f\left(B\right) dt$.

We use a similar argumentation scheme as in the paper by Merkulov et al.\cite{merkulov2002} Without external magnetic fields, the time evolution is completely determined by the fluctuating magnetic field ($\mathcal{H}_B=\frac{B}{2}\sigma_z$). When we calculate the time evolution and average it over the magnetic field distribution $f\left(B\right)$, we see that the component perpendicular to the magnetic field decreases exponentially:
\begin{align}
	\left\langle S_x\left(t\right)\right\rangle&=e^{-\frac{\delta E_{nuc}^2 t^2}{2}}S_x\left(0\right).
\end{align}
Since there is no fixed quantization axis of the qubit, one expects that all components of the spin decrease. The semiclassical analysis predicts therefore a Gaussian decay with a time constant $\left(\delta E_{nuc}\right)^{-1}$. The DM describes this behavior by a transition rate $\Upsilon_i^{z,x}=\delta E_{nuc}$. This is the value of \equationshortname~\eqref{eq:ratehyperfine} at $\omega=0$. The energy difference $\omega$ is determined by the external magnetic field strength. In the absence of magnetic fields $\Upsilon_i^{z}$ and $\Upsilon_i^{x}$ must be indistinguishable.

A fixed external magnetic $E_z$ (cf. \figureshortname~\ref{fig:semiclpicturenoise}) and a fluctuating parallel magnetic field ($\mathcal{H}_B=\frac{E_z+B}{2}\sigma_z$) generates similarly also dephasing of the transverse spin components. Now the single spin precesses, however, with the angular frequency $E_z$ around the $z$ axis:
\begin{align}
\left\langle S_x\left(t\right)\right\rangle&=e^{-\frac{\delta E_{nuc}^2 t^2}{2}}\left[
S_x\left(0\right)\cos\left(E_z t\right)-S_y\left(0\right)\sin\left(E_z t\right)\right].
\end{align}
This is also reflected in the DM. Dephasing transition rates for single spins are independent of the energy difference. $\sigma_z^i$ causes only transitions between identical spin states, which is reflected by the transition rate $h\left(\sigma_z^i,0\right)$.\\
At finite external magnetic fields the relaxations decrease with $\left(\frac{\omega}{\delta E_{nuc}}\right)^2$. We can see this when calculating the time evolution for a fluctuating perpendicular magnetic field for an initially spin-up polarized particle ($Z_0=1$ and $\mathcal{H}_B=\frac{\omega}{2}\sigma_z+\frac{B}{2}\sigma_x$):
\begin{equation}
	Z_B\left(t\right)=\frac{E_z^2+B^2\cos\left(E_zt \right)}{E_z^2+B^2}.
\end{equation}
When expanding in $\left(B/E_z\right)$ and averaging over the field distribution $f\left(B\right)$, we get
\begin{equation}
	\left\langle Z\left(t\right)\right\rangle\approx Z_0-\frac{1-\cos\left(\omega t\right)}{\left(\frac{\omega^2}{E_{nuc}^2}\right)^2}.
\end{equation}
Since in the DM relaxation always contain the parameter $h\left(\sigma^i_{x},\omega\right)$, we use a Gaussian dependence on the energy difference. It describes a quadratic dependence on $\left(\frac{\omega}{\delta E_{nuc}}\right)$ for finite $\omega$, when $\left(\frac{\omega}{\delta E_{nuc}}\right)^2\ll 1$.

To extract parameters for $h\left(\sigma_z^i,\omega\right)$ at finite energy differences, we can consider double dot experiments. A fluctuating local magnetic field parallel to the static external field causes transitions of the singlet-triplet qubit between the singlet state $S_0$ and the $s_z=0$ triplet state $T_0$. On the relevant subspace it acts like a perpendicular magnetic field:
\begin{equation}
B\sigma_z^i=B\left(\sigma_x\right)_{\left\{S_0,T_0\right\}}.
\end{equation}
In the DM the fluctuating parallel magnetic field involves quantum jumps between the $S_0$ and $T_0$ states at the energy difference of these two levels: $h\left(\sigma_z^i,\omega\right)$. Using the same description as for single quantum dots before, the transition rates exponentially decrease with the energy difference of the levels.

\begin{figure}
\centering
\includegraphics[width=0.49\textwidth]{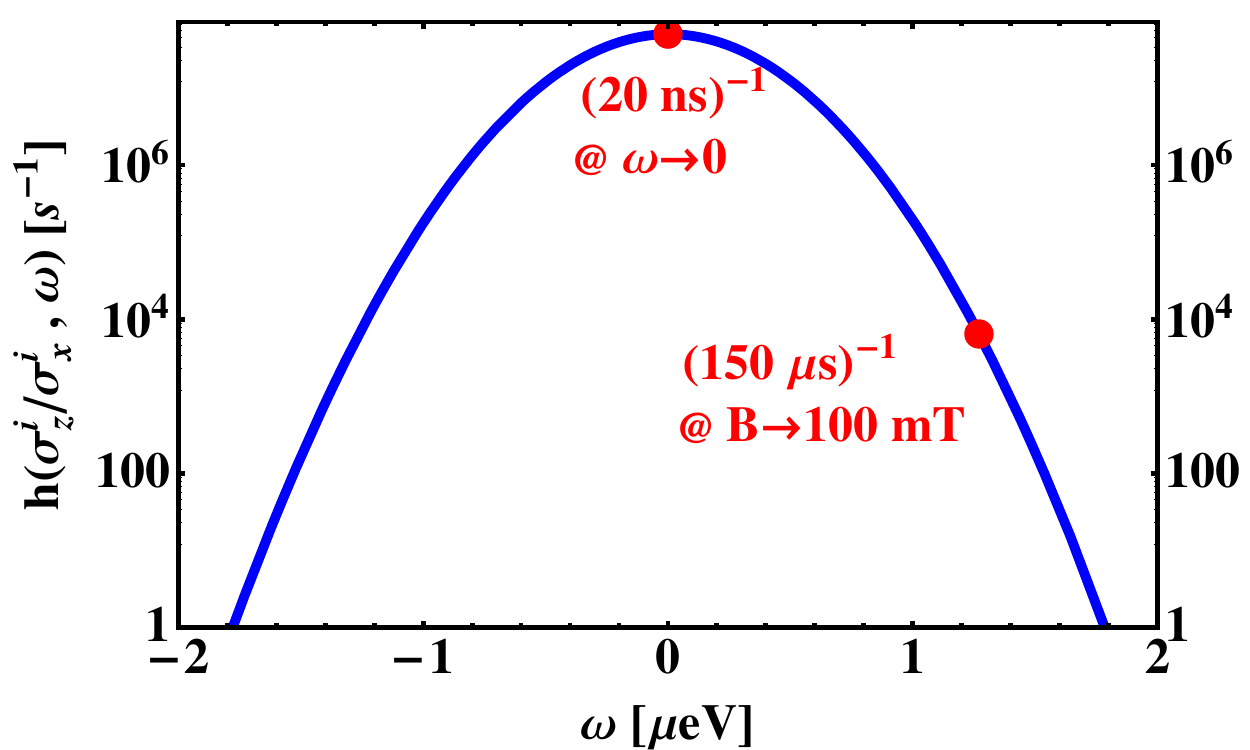}
\caption{\label{fig:Trans1}
Hyperfine interaction causes dephasing and relaxation through local magnetic field fluctuations on each quantum dot. Dephasing and relaxation is generated by the transition rates $h\left(\sigma^i_{z}/ \sigma^i_{x},\omega\right)$ according to \equationshortname~\eqref{eq:ratehyperfine}. The marked points refer to results from two qubit experiments, described in the main text.}
\end{figure}

Transitions between the $s_z=0$ states with the $s_z\ne0$ states are possible through local raising and lowering operators. They are generated through perpendicular magnetic field fluctuations $h\left(\sigma_x^i,\omega\right)$ at the energy difference $\omega$ of the $s_z=0$ and $s_z\ne0$ states. The transition rates are highly sensitive to $\omega$. This result was also observed in experiments on double dot qubits. The transition rates of $\left(20\ \text{ns}\right)^{-1}$ at zero magnetic fields and $\left(150\ \text{ns}\right)^{-1}$ at $100\ \text{mT}$ in our model (cf. \figureshortname~\ref{fig:Trans1}) match approximately these results.\cite{johnson2005} For weakly coupled singlet-triplet qubits the energy difference $\omega$ is determined directly by the external magnetic field.

Interestingly, we automatically describe in the DM the large transition rates at the crossings of levels with different spin quantum numbers. For double quantum dots the $S_0$-$T_+$ crossing is extensively used in experiments.\cite{petta2005} In triple dots we will observe doublet-quadruplet crossings, which were analyzed in the experiment by Gaudreau et al.\cite{gaudreau2012} In experiments a large enhancement of transition rates near these level crossings were observed, with transition rates in the nanosecond regime. These transition rates decrease quickly when going away from the level crossing.

\subsubsection{\label{sssec:RatePhonon}
Interactions with Phonons}

\begin{figure}
\centering
\includegraphics[width=0.49\textwidth]{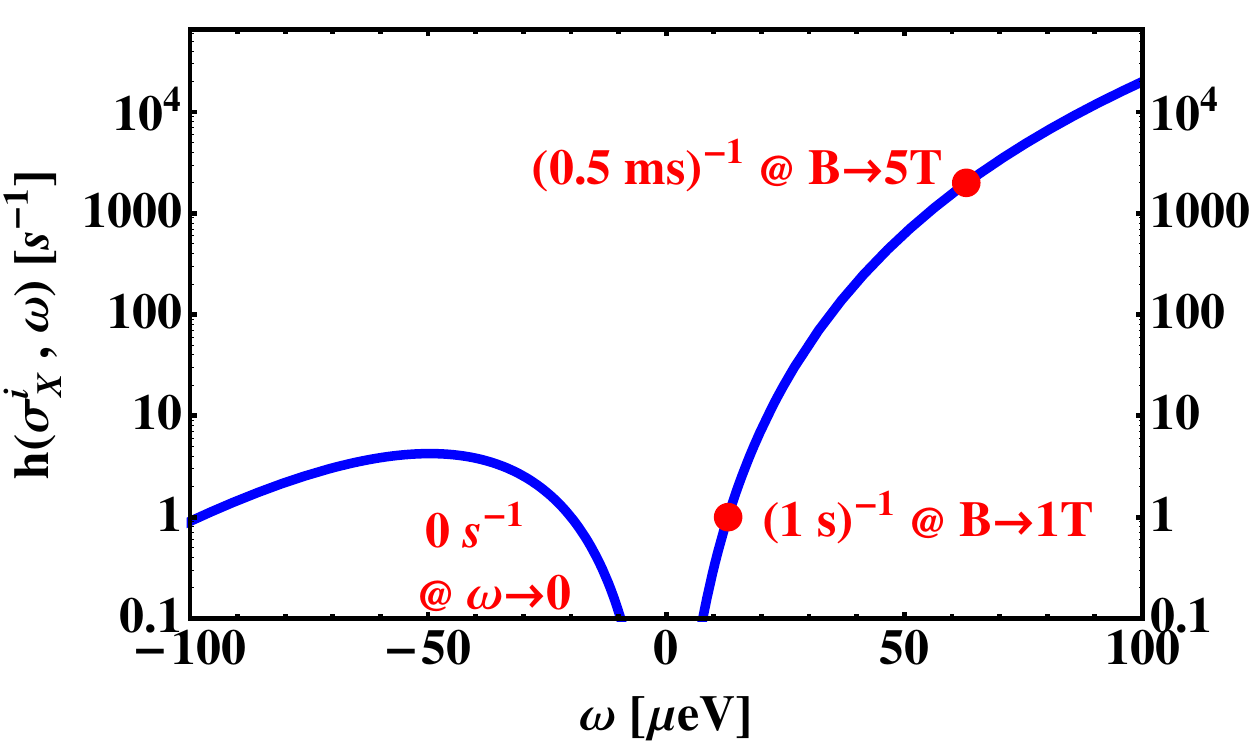}
\caption{\label{fig:Trans2}
Relaxation rates $h\left(\sigma_{x}^{i},\omega\right)$ generated by piezoelectric phonons according to \equationshortname~\eqref{eq:ratephonon}. Experimentally observed time scales from qubit experiments\cite{amasha2008} are included in this picture.}
\end{figure}

Additionally, relaxations are generated by interactions of the quantum dot with electric fields, e.g. from polar phonons. However, direct transitions are forbidden between different spin states of the same orbital level by the dipole selection rule. They must be mediated by another process like spin-orbit interaction. In single quantum dot experiments, relaxation times of about $1\ \text{s}$ at magnetic fields of $1\ \text{T}$ and $0.5$ ms for a magnetic field of $5$ T have been identified.\cite{amasha2008} The scaling law of $\omega ^3 E_z^2$ governs transitions by piezoelectric phonons between Zeeman split quantum dot eigenstates for single qubit experiments. The phonon energy must match the energy difference of the states, resulting in the $\omega^3$ scaling law. The Zeeman split eigenstates are mixed by spin orbit interaction, which causes the $E_z^2$ dependence. Experiments and theory suggest, therefore, relaxation rates modeled as
\begin{equation}
h\left(\sigma_{x}^{i},\omega\right)=\Xi_i^{x}\left|\frac{\omega^3 E_z^2}{1-e^{-\frac{\omega}{T_K}}}\right|.
\label{eq:ratephonon}
\end{equation}
This should also apply for triple dot setups. The coupling operators are again local spin flip operators $\sigma_x^i=\sigma_i^{+}+\sigma_i^{-}$ at quantum dot $i$. A picture of the generated transition rates, including the results from single qubit experiments, is shown in \figureshortname~\ref{fig:Trans2}.

For double dot experiments in the weak coupling regime, transition rates have been identified that are consistent with this picture.\cite{nowack2011} At high bias (readout regime) other effects cause transitions between different charge states.\cite{taylor2007} We do not include these effects in our model since we do not analyze the readout regime ($\epsilon>\epsilon_+$ or $\epsilon<\epsilon_-$). Electron-phonon interactions can also lead to pure spin dephasing,\cite{hu2011,gamble2012} but so far this effects has never been observed experimentally. It should be weaker than dephasing due to hyperfine fields, as described in the previous section.

\begin{table}
\centering
\begin{tabular}{lc c}
\hline
\hline
Mechanism & Constant & Value\\
\hline
Local dephasing and relaxation& \multirow{2}{*}{$\Upsilon_i^{z,x}$} & \multirow{2}{*}{$\frac{1}{20\ \text{ns}}$} \\
through hyperfine fields& &\\ \\
Local spin relaxation& \multirow{2}{*}{$\Xi_i^{x}$} & \multirow{2}{*}{$2\cdot 10^{-6}\frac{1}{\text{s}\ \mu\text{eV}^5}$} \\
through phonons& &\\
\hline
\hline
\end{tabular}
\caption{\label{tab:characteristicparameters}
Transition mechanisms describing noise on the triple quantum dot and estimates of parameters. The different mechanisms are described in the text.}
\end{table}

\subsubsection{\label{sssec:RateAll}
Summary and Description of Noise Regimes}

In the previous analysis we identified two influences of the qubit environment, which should be most important for triple dot experiments. Hyperfine interactions cause large error rates through the random distribution of their magnetic moments. Direct spin flips are generated by phonons, which couple to the spin indirectly via spin orbit interaction. We want to point out that charge fluctuations might be an additional noise term in triple quantum dot experiments. Through the exchange interaction electrostatic interactions are used to manipulate the qubit. Therefore, charge fluctuations will be important. This problem was already pointed out for double quantum dots.\cite{hu2006} Especially for strong electrostatic bias, charge dipoles are created and charge fluctuations will gain influence. Here we study only the weak bias regime and do not take charge fluctuations into account.

When we analyze the qubit, we find that the described noise mechanism is relevant for distinguishable parameter sets (see \tableshortname~\ref{tab:characteristicparameters}). A summary of the three most important regimes is given in \tableshortname~\ref{tab:NoiseRegimes}. For phase noise we describe only the influence of hyperfine interaction through the transition rates $h\left(\sigma_z^i,\omega\right)$ from \equationshortname~\eqref{eq:ratehyperfine}. This interaction causes major errors for energy differences smaller and in the range of the rms of the hyperfine energy (cf. \figureshortname~\ref{fig:Trans1}). This regime is the most critical one for experiments in semiconducting spin qubits. We call it ``Regime 1'' in the following.

\begin{figure}
\centering
\includegraphics[width=0.49\textwidth]{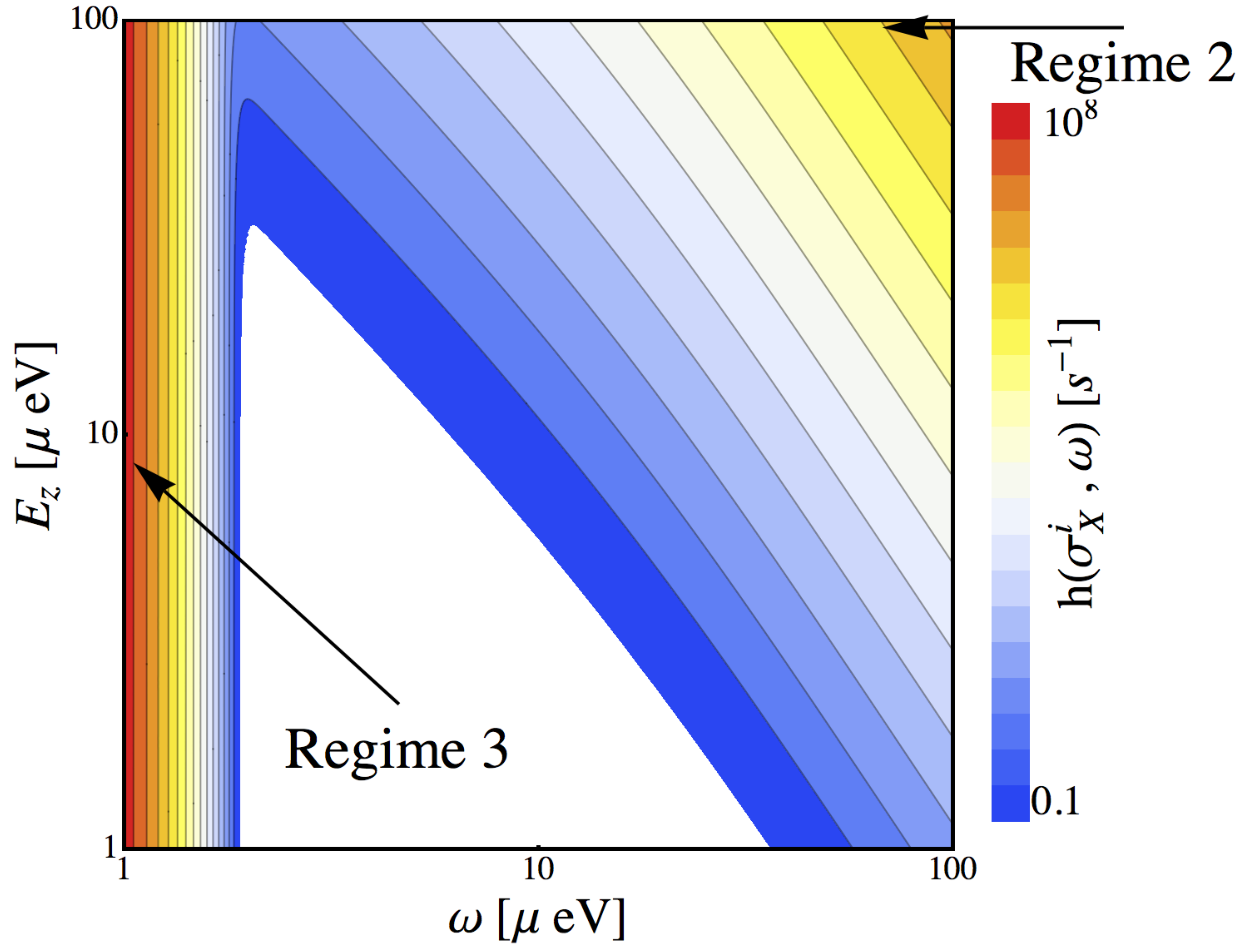}
\caption{\label{fig:NoiseRegimes}
Density plot of the transition rates for local spin relaxation $h\left(\sigma_x^i,\omega\right)$ as a function of the energy difference $\omega$ and the external magnetic field $E_z$. The influence of hyperfine interaction (Regime 2) and phonon interaction (Regime 3) can be separated due to the different parameter ranges when the effects are dominant. The hyperfine mechanism is independent of $E_z$.}
\end{figure}

Local spin relaxation can cause large transition rates in two completely different regimes (see transition rates in \figureshortname~\ref{fig:NoiseRegimes}). For large energy differences and external magnetic fields the interaction with phonons will be dominant, while at small energy differences the hyperfine interaction will determine the relaxation process. Since these two effects are important in different parameter ranges, we can easily separate their influence. We call the two parameter ranges ``Regime 2'' and ``Regime 3'', respectively.

\begin{table}
\centering
\begin{tabular}{c l}
\hline
\hline
\multirow{3}{*}{Regime 1} & Local phase noise $h\left(\sigma_z^i,\omega\right)$ generated from\\&
fluctuating hyperfine fields. Strong influence\\& at small energy differences.\\ \\
\multirow{3}{*}{Regime 2} & Local spin relaxation $h\left(\sigma_x^i,\omega\right)$ from the interaction\\
& with phonons. Strong influence at large energy\\& differences and large external magnetic fields.\\ \\
\multirow{4}{*}{Regime 3} & Local spin relaxation $h\left(\sigma_x^i,\omega\right)$ from the interaction\\&
with hyperfine fields. Dominant relaxation\\&
mechanism at small energy differences, independent\\&
of the external magnetic field.\\
\hline
\hline
\end{tabular}
\caption{\label{tab:NoiseRegimes}
Dominant noise regimes for qubit experiments.}
\end{table}

\section{\label{sec:FullTime}
Analysis of Time Evolution}

To analyze the realization of qubits in a subspace and a subsystem of triple quantum dots, we discuss the time in the noisy environment introduced in \autoref{ssec:Rates}. We are be able to show that the early time evolution on the interval  $\left[0,\delta t\right]=\left[0,10\right]\ \text{ns}$ (see choice of parameter in \autoref{ssec:Parameter}) can be described with different means as the long time evolution of the qubit. The technical details of the description of the initial and the long time evolution are summarized in Appendices \ref{sec:Init} and \ref{sec:LongTime}.

We numerically calculate the time evolution for a subspace and subsystem qubit on the full eight-dimensional Hilbert space. Since we are only interested in the noisy part of the evolution, we solve the full master equation in the rotating frame with respect to the ideal Hamiltonian from \equationshortname~\eqref{eq:Hamiltonian}:
\begin{equation}
	\dot{\rho}\left(t\right)=\mathcal{L}^{rot}\rho\left(t\right).
	\label{eq:MasterEqFinal}
\end{equation}
The Lindbladian $\mathcal{L}^{rot}$ is given by $\mathcal{L}_D$ of the DM from \equationshortname~\eqref{eq:DaviesModel}, as described in Appendix \ref{ssec:SimpRot}.

From the time evolution of the full density matrix $\rho\left(t\right)$, we calculate the time evolution of the qubit's population $O\left(t\right)=Tr\left(\mathcal{P}\rho\left(t\right)\right)$. $\mathcal{P}$ is a linear map, which constructs from $\rho\left(t\right)$ only the relevant part describing the qubit (see Appendix \ref{sec:Init}). Additionally, we extract the trajectory on the Bloch sphere $\mathbf{P}\left(t\right)=\left(X\left(t\right),Y\left(t\right),Z\left(t\right)\right)$, with $P_i\left(t\right)=Tr\left(\sigma_i\mathcal{P}\rho\left(t\right)\right)$. For the subspace qubit, we use the map $\mathcal{P}_P$ from \equationshortname~\eqref{eq:ProjMap}. It projects from the full eight-dimensional Hilbert space on the two-dimensional Hilbert space defining the subspace qubit. To describe the subsystem qubit, we use the combination $\mathcal{P}=\mathcal{P}_S\mathcal{P}_P$. $\mathcal{P}_P$ is a projective map, similar to the construction of the subspace qubit. For the subsystem qubit it projects, however, on the four-dimensional subspace $span\left\{\Delta_{\frac{1}{2}},\Delta^\prime_{\frac{1}{2}},\Delta_{-\frac{1}{2}},\Delta^\prime_{-\frac{1}{2}}\right\}$ (cf. description in Appendix \ref{ssec:InitSubsystem}). $\mathcal{P}_S$ is defined in \equationshortname~\eqref{eq:MapSubsystem}.

Our aim is to show that we can describe the initial time evolution closely by an effective description derived in Appendix \ref{sec:Init}. We use a Nakajima-Zwanzig approach, where the full details of the derivation are given in the appendices. We compare the numerical solution of the full master equation with the solution of this effective description.  For the subspace qubit we use \equationshortname~\eqref{eq:NZBornSubspace}; for the subsystem qubit we use \equationshortname~\eqref{eq:NZBornSubsys}. We solve both Nakajima-Zwanzig equations numerically. We keep noise terms in first (second) order for the first (second)-order Born approximation.

We discuss the subspace and subsystem qubit separately. Even though we model the time evolution quite generally, it will turn out that the environment influences the qubit evolution dominantly through transition rates from three different noise regimes. We have specified them already in \autoref{sssec:RateAll} and refer to them in the following (cf. \tableshortname~\ref{tab:NoiseRegimes}).

\subsection{\label{ssec:FTSubspace}
Subspace Qubit}

The subspace qubit is defined on the subspace $span\left\{\Delta_{\frac{1}{2}},\Delta^\prime_{\frac{1}{2}}\right\}$ (see description in \autoref{ssec:Hamiltonian}). We discuss it using all parameters from \autoref{ssec:Parameter}. It constitutes a decoherence-free subspace with respect to global phase noise $\mathcal{D}\left[Z\right]$ (weak collective decoherence; cf. \autoref{sec:Intro}). Other global noise (through the action of the Lindblad operators $\mathcal{D}\left[X\right]$ and $\mathcal{D}\left[Y\right]$, with $X,Y\equiv \sum_{i=1,2,3}\sigma_{x,y}^i$) will, however, lead to leakage from the computational subspace. This leakage will separately bring the $l=1$ and $l=0$ states to thermal equilibrium. However, since the subspace qubit can be operated only in the limit of large external magnetic fields, the resulting leakage will be negligible. For $E_z=100\ \mu\text{eV}$ and $T_K=10\ \mu\text{eV}$ only a fraction $e^{-\frac{Ez}{T_K}}/\left(1+e^{-\frac{Ez}{T_K}}\right) \approx 4.5\cdot 10^{-5}$ of the probability will leak out of the computational subspace. In any case, we argued in \autoref{ssec:Rates} that the external environment couples dominantly through local interactions.

\subsubsection{\label{sssec:FTR1}
Regime 1}

Local phase noise turns out to be critical for the subspace qubit. We discuss in the following an unbiased triple dot ($\epsilon=0$). Additionally, we include phase noise generated from fluctuating hyperfine fields [cf. $h\left(\sigma_z^i,\omega\right)$ from \equationshortname~\eqref{eq:ratehyperfine}]. We include phase noise on all three quantum dots. The parameters $\Upsilon_1^z=\left(20\ \text{ns}^{-1}\right)$, $\Upsilon_2^z=\left(30\ \text{ns}^{-1}\right)$, and $\Upsilon_3^z=\left(40\ \text{ns}^{-1}\right)$ describe the common noise strength for fluctuating hyperfine fields (see \tableshortname~\ref{tab:characteristicparameters}). Local phase noise generates large transition rates only at small energy differences (Regime 1, cf. \tableshortname~\ref{tab:NoiseRegimes}).

A qubit initialized in the $s_z=\frac{1}{2}$ subspace will only undergo transitions within this subspace. Since the energy differences of all three energy levels are comparable to the thermal energy [see \equationshortname~\eqref{eq:Energy1}-\eqref{eq:Energy3}], a considerable part of the population leaks into the $Q_{\frac{1}{2}}$ subspace. The finite energy difference of the two qubit levels leads in the long time limit to a finite polarization on the $z$ axis: $Z_\infty$.

\begin{figure}
\subfigure[$\mathbf{P}\left(0\right)=\left(0,0,1\right)$]{\label{fig:FT1-1}\includegraphics[width=0.49\textwidth]{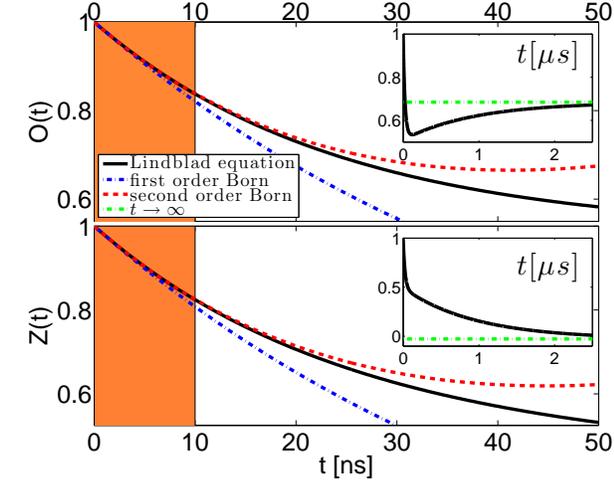}}
\subfigure[$\mathbf{P}\left(0\right)=\left(0,1,0\right)$]{\label{fig:FT1-2}\includegraphics[width=0.49\textwidth]{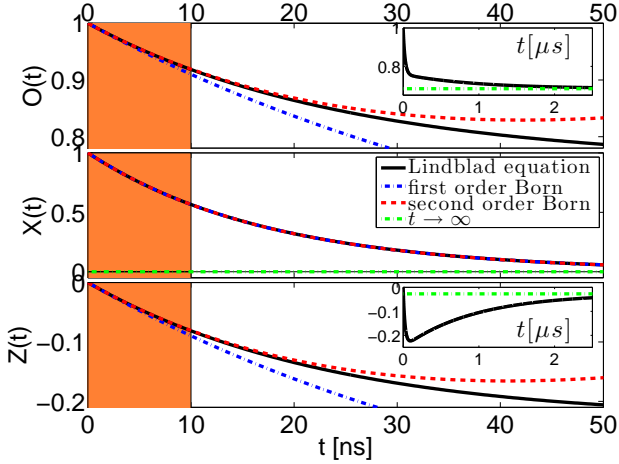}}
\caption{\label{fig:FullTime1}
Time evolution of the subspace qubit without electric bias ($\epsilon=0\ \mu\text{eV}$) at $E_z=100\ \mu\text{eV}$. The qubit is subjected to local phase noise generated by fluctuating hyperfine fields [$\Upsilon_1^z=\left(20\ \text{ns}^{-1}\right)$, $\Upsilon_2^z=\left(30\ \text{ns}^{-1}\right)$, $\Upsilon_3^z=\left(40\ \text{ns}^{-1}\right)$]. The orange region marks a typical time interval for qubit experiments: $\left[0,10\right]\ \text{ns}$. Blue lines represent the results from the first order Born approximation, while red lines are calculated in the second-order Born approximation. The insets show the evolution on a longer time scale. Green lines represent the long time limit of the evolution as discussed in the text.}
\end{figure}

We show in \figureshortname~\ref{fig:FullTime1} the time evolution for the unbiased subspace qubit ($\epsilon=0$). It is initially in a pure excited state [$\mathbf{P}\left(0\right)=\left(0,0,1\right)$, see \figureshortname~\ref{fig:FT1-1}] or in a superposition of $\Delta_\frac{1}{2}$ and $\Delta^\prime_\frac{1}{2}$ [see \figureshortname~\ref{fig:FT1-2}]. For the initial time evolution (orange interval) the Born approximation in \equationshortname~\eqref{eq:NZBornSubspace} gives a highly accurate description. One can see that the thermalization happens on a microsecond time scale. Simple arguments can be used to predict the long time limit of the time evolution. The arguments in Appendix \ref{sec:LongTime} indicate $O_\infty\approx 0.66$ and $\mathbf{P}_\infty\approx \left(0, 0, -0.03\right)$. The small value of the polarization in the long time limit ($Z_\infty\approx-0.03$) reflects that the thermal energy $T_K$ is large compared to the energy splitting of the qubit levels [$T_K=10\ \mu\text{eV}$; cf. energy diagram in \figureshortname~\ref{fig:EnergyDiag1}].

\subsubsection{\label{sssec:FTR2}
Regime 2}

\begin{figure}
\subfigure[$\mathbf{P}\left(0\right)=\left(0,0,1\right)$]{\label{fig:FT2-1}\includegraphics[width=0.49\textwidth]{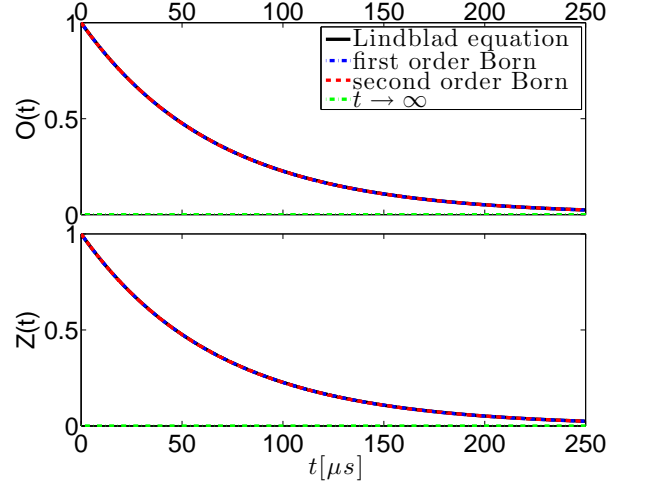}}
\subfigure[$\mathbf{P}\left(0\right)=\left(0,1,0\right)$]{\label{fig:FT2-2}\includegraphics[width=0.49\textwidth]{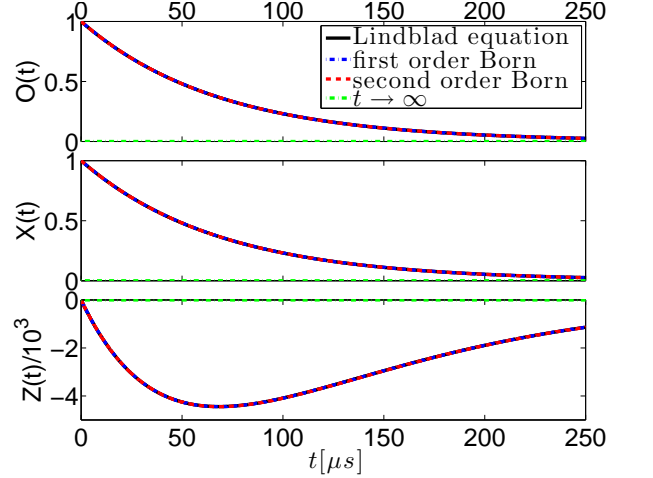}}
\caption{\label{fig:FullTime2}
Time evolution from local spin relaxation for the unbiased subspace qubit generated by the interaction with piezoelectric phonons for $\epsilon=0$ and $E_z=100\ \mu\text{eV}$. The transition parameters are chosen to $\Xi^x=2\cdot 10^{-6}\frac{1}{s\ \mu\text{eV}^5}$, $\Xi^x=1.5\cdot 10^{-6}\frac{1}{s\ \mu\text{eV}^5}$, and $\Xi_3^x=1\cdot 10^{-6}\frac{1}{s\ \mu\text{eV}^5}$. The numerical solution of the full master equation (black lines) and the first-order (blue lines) and second-order (red lines) Born approximations are identical. The qubit depopulates completely in the long time limit.}
\end{figure}

Local spin relaxation of the subspace qubit at high magnetic fields will predominantly depopulate the qubit by the transition from the qubit eigenstates $\Delta_{\frac{1}{2}}$ and $\Delta^\prime_{\frac{1}{2}}$ to states with different $s_z$-quantum numbers. We simulate the time evolution for the unbiased subspace qubit $\left(\epsilon=0\right)$ at $E_z=100\ \mu\text{eV}$ (see \figureshortname~\ref{fig:FullTime2}). Large transition rates are generated from the interaction with piezoelectric phonons. These rates are highly enhanced for high energy differences and large external magnetic fields (Regime 2, cf. \tableshortname~\ref{tab:NoiseRegimes}). We use the following transition parameters: $\Xi^x=2\cdot 10^{-6}\frac{1}{s\ \mu\text{eV}^5}$, $\Xi^x=1.5\cdot 10^{-6}\frac{1}{s\ \mu\text{eV}^5}$, and $\Xi_3^x=1\cdot 10^{-6}\frac{1}{s\ \mu\text{eV}^5}$ (cf. \tableshortname~\ref{tab:characteristicparameters}). At low electric bias we only see phonon-generated transitions to the $Q_\frac{3}{2}$ state.

Local spin relaxation from fluctuating hyperfine fields can be neglected. This mechanism is only dominant for small energy differences (Regime 3; cf. \tableshortname~\ref{tab:NoiseRegimes}). For large external magnetic fields, the unbiased subspace qubit does not have a state of different $s_z$-quantum number, which is close to the qubit states (cf. \figureshortname~\ref{fig:EnergyDiagram}).

It can be seen that the transitions empty the qubit's population completely, but the time evolution is very slow. Overall, phonons generate considerable changes only on microsecond time scales. The first order Born approximation is already sufficient for the description of the qubit dynamics.

\subsection{\label{ssec:FTSubsys}
Subsystem Qubit}

A subsystem qubit is implemented in the formally introduced $l$-quantum number. We define the $\Delta_{\pm 1/2}$ and $\Delta^\prime_{\pm 1/2}$ states as the qubit levels (see description in \autoref{ssec:Hamiltonian}). It operates in a decoherence-free subsystem to all interactions acting globally on its defining system (strong collective decoherence; cf. \autoref{sec:Intro}). All global noise mechanisms will hence be irrelevant. In any case, we argued in \autoref{ssec:Noise} that local interactions dominate the noise properties in triple quantum dot experiments.

Local phase noise will have the same effect on the subsystem qubit and the subspace qubit. As described in the symmetry discussion in Appendix \ref{ssec:SimpPhase}, phase noise will act separately on the $s_z=\pm\frac{1}{2}$ subspaces. It will, however, never mix them. Additionally, the action is the same on both subspaces. Since the transition rates from the interactions with nuclear spins are assumed to be independent of the magnetic field strength [cf. \equationshortname~\eqref{eq:ratehyperfine}], the time evolution is also identical. Large transition rates are generated for energy levels that are close in energy (Regime 1; cf. \tableshortname~\ref{tab:NoiseRegimes}).

For local spin relaxations the description of the subsystem qubit is comparable to the subspace qubit for large external magnetic fields. We can especially see that the $s_z$ distribution \eqref{eq:subsysbath} is close to a pure $s_z=+\frac{1}{2}$ state.

\subsubsection{\label{sssec:FTR3}
Regime 3}

Local spin relaxation will gain in importance for small magnetic fields. In the low-bias regime (i.e., small $\left|\epsilon\right|$) relaxation rates generated from the interaction with phonons are small. We can see this by inspecting the transition rates $h\left(\sigma_x^i,\omega\right)$ for small magnetic fields $E_z$ [cf. \equationshortname~\eqref{eq:ratephonon}]. On the other hand, fluctuating hyperfine fields can strongly mix different $s_z$ states. Especially at the points of level crossings this effect will be critical. The relaxation effects through nuclear magnetic fields are highly enhanced at small energy differences [see transition rates $h\left(\sigma_x,\omega\right)$ in \equationshortname~\eqref{eq:ratehyperfine}]. This dominant noise mechanism is summarized in Regime 3 of \tableshortname~\ref{tab:NoiseRegimes}.

\begin{figure}
\subfigure[$\mathbf{P}\left(0\right)=\left(0,0,1\right)$]{\label{fig:FT3-1}\includegraphics[width=0.49\textwidth]{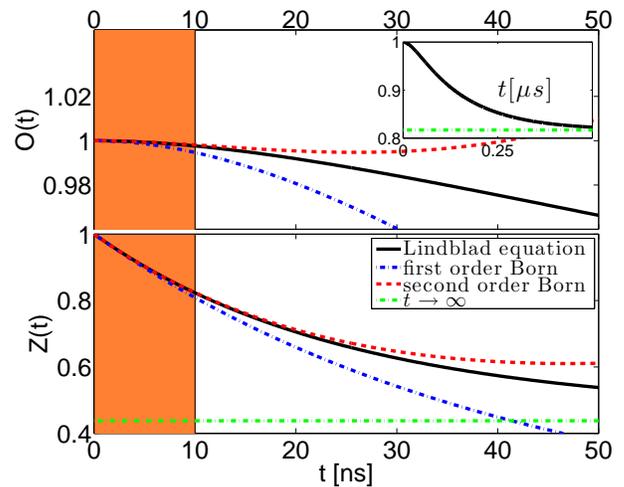}}
\subfigure[$\mathbf{P}\left(0\right)=\left(0,1,0\right)$]{\label{fig:FT3-2}\includegraphics[width=0.49\textwidth]{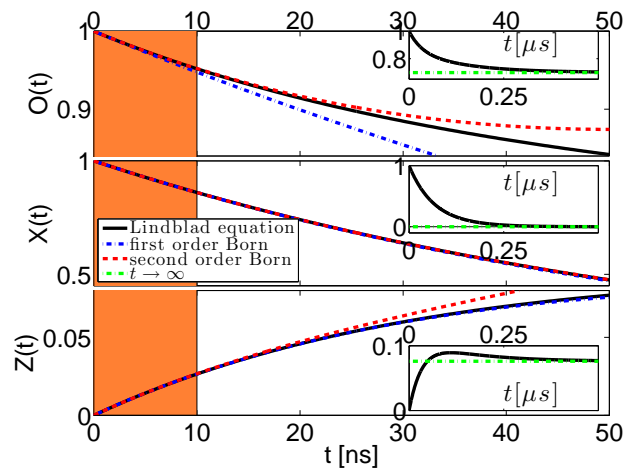}}
\caption{\label{fig:FullTime3}
Time evolution of the subsystem qubit at $E_z=2.5\ \mu\text{eV}$ through local spin relaxations at $\epsilon=354.6\ \mu\text{eV}$ [compare energy diagram in \figureshortname~\ref{fig:EnergyDiag2}]. The major influence is from local spin relaxation, generated from hyperfine interactions. We use $\Upsilon_1^x=\left(20\ \text{ns}^{-1}\right)$, $\Upsilon_2^x=\left(30\ \text{ns}^{-1}\right)$ and $\Upsilon_3^x=\left(40\ \text{ns}^{-1}\right)$. The orange region marks a typical time scale of qubit experiments. Here the second-order Born approximation and the results from the full master equation match very closely. The insets show the long time evolution of the qubit. We see thermalization occurring within microseconds.}
\end{figure}

We simulate the qubit evolution at $\epsilon=354.6\ \mu\text{eV}$ and $E_z=2.5\ \mu\text{eV}$. Here two different doublet levels and also doublet and quadruplet levels are close in energy [see the orange line in \figureshortname~\ref{fig:EnergyDiag2}]. We simulate the time evolution in the rotating frame [cf. \equationshortname~\eqref{eq:MasterEqFinal} with $\mathcal{L}^{rot}=\mathcal{L}_D$ from \equationshortname~\eqref{eq:DaviesModel}] and compare it with the results from the Born approximation in \equationshortname~\eqref{eq:NZBornSubsys}. The results are shown in \figureshortname~\ref{fig:FullTime3}. We use following transition parameters to model the hyperfine interaction: $\Upsilon_1^x=\left(20\ \text{ns}^{-1}\right)$, $\Upsilon_2^x=\left(30\ \text{ns}^{-1}\right)$, $\Upsilon_3^x=\left(40\ \text{ns}^{-1}\right)$ (cf. \tableshortname~\ref{tab:characteristicparameters}).

Local spin relaxation, generated from fluctuating hyperfine fields, mixes especially two subspaces. First of all, the subspace $span\left\{Q_{1/2},\Delta_{1/2},\Delta^\prime_{-1/2}\right\}$ is mixed strongly. Second, also transitions between the levels $\Delta_{1/2}^\prime$ and $Q_{3/2}$ are strong. In the long time dynamics, we see the thermalization of both subspaces within microseconds. One can calculate the final occupation of the computational subspace from the initial density matrix (see description in Appendix \ref{sec:LongTime}). For the time evolution of the excited state, we calculate $O_\infty\approx 0.82$ and $\mathbf{P}_\infty\approx \left(0, 0, 0.44\right)$ [cf. \figureshortname~\ref{fig:FT3-1}]. A superposition of $\Delta_\frac{1}{2}$ and $\Delta^\prime_\frac{1}{2}$ evolves to $O_\infty\approx 0.70$ and $\mathbf{P}_\infty\approx \left(0, 0, 0.08\right)$ [cf. \figureshortname~\ref{fig:FT3-2}]. It is again seen explicitly that the long time evolution can be separated from the initial time evolution (orange region). The initial time evolution on the interval $\left[0,10\right]\ \text{ns}$ can be described accurately by the second-order Born approximation.

\section{\label{sec:Errors}
Effective Errors}

Our main goal of the current analysis is to extract and quantify the coherence properties of the subspace and subsystem implementation in triple quantum dots. In qubit experiments one is commonly interested in the time evolution of the qubit on short time scales, not on the equilibration properties of the system. We saw in the previous section that we can develop an effective description of the initial time evolution, as derived in Appendix \ref{sec:Init}. We use this description to extract errors for the initial time evolution. An introduction in the error analysis of the single qubit time evolution is given in Appendix \ref{sec:EffRates}.

We numerically simulate the time evolution of the qubit for different initial density matrices $\rho\left(0\right)$. We label the initial density matrix with the corresponding point $\mathbf{P}\left(0\right)=Tr\left(\bm{\sigma} \rho\left(0\right)\right)$ on the Bloch sphere, with $\bm{\sigma}=\left(\sigma_x,\sigma_y,\sigma_z\right)$. All analysis is done in the rotating frame with respect to $\mathcal{H}$ from \equationshortname~\eqref{eq:Hamiltonian} (cf. Appendix \ref{ssec:SimpRot}).  We extract errors from simulations of the full master equation (cf. \equationshortname~\eqref{eq:MasterEqFinal} with $\mathcal{L}^{rot}=\mathcal{L}_D$ from \equationshortname~\eqref{eq:DaviesModel}) and compare them with the results of the corresponding Nakajima-Zwanzig equation in second-order Born approximation (\equationshortname~\eqref{eq:NZBornSubspace} for the subspace qubit and \equationshortname~\eqref{eq:NZBornSubsys} for the subsystem qubit). We use the parameters of \autoref{ssec:Parameter}. We find that the result of both descriptions is equivalent. We saw in the previous section, that on the interval $\left[0,\delta t\right]$ the time evolutions of these equations match very closely.

Following our discussion of the qubit's time evolution in Appendix \ref{ssec:SimpRate}, we need only 7 parameters to describe the time evolution of the qubit. As a consequence of the analyzed error model, as well as from the analysis in the DM, the path of the trajectory is very restricted. We derive the relaxation rates $\Gamma_{0,1,2}^{P_i}$ at $\mathbf{P}_1=\left(0,0,1\right)$, $\mathbf{P}_2=\left(1,0,0\right)$ and $\mathbf{P}_3=\left(0,0,-1\right)$ (cf. Appendix \ref{sec:EffRates}). We use in the following also the terminology ``upper pole'', ``equator'' and ``lower pole'' for these three points. The trajectory starting at the upper and the lower poles will be restricted to the $z$ axis, which sets $\Gamma_{2}^{\mathbf{P}_{1},\mathbf{P}_{3}}=0$ (cf. Appendix \ref{ssec:SimpRate}).

To quantify the influence of noise on the qubit, we need to compare the error rates $\Gamma_i^{\mathbf{P}_j}$ at $\mathbf{P}_j$, as defined in Appendix \ref{sec:EffRates}, to the time of the experiment. The product $\left(\Gamma_{i}^{\mathbf{P}_j}\delta t\right)$ describes an error, which measures the leakage probability $\left(\Gamma_0^{\mathbf{P}_j}\delta t\right)$, the relaxation probability $\left(\Gamma_{1}^{\mathbf{P}_j}\delta t\right)$ and the dephasing probability $\left(\Gamma_{2}^{\mathbf{P}_j}\delta t\right)$. The entanglement fidelity $F_e$ from \equationshortname~\eqref{eq:entfid} describes the effect of noise for all initial density matrices. We use the deviation of the entanglement fidelity from its ideal value, $1-F_e$, as a measure to quantify the overall error.

We find that all error rates can be described by only four toy models, introduced in Appendix \ref{sec:MS}. For these models we can calculate the time evolution analytically. We can describe with the extracted errors the error probabilities of the triple dot qubit.

\subsection{\label{ssec:ErrSubspace}
Subspace Qubit}

\subsubsection{\label{sssec:ErrR1}
Regime 1}

\begin{figure*}
\subfigure[$\Upsilon_1^z=\left(20\ \text{ns}\right)^{-1}$, $\Upsilon_2^z=\Upsilon_3^z=0$]{\label{fig:subspacephase1}\includegraphics[width=.75\textwidth]{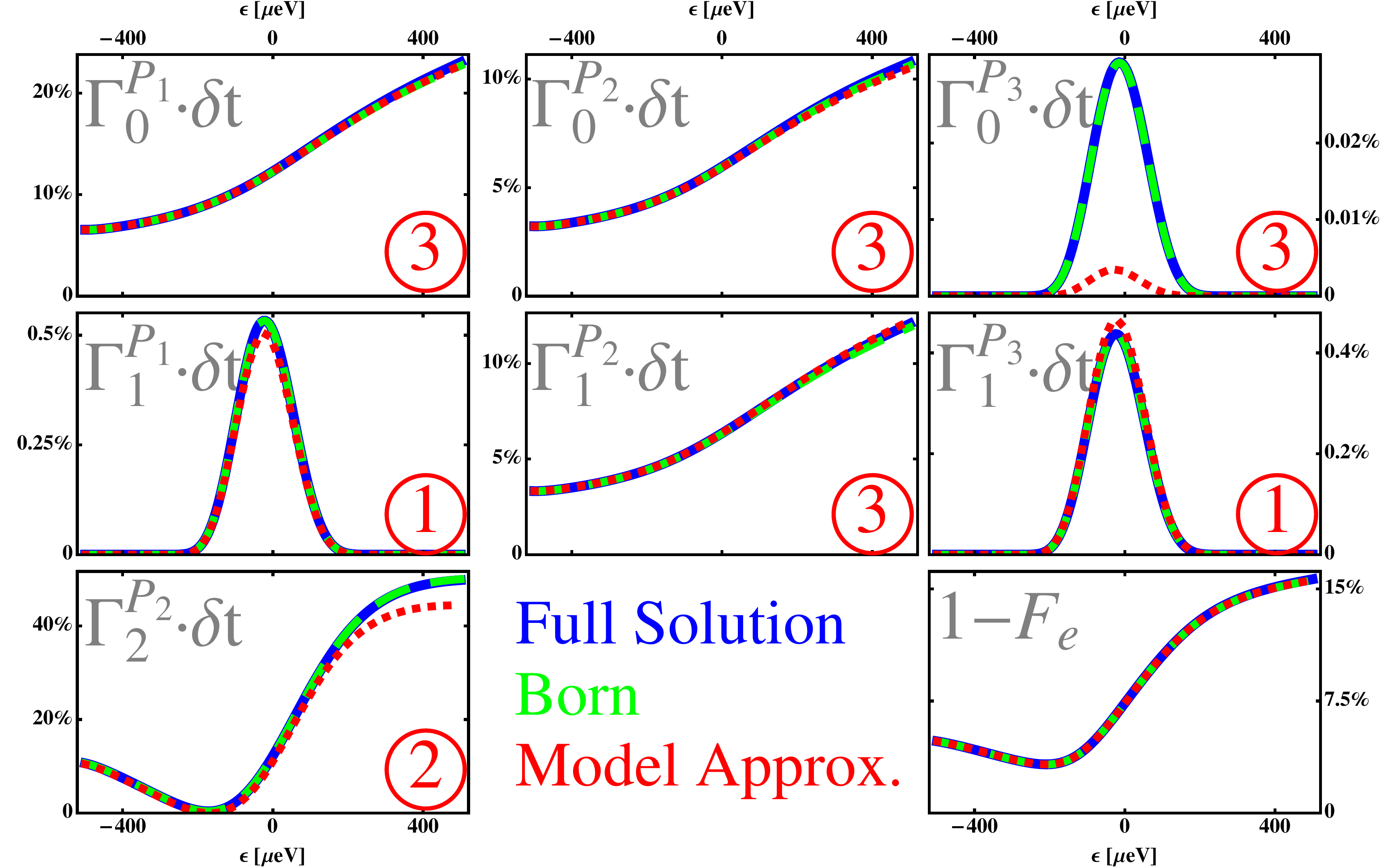}}
\subfigure[$\Upsilon_2^z=\left(20\ \text{ns}\right)^{-1}$, $\Upsilon_1^z=\Upsilon_3^z=0$]{\label{fig:subspacephase2}\includegraphics[width=.75\textwidth]{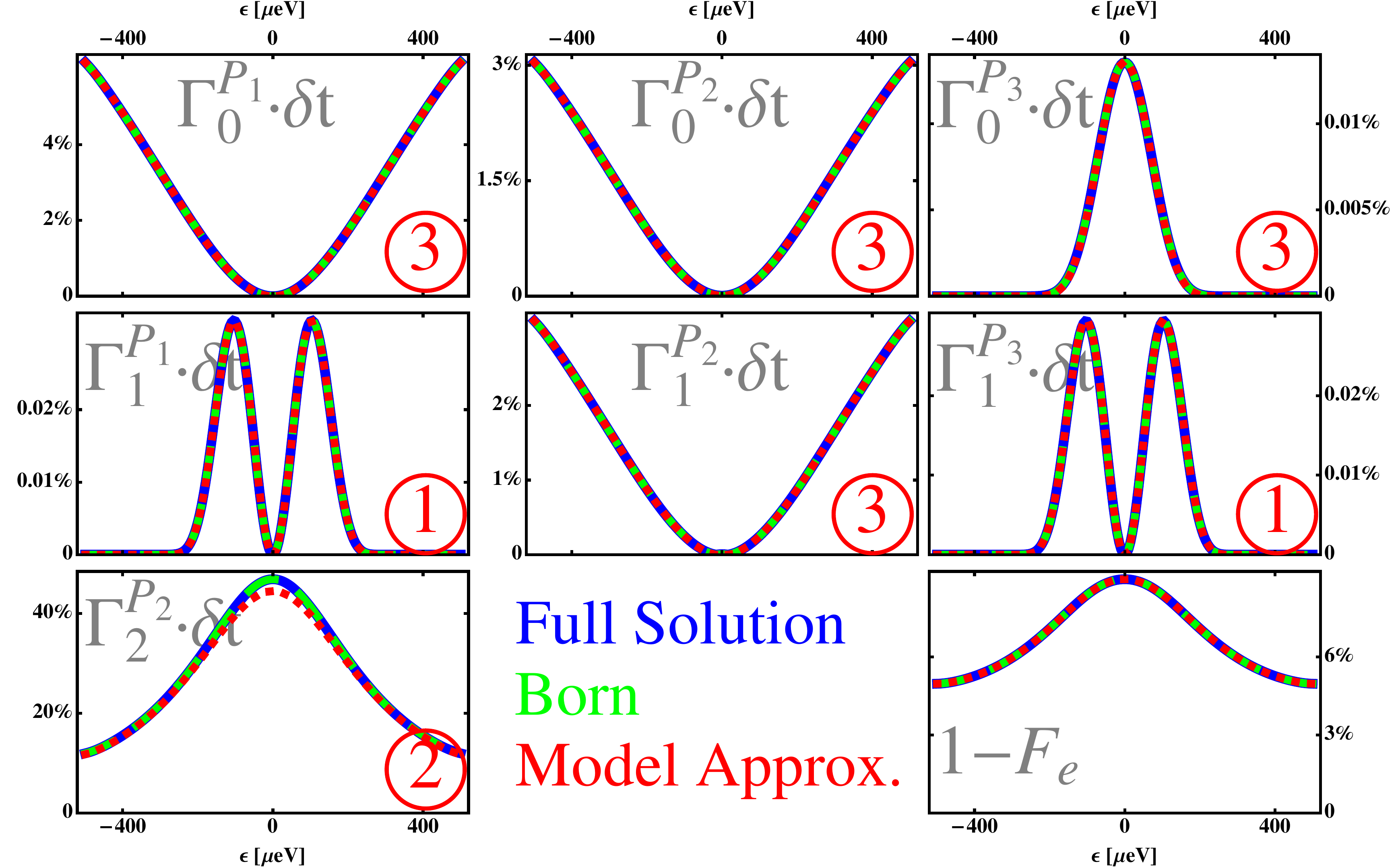}}
\caption{\label{fig:subspacephase}
Errors from local phase noise on the subspace qubit at $E_z=100\ \mu\text{eV}$. We take into account the influence of fluctuating hyperfine fields through the transition rates from \equationshortname~\eqref{eq:ratehyperfine}. The effective error rates are extracted from the numerical simulation of the full master equation (blue lines) and of the second-order Born approximation (green lines). Red lines describe effective errors from simple toy models, as described in the text. For each extracted error rate we add with a red number the specific model system under consideration from Appendix \ref{sec:MS}.}
\end{figure*}

First, we analyze the subspace qubit in the $\left(1,1,1\right)$ regime (small $\epsilon$) with local phase noise from the interaction with nuclear magnetic fields. As discussed in \autoref{sssec:FTR1}, phase noise generates large error rates only for small energy differences (Regime 1 in \tableshortname~\ref{tab:NoiseRegimes}). Simulations for local phase noise are shown in \figureshortname~\ref{fig:subspacephase}. We analyze phase noise on dot 1 and dot 2 separately and use $\Upsilon_{1,2}^z=\left(20\ \text{ns}\right)^{-1}$ (cf. \tableshortname~\ref{tab:characteristicparameters}).

\begin{figure}
\centering
\includegraphics[width=.49\textwidth]{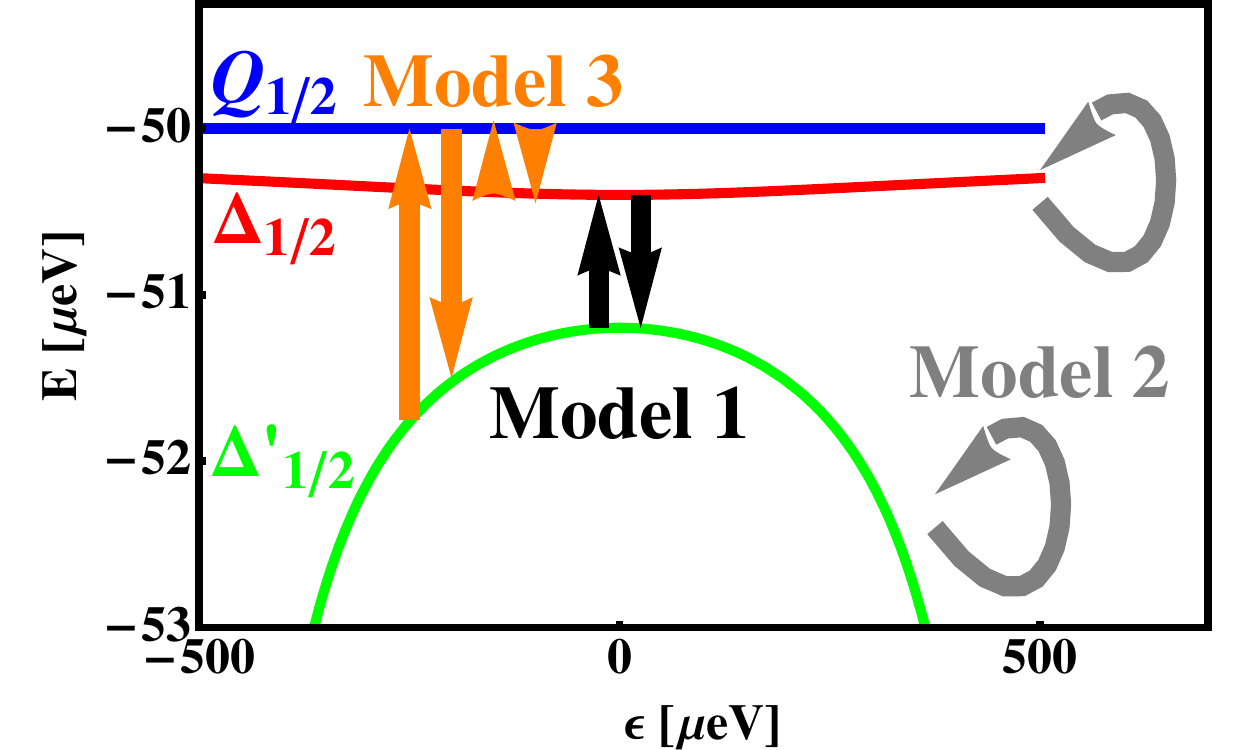}
\caption{\label{fig:TransPict1}
Energy diagram of the subspace qubit with major transition rates generated from local phase noise. The energy diagram shows only the relevant energy levels for this description [cf. \figureshortname~\ref{fig:EnergyDiag1}]. The transition rates can be grouped into three sets, which are described by the toy models analyzed in Appendix \ref{sec:MS}. Model 1 describes pure relaxation of the qubit states (black arrows; see Appendix \ref{ssec:MS1}), model 2 describes pure dephasing of the qubit states (gray arrows; see Appendix \ref{ssec:MS2}). Model 3 characterizes leakage of the qubit states to one state in the surroundings (orange arrows; see Appendix \ref{ssec:MS3}).}
\end{figure}

The effective error rates can be understood when analyzing the transition behavior in the DM. Quantum jump operations between the energy levels are only possible for the same $s_z$ eigenstates. Since the interactions are local, the total spin quantum number $S$ is not preserved. Initially, only the $s_z=\frac{1}{2}$ doublet levels are occupied. Phase noise will mix within the $s_z=\frac{1}{2}$ subspace (see sketch in \figureshortname~\ref{fig:TransPict1}). One should notice that all energy differences in the $s_z=\frac{1}{2}$ subspace are small or comparable to the thermal energy. This makes transition rates for positive and negative energy differences similar. We can group all transition rates into three sets; each set corresponds to the error processes of a toy models from Appendix \ref{sec:MS}. The transition rates of these toy models match to a high degree the results of the numerical solution of the full master equation, and of the calculation using the second-order Born approximation.

We want to discuss the results from \figureshortname~\ref{fig:subspacephase} in detail and start with an analysis of local phase noise on the first quantum dot [cf. \figureshortname~\ref{fig:subspacephase1}]. Model 1 of pure relaxation will mainly determine the $T_1$ behavior of the qubit  (compare model 1 in Appendix \ref{ssec:MS1}). The interaction with hyperfine fields generates direct transitions between the qubit levels. These rates are large if the states are close in energy [compare $h\left(\sigma_z^i,\omega\right)$ in \equationshortname~\eqref{eq:ratehyperfine}]. Hence, the error rates $\Gamma_{1}^{\mathbf{P}_1}$ and $\Gamma_{1}^{\mathbf{P}_3}$ vanish quickly when increasing the bias on the dots. Only a large difference of the transition rate from the excited qubit level to the ground state compared to the reversed effect would cause large error rates $\Gamma_{1}^{\mathbf{P}_2}$ and $\Gamma_{2}^{\mathbf{P}_2}$ in model 1. Here both rates are either very similar (for small $\epsilon$) or both small (for finite $\epsilon$). $\Gamma_{1}^{\mathbf{P}_2}$ and $\Gamma_{2}^{\mathbf{P}_2}$ are therefore not described in the model of pure relaxation.\\
Transitions between the qubit subspace $span\left\{\Delta_{\frac{1}{2}},\Delta_{\frac{1}{2}}^\prime\right\}$ and the quadruplet state $Q_{\frac{1}{2}}$ will determine leakage and also the error rate $\Gamma_{1}^{\mathbf{P}_2}$. We describe the effective error rates by the formulas from model 3 in Appendix \ref{ssec:MS3}. We point out two important characteristics of this process. First of all, the transition rate from the $\Delta_{\frac{1}{2}}$ level to the $Q_{\frac{1}{2}}$ level is larger than the rate from  the $\Delta^\prime_{\frac{1}{2}}$ level. The smaller energy differences enhance the transition rates [compare $h\left(\sigma_z^1,\omega\right)$ in \equationshortname~\eqref{eq:ratehyperfine}]. Second, $\Gamma_{0}^{\mathbf{P}_{1},\mathbf{P}_{2}}$ and $\Gamma_{1}^{\mathbf{P}_{2}}$ are larger for positive bias than for negative bias. This is because the transition amplitude $\left|\Dirac{Q_{\frac{1}{2}}}{\sigma_z^1}{\Delta_{\frac{1}{2}}}\right|$ is larger at positive bias than at negative bias. The transition amplitude can be read off directly when comparing the eigenstates for $\epsilon=\pm\infty$ with the eigenstate at $\epsilon=0$ in \figureshortname~\ref{fig:HighSymmetryRegimes}.\\
The error rate $\Gamma_{2}^{\mathbf{P}_2}$ has additionally a very interesting behavior. It can be described mainly by pure dephasing (see model 2 in Appendix \ref{ssec:MS2}). We especially find a point where the dephasing rate has a minimum. When the energy level fluctuations at both levels are equal, $\Gamma_2^{\mathbf{P}_2}$ vanishes [cf. \equationshortname~\eqref{eq:cancpuredeph}]. We can determine this point analytically $\left(\Dirac{\Delta_{\frac{1}{2}}}{\sigma_z^1}{\Delta_{\frac{1}{2}}}=\Dirac{\Delta^\prime_{\frac{1}{2}}}{\sigma_z^1}{\Delta^\prime_{\frac{1}{2}}}\right)$ and find $J_{12}=2 J_{23}$. For the chosen parameters in our calculation, we can approximate $\epsilon\approx-\frac{\epsilon_+}{3}$.\\
For phase noise on dot 1 the behavior of $\Gamma_{0}^{\mathbf{P}_3}$ is not captured by the error rates from all three error models. This is caused by the large dependency of the transition rates on the energy differences [compare $h\left(\sigma_z^1,\omega\right)$ in \equationshortname~\eqref{eq:ratehyperfine}]. Instead of a direct transition $\left(\Delta^\prime_{\frac{1}{2}}\rightarrow Q_{\frac{1}{2}}\right)$, we observe rather a two-step process $\left(\Delta^\prime_{\frac{1}{2}}\rightarrow \Delta_{\frac{1}{2}}\rightarrow Q_{\frac{1}{2}}\right)$.\\
We can summarize the contribution of all error rates to the deviation of the entanglement fidelity from its ideal value $\left(1-F_e\right)$. The overall behavior is determined mainly by the dominant error rate, which is in this case dephasing at the equator. We will, however, never find a value $\epsilon$ for which the errors due to phase noise on dot 1 have a vanishing effect.

Local phase noise on dot 2 does not allow any transitions from the $\Delta_{\frac{1}{2}}$ eigenstate if there is no external bias. This can be understood when analyzing the eigenstates of Hamiltonian \eqref{eq:Hamiltonian} at $\epsilon=0$ in \figureshortname~\ref{fig:HighSymmetryRegimes}. The $\Delta_{\frac{1}{2}}$ state involves a singlet state on the outer dots, while the remaining eigenstates contain only triplet states. Since interactions on the middle dot leave the states on the two outer dots untouched, the $\Delta_{\frac{1}{2}}$ state is protected from any local noise on the middle dot. This directly forbids leakage and relaxations from the upper pole at $\epsilon=0$. Further, $\Gamma_{1}^{\mathbf{P}_3}\left(\epsilon=0\right)=0$, since the lower pole never relaxes to the upper one.\\
The remaining features of the transition rates can be understood from their strong energy dependence [compare $h\left(\sigma_z^2,\omega\right)$ in \equationshortname~\eqref{eq:ratehyperfine}]. Transitions from the qubit states to the $Q_{\frac{1}{2}}$-quadruplet states describe leakage errors $\Gamma_0^{\mathbf{P}_1,\mathbf{P}_2,\mathbf{P}_3}$ through model 3 [cf. \equationshortname~\eqref{eq:model3-1}-\eqref{eq:model3-3}]. Leakage from the $\Delta^\prime_{\frac{1}{2}}$ state $\left(\Gamma_0^{\mathbf{P}_3}\right)$ decreases strongly at finite bias, since the energy difference to the $Q_{\frac{1}{2}}$ state increases. The leakage error from $\Delta_{\frac{1}{2}}$ $\left(\Gamma_0^{\mathbf{P}_1}\right)$ has the opposite characteristic. $\Gamma_0^{\mathbf{P}_2}$ and $\Gamma_1^{\mathbf{P}_2}$ are mainly determined by the average value of $\Gamma_0^{\mathbf{P}_1}$ and $\Gamma_0^{\mathbf{P}_3}$. Since $\Gamma_0^{\mathbf{P}_1}\gg\Gamma_0^{\mathbf{P}_3}$, we get approximately $\Gamma_0^{\mathbf{P}_2}=\Gamma_1^{\mathbf{P}_2}\approx\frac{\Gamma_1^{\mathbf{P}_1}}{2}$.\\
Model 1 describes relaxation of the states at the upper and the lower pole $\left(\Gamma_1^{\mathbf{P}_1,\mathbf{P}_3}\right)$. The energy difference of the qubit states $\Delta_{\frac{1}{2}}$ and $\Delta^\prime_{\frac{1}{2}}$ is minimal at $\epsilon=0$ (cf. \figureshortname~\ref{fig:TransPict1}). The relaxation rate $h\left(\sigma_z^2,\omega\right)$ is large for small $\epsilon$. Only the special symmetry at $\epsilon=0$ causes the large decrease of $\Gamma_1^{\mathbf{P}_1,\mathbf{P}_3}$ directly around $\epsilon=0$, where the transition amplitude vanishes.\\
Dephasing errors $\left(\Gamma_{2}^{\mathbf{P}_2}\delta t\right)$ are described by the pure dephasing mechanism of model 2 (see Appendix \ref{ssec:MS2}). We can calculate the difference in the energy-level fluctuation of the two states:
\begin{align}
\Dirac{\Delta_{\frac{1}{2}}}{\sigma_z^2}{\Delta_{\frac{1}{2}}}-\Dirac{\Delta^\prime_{\frac{1}{2}}}{\sigma_z^2}{\Delta^\prime_{\frac{1}{2}}}=\frac{2}{3}\frac{J_{12}+J_{23}}{\sqrt{J_{12}^2-J_{12}J_{23}+J_{23}^2}}.
\label{eq:SimplePict}
\end{align}
\equationshortname~\eqref{eq:SimplePict} has the limit $\frac{2}{3}$ for $J_{12}\substack{\gg\\ \ll} J_{23}$ and $\frac{4}{3}$ for $J_{12}= J_{23}$. These limits determine for the most part the overall error of the qubit $\left(1-F_e\right)$.

We find that local phase noise induces large errors to the time evolution of the subspace qubit. Especially pure dephasing, as described by toy model 2 (see Appendix \ref{ssec:MS2}), limits the performance of the triple quantum dot. Large errors are generated via phase noise on dot 1 for strong external bias, while phase noise on dot 2 is critical for the unbiased dot. This effect can be understood when considering the high-symmetry regimes of the Hamiltonian in \figureshortname~\ref{fig:HighSymmetryRegimes}. Phase noise is always most critical when it acts on the eigenstates of a single quantum dot.

\subsubsection{\label{sssec:ErrR2}
Regime 2}

\begin{figure}
\centering
\subfigure{\label{fig:TransPict2-1}\includegraphics[width=.49\textwidth]{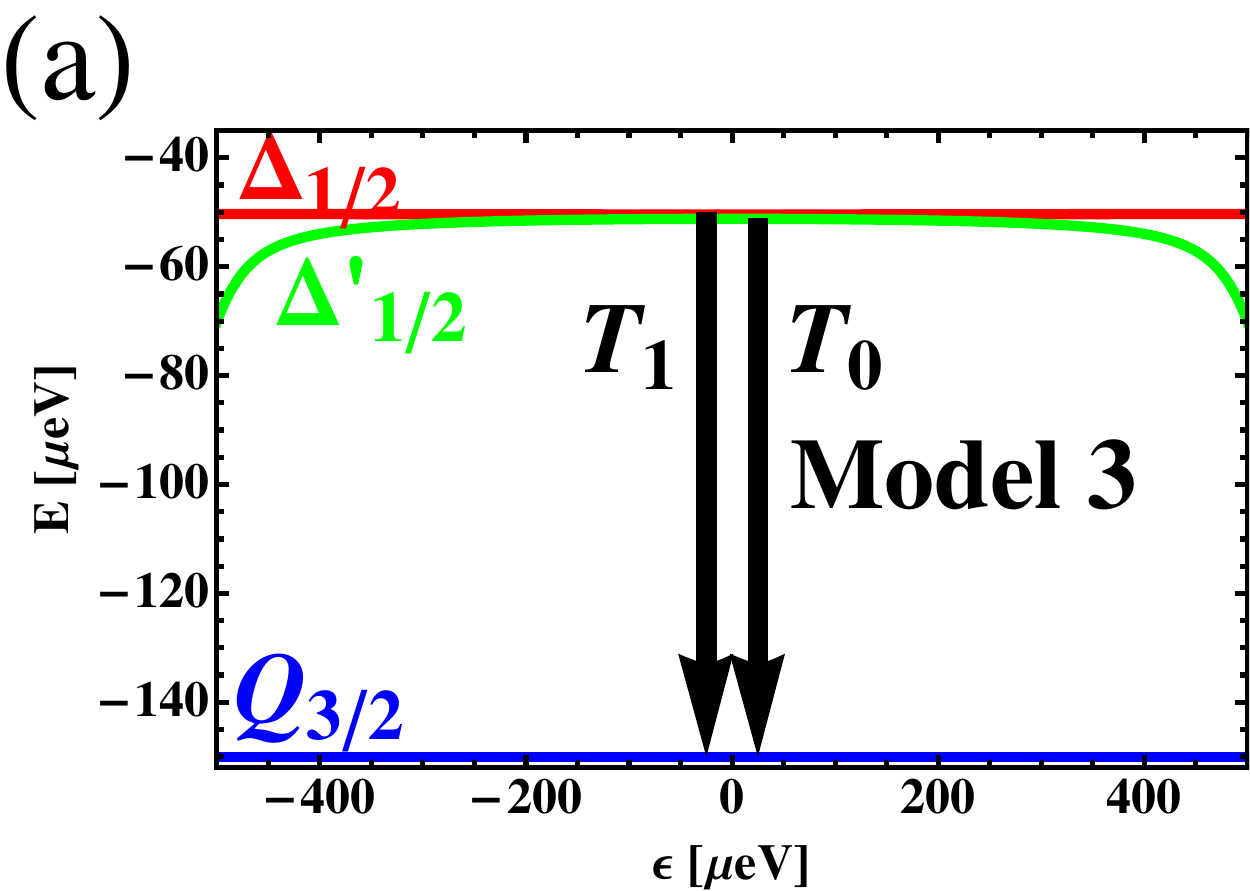}}
\subfigure{\label{fig:TransPict2-2}\includegraphics[width=.49\textwidth]{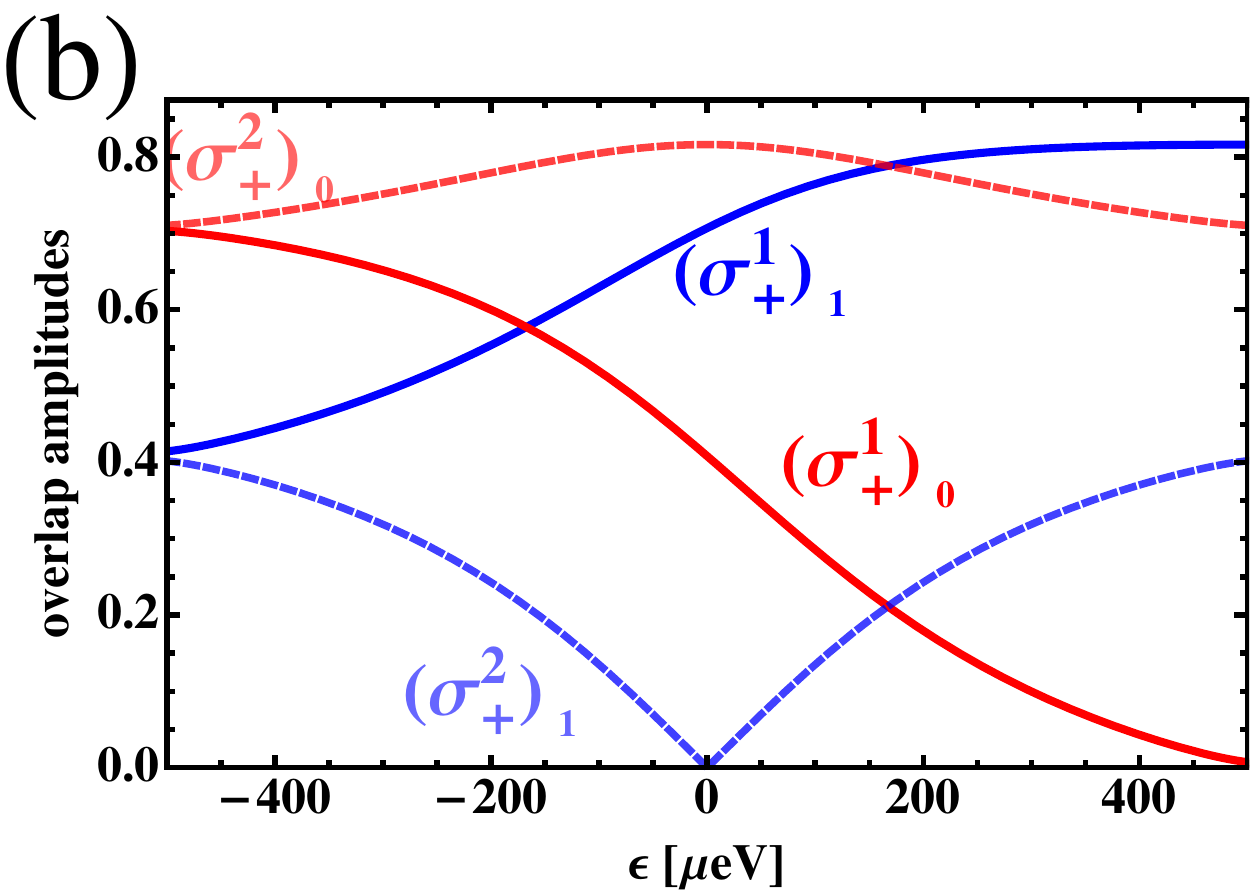}}
\caption{Description of local spin relaxation for the subspace qubit at large external magnetic fields ($E_z=100\ \mu\text{eV}$). Only local spin relaxations from the interaction with phonons are significant. (a) Sketch of energy diagram of the subspace qubit to describe the time evolution under local spin relaxation. $T_1$ and $T_0$ are the dominant transition rates from the qubit levels to the $s_z=\frac{3}{2}$ quadruplet state. (b) Transition amplitudes for the qubit state $i$ through the noise operator $\sigma_+$ acting on dot $j$: $\left(\sigma_+^j\right)_i=\left|\Dirac{Q_\frac{3}{2}}{\sigma_+^j}{W_i}\right|$ ($W_1=\Delta_{\frac{1}{2}}$, $W_0=\Delta^\prime_{\frac{1}{2}}$).}
\end{figure}

\begin{figure*}
\subfigure[$\Xi_{1}^x=2\cdot10^{-6}\frac{1}{s\ \mu\text{eV}^5}$, $\Xi_{2}^x=\Xi_{3}^x=0$]{\label{fig:subspacedec1}\includegraphics[width=.7\textwidth]{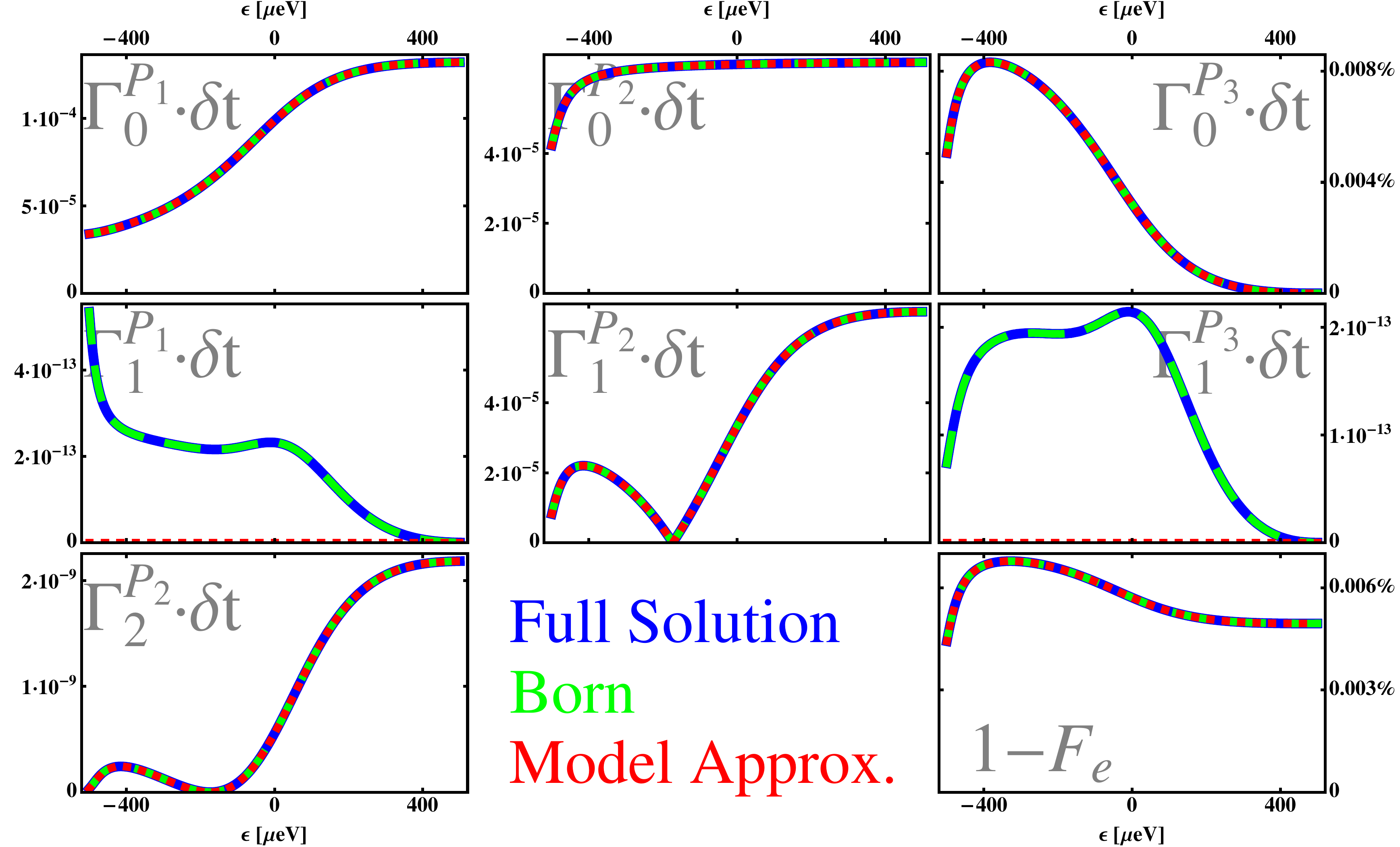}}
\subfigure[$\Xi_{2}^x=2\cdot10^{-6}\frac{1}{s\ \mu\text{eV}^5}$, $\Xi_{1}^x=\Xi_{3}^x=0$]{\label{fig:subspacedec2}\includegraphics[width=.7\textwidth]{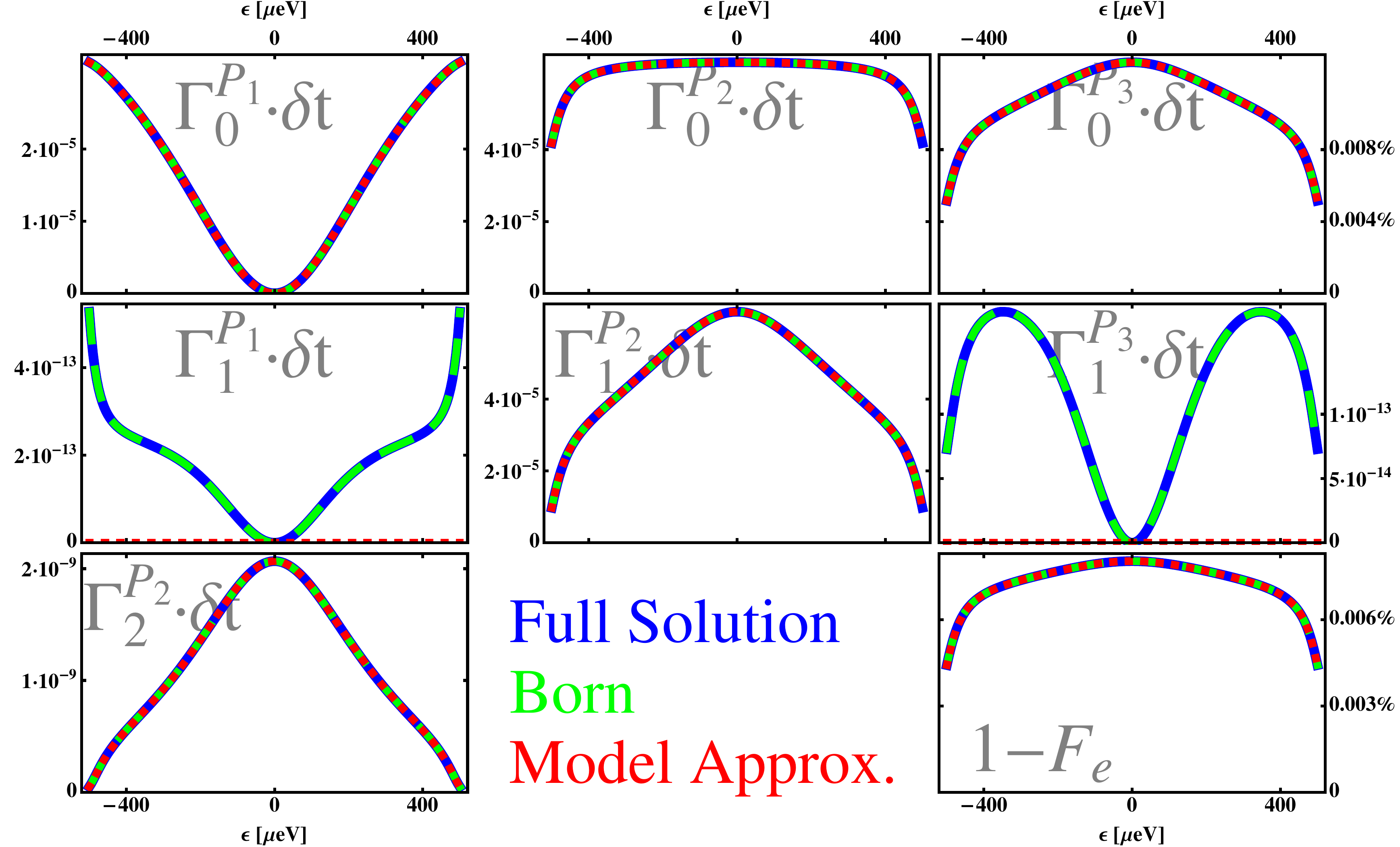}}
\caption{\label{fig:subspacedec}
Errors from local spin relaxation generated from interactions with phonons for the subspace qubit in the $\left(1,1,1\right)$ regime at $E_z=100\ \mu\text{eV}$. Blue lines are calculated from the numerical simulation of the full master equation. Green lines are obtained from the second order Born approximation. The red lines represent the results from the analysis of model 3, which involve only two transition rates (compare description in the main text). We can see that all descriptions are matching well.}
\end{figure*}

Next we analyze errors of the subspace qubit though local spin relaxation in the low-bias regime ($\epsilon$ small). We consider only electric bias $\epsilon$ in the range $\left[\epsilon_-,\epsilon_+\right]$. Local spin flip operators can generate transitions, changing the angular momentum quantum number ($\Delta s_z=\pm 1$). The total spin quantum number $S$ is not necessarily preserved. Transition rates through hyperfine interactions are highly suppressed due to the large energy difference of states with different $s_z$-quantum number (cf. \figureshortname~\ref{fig:EnergyDiag1} and the transition rates in \equationshortname~\eqref{eq:ratehyperfine}; the main effects are captured in Regime 2 of \tableshortname~\ref{tab:NoiseRegimes}). For phonon-mediated transitions the level splitting must be quite large to see transitions in the nanosecond regime [cf. \equationshortname~\eqref{eq:ratephonon}]. At $E_z=100\ \mu\text{eV}$ we only see the effect of two transitions rates. All other transition rates are greatly suppressed [cf. \figureshortname~\ref{fig:TransPict2-1}]. In \figureshortname~\ref{fig:subspacedec} we show the error probabilities for this parameter regime.

To explain the error probabilities in detail, we only need to discuss two properties of the relevant transition rates. First of all we need to analyze the transition amplitude from the qubit levels $\left\{\Delta_{\frac{1}{2}},\Delta^\prime_{\frac{1}{2}}\right\}$ to the $s_z=\frac{3}{2}$-quadruplet state. These amplitudes are drawn as a function of the external bias $\epsilon$ in \figureshortname~\ref{fig:TransPict2-2}. For relaxations on dot 1 the transition amplitude $\left|\Dirac{Q_\frac{3}{2}}{\sigma_+^1}{\Delta_\frac{1}{2}}\right|$ increases steadily from negative to positive bias. For the relaxations from the $\Delta^\prime_\frac{1}{2}$ state this effect is reversed. The transition amplitudes are equal at $J_{12}=2 J_{23}$. One can proove this immediately, when looking at the eigenstates $\Delta_{\frac{1}{2}}$ and $\Delta^{\prime}_{\frac{1}{2}}$ of the triple dot Hamiltonian $\mathcal{H}$ in \equationshortname~\eqref{eq:state3}-\eqref{eq:state4}. For spin relaxation on dot 2 the transition amplitude from $\Delta^\prime_{\frac{1}{2}}$ is always greater than the transition amplitude from $\Delta_{\frac{1}{2}}$. Additionally, we see a maximum of $\left|\Dirac{Q_\frac{3}{2}}{\sigma_+^2}{\Delta^\prime_\frac{1}{2}}\right|$ at zero bias. $\left|\Dirac{Q_\frac{3}{2}}{\sigma_+^2}{\Delta_\frac{1}{2}}\right|$ vanishes at this point. Second, the transition rates depend on the energy difference between the quadruplet state and the doublet states. We do not see any effect from transitions involving the $\Delta_{\frac{1}{2}}$ state, since its eigenenergy is only weakly dependent on $\epsilon$. For the $\Delta^\prime_{\frac{1}{2}}$ state the eigenenergy is influenced strongly by $\epsilon$.

Having these two discussions in mind we can understand the effective error rates of \figureshortname~\ref{fig:subspacedec}. We only need to compare the results with the analysis from model 3 in Appendix \ref{ssec:MS3}. Here the special case of transitions from the qubit to the surroundings, without its opposite effect, applies [cf. \equationshortname~\eqref{eq:model2-b}-\eqref{eq:model2-e}].

Leakage at the upper pole $\left(\Gamma_0^{\mathbf{P}_1}\right)$ is only dependent on the transition rate from $\Delta_{\frac{1}{2}}$ to $Q_{\frac{3}{2}}$ [cf. \equationshortname~\eqref{eq:model2-b}]. For relaxation at dot 1 the leakage rate therefore steadily increases, while relaxation on dot 2 has a local minimum at zero bias. For the lower pole $\mathbf{P}_3$ the effect is reversed. Leakage on dot 1 steadily decreases with $\epsilon$, while it has a maximum at $\epsilon=0$ for noise acting on dot 2. We also see the dependence on the transition energy at large bias. Here the error rate $\Gamma_{0}^{\mathbf{P}_3}$ decreases further. The leakage rate at the equator $\left(\Gamma_0^{\mathbf{P}_2}\right)$ is represented in leading order by the average leakage rates at $\mathbf{P}_1$ and $\mathbf{P}_3$ [cf. \equationshortname~\eqref{eq:model2-2}].

Relaxation at $\mathbf{P}_1$ and $\mathbf{P}_3$ vanishes in model 3 in leading order [cf. \equationshortname~\eqref{eq:model2-4} and \eqref{eq:model2-6}]. The results from the numerical analysis are not obtained in model 3. However, note that their magnitude is very small. Relaxation at the equator $\left(\Gamma_{1}^{\mathbf{P}_2}\right)$ is determined in leading order by the difference in transition rates from $\Delta_{\frac{1}{2}}$ to $Q_{\frac{3}{2}}$, compared to the rate from $\Delta^\prime_{\frac{1}{2}}$ [cf. \equationshortname~\eqref{eq:model2-5}]. We see vanishing relaxations at $J_{12}=2J_{23}$ for noise acting on dot 1. Dephasing at the equator $\Gamma_{2}^{\mathbf{P}_2}$ shows a very similar characteristic. It is also dependent on the difference in the transition rates [cf. \equationshortname~\eqref{eq:model2-e}].

The detected entanglement fidelity for local spin relaxation is close to 1. For local noise on dot 1 or 2 it does not show a characteristic dependence on the bias parameter. In total, all resulting errors are much smaller compared to the influence of phase noise discussed in the last section.

\subsection{\label{ssec:ErrSubsys}
Subsystem Qubit}

The subsystem qubit is for large external magnetic fields equivalent to the subspace qubit. We do not add any results for the subsystem qubit in this case. As described in Appendix \ref{ssec:SimpPhase}, phase noise also acts on the subsystem qubit identically at small external magnetic fields. The subsystem qubit differs from the subspace qubit only at small external fields and for local spin relaxations. We analyze the subsystem qubit only for magnetic field strengths comparable to the thermal energy. Here both the $s_z=\frac{1}{2}$ and $s_z=-\frac{1}{2}$ subspace is initially occupied [see distribution function in \equationshortname~\eqref{eq:subsysbath}]. The qubit levels are characterized by the formally introduced $l$-quantum number, as described in \autoref{ssec:SsubAndSsys}.

\subsubsection{\label{sssec:ErrR3}
Regime 3}

\begin{figure*}
\subfigure[$\Upsilon^{x}_1=\left(20\ \text{ns}\right)^{-1}$, $\Upsilon^x_2=\Upsilon^x_3=0$]{\label{fig:subsysdec1}\includegraphics[width=.7\textwidth]{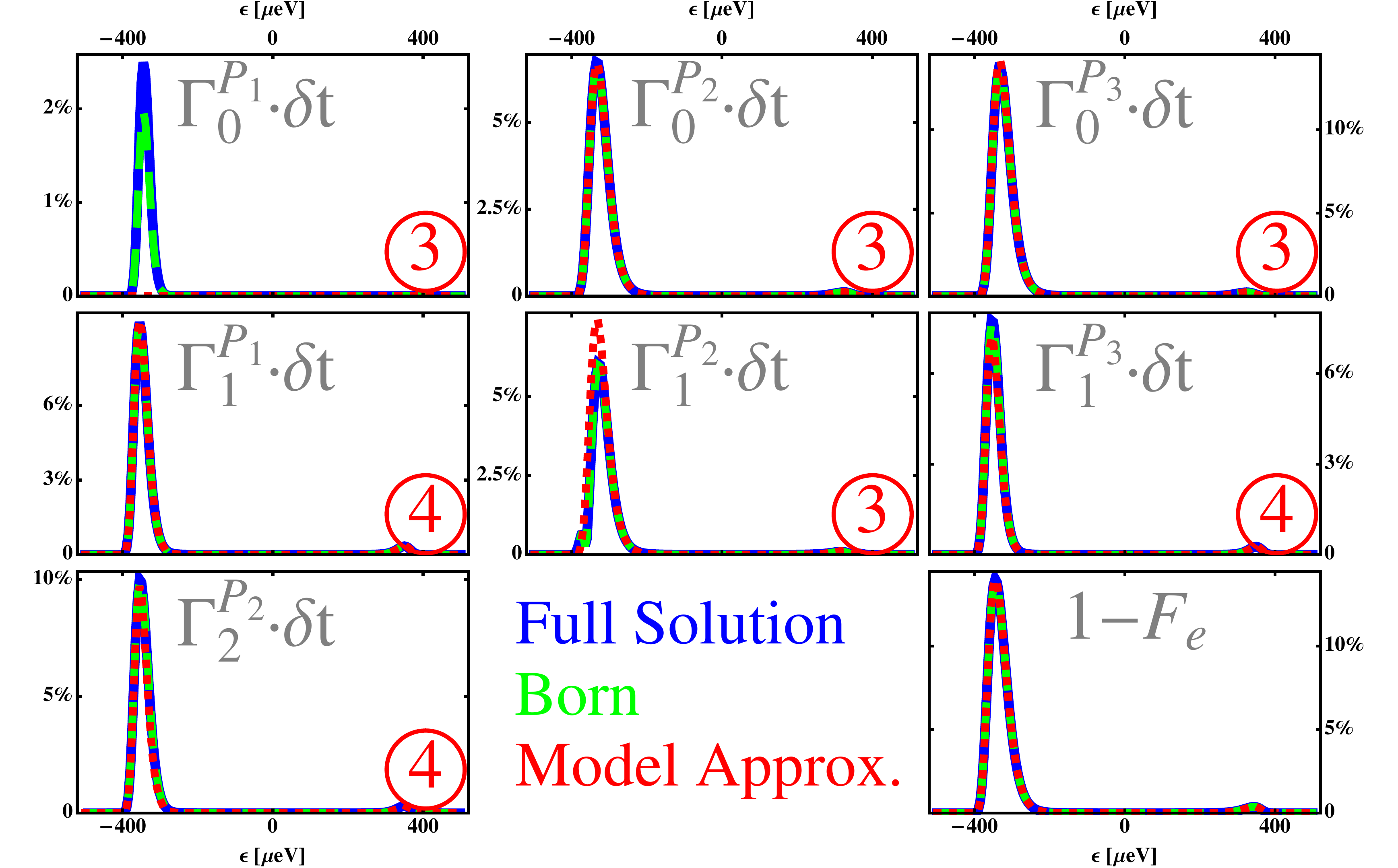}}
\subfigure[$\Upsilon^{x}_2=\left(20\ \text{ns}\right)^{-1}$, $\Upsilon^x_1=\Upsilon^x_3=0$]{\label{fig:subsysdec2}\includegraphics[width=.7\textwidth]{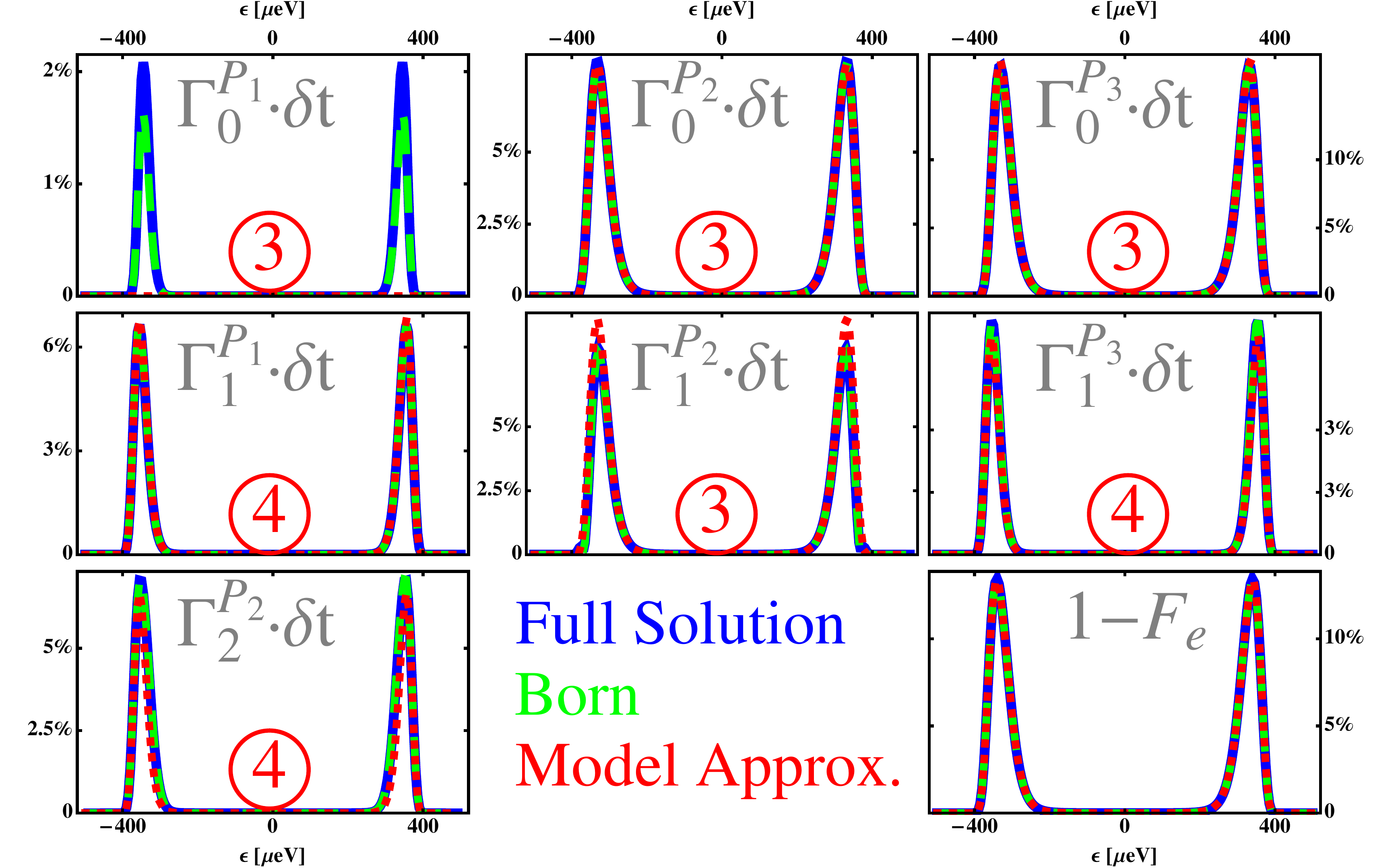}}
\caption{\label{fig:subsysdec}
Errors of the subsystem qubit  generated by local spin relaxations for small external magnetic fields $\left(E_z=2.5\ \mu\text{eV}\right)$. Large error rates are only seen at the point of level crossings. For noise acting on dot 1 the errors at positive detuning are highly suppressed. Blue lines represent the calculation via the simulation of the full master equation. Green lines represent the second-order Born approximation. Red lines show results from the analysis of model system 3 (see Appendix \ref{ssec:MS3}) and model system 4 (see Appendix \ref{ssec:MS4}).}
\end{figure*}

Already in the small detuning regime, levels of different $s_z$-quantum numbers cross. Hyperfine interactions can generate transitions between these levels through local spin flips. The total spin quantum number $S$ is not preserved by local interactions. In \figureshortname~\ref{fig:subsysdec} we extract the error rates generated by local spin relaxation on dots 1 and 2. Relaxation due to phonons is not detectable for the subsystem qubit in the regime of nanoseconds. Larger energy differences are needed to see strong effects [see transition rates in \equationshortname~\eqref{eq:ratephonon}].

\begin{figure}
\centering
\includegraphics[width=.49\textwidth]{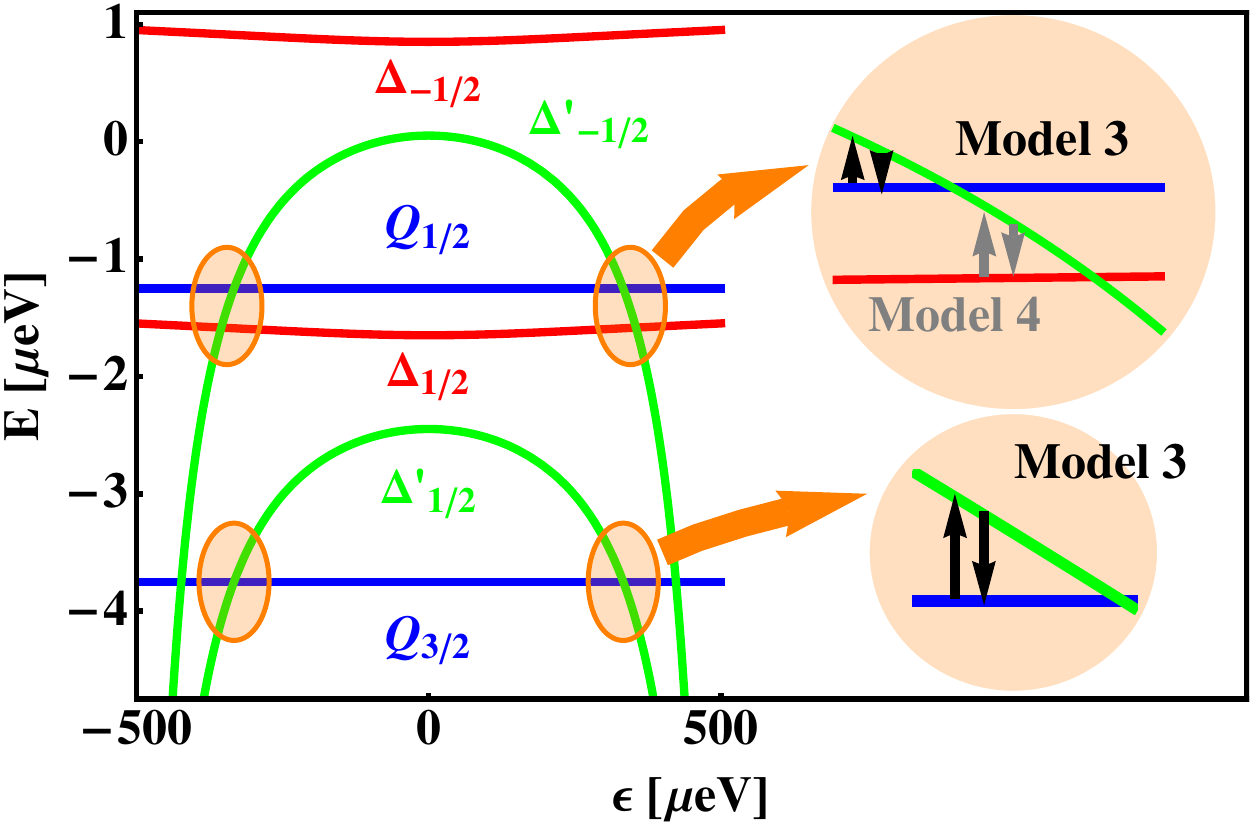}
\caption{\label{fig:TransPict3}
Transition diagram for subsystem qubit when hyperfine interactions generate local spin flips. Large error rates are observed only at the region of level crossings. They can be described by leakage transitions to a quadruplet state (model 3 from Appendix \ref{ssec:MS3}) or internal transitions between two states of the subsystem qubit (model 4 from Appendix \ref{ssec:MS4}).}
\end{figure}

Phase noise from hyperfine interaction generates large transition rates only if energy levels of different $s_z$-quantum numbers are close in energy (compare $h\left(\sigma_x^i,\omega\right)$ in \equationshortname~\eqref{eq:ratehyperfine}; Regime 3 in \tableshortname~\ref{tab:NoiseRegimes}). This is the case at the level crossings of the energy diagrams (cf. \figureshortname~\ref{fig:TransPict3}). Essentially, two sets of transitions are important. First of all, leakage of the doublet levels to quadruplet states plays an important role (model 3 from Appendix \ref{ssec:MS3}). Since local spin flip operators always change the $s_z$-quantum number, only transitions from the $l=1$ states to the quadruplet levels are possible. Leakage from the $l=0$ states is highly suppressed. Their energy differs significantly from the quadruplet levels of different $s_z$-quantum number. Second, there are internal transitions between the subsystem states $\Delta_{\frac{1}{2}}$ and $\Delta^\prime_{-\frac{1}{2}}$, as described with model 4 from Appendix \ref{ssec:MS3}.

All effects can be summarized easily. Model 4 determines the relaxation errors at the upper and lower poles as well as the dephasing properties. Model 3 describes the leakage behavior. Since the internal transition rates are very similar at the equator, effectively no relaxation is generated internally by transitions between the $\Delta_{\frac{1}{2}}$ and $\Delta^\prime_{-\frac{1}{2}}$ states. Relaxation is rather determined by the indirect process of leakage to the quadruplet levels.

For relaxations acting on dot 2 we see the symmetric error rates at positive and negative bias. For relaxations on dot 1 the error rates for positive bias are greatly suppressed. $\Delta^\prime_{\frac{1}{2}}$ approaches for positive bias the eigenstate from $\epsilon\rightarrow\infty$ in \figureshortname~\ref{fig:HighSymmetryRegimes}. It involves a singlet state on the strongly coupled dots 2 and 3. Noise on dot 1 leaves this singlet untouched. This state does not couple to any quadruplet state or $l=0$ state.

Finally, we want to explain why $\Gamma_{0}^{\mathbf{P}_1}$ is not correctly described by our simple model analysis. Leakage at the upper pole is determined by a second-order process of internal transitions followed by leakage ($\Delta_{\frac{1}{2}}\rightarrow\Delta^\prime_{-\frac{1}{2}}\rightarrow Q^\prime_{\frac{1}{2}}$). We describe this process neither by model 3 nor by model 4.

Overall, we find that at the point of level crossings fluctuating hyperfine fields can generate major errors of the subsystem qubit. The asymmetry of the error rates for spin relaxation on dot 1 between positive and negative bias is a very interesting result. For positive bias fluctuating hyperfine fields generate nearly no errors, since the Hamiltonian \eqref{eq:Hamiltonian} is approaching a high-symmetry regime as discussed in Appendix \ref{ssec:SimpRegimes}.

\section{\label{sec:Summary}
Summary and Outlook}

The exchange-only qubit has been analyzed with two different schemes for implementing a coded qubit in the Hilbert space of three single occupied quantum dots, resulting in the subspace and the subsystem qubit.  The relaxation and decoherence dynamics of both these qubits have been calculated, with both nuclear spins and phonons taken into account. These interactions are described in the DM, a particular Markov approximation with a transparent quantum-jump interpretation and consistent long-time behavior. The systematics of the early time dynamics, which is of most interest for qubit experiments, is quite distinct from the systematics of the long time evolution. We have focused on the initial time evolution and have extracted errors for the subspace and the subsystem qubit. We can describe all results by relating them to just four toy models, whose time evolution can be calculated analytically.

For local phase noise, arising from the interaction with fluctuating hyperfine fields, the influence on the the subsystem and the subspace qubit are identical. Local phase noise is critical for GaAs systems; it is the strongest mechanism for the loss of phase coherence. Sizable phase errors for both the subspace and subsystem arise after just $10$ ns of evolution.

The influence of local spin relaxation is different for the subspace and the subsystem qubit. Since the subspace qubit is always operated at large external magnetic fields, spin relaxations from the interactions with phonons need to be considered. This effect generates large transition rates only between energy levels with large energy difference. Our analysis shows that in GaAs systems, operated at large external magnetic fields, only small errors are generated. For the subsystem qubit an operation at small magnetic fields is also possible. Here, only local spin relaxations from the interactions with nuclear magnetic fields are important. These interactions generate large errors at the crossing of energy levels of different $s_z$-quantum numbers. This process has a very interesting property for phase noise acting locally on one of the outer quantum dots. Errors can be highly suppressed depending on the sign of the bias parameter $\epsilon$.

To state our results briefly, our analysis shows that in GaAs samples (large nuclear bath) the subsystem and the subspace qubit have about equally good coherence properties. Both qubit implementation schemes suffer from local phase noise, generated from fluctuating hyperfine fields. Spin relaxation from hyperfine fields will be important only at the point of level crossings. If these points can be avoided when manipulating the qubit, spin relaxations induced by fluctuating hyperfine fields are negligible. If one attempts to use the crossing points in the energy level diagram for qubit manipulations (cf. the attempt to manipulate a singlet-triplet qubit at crossing points in the energy diagram\cite{ribeiro2010}), one has to pay attention to fluctuating hyperfine fields. Interactions with phonons will usually be less important. This mechanism will only be significant if there are strong external magnetic fields and large energy differences. The interaction with phonons can completely depopulate the qubit, but in GaAs systems this evolution only occurs on the microsecond time scale.

A way to suppress the influence of hyperfine spins in GaAs triple quantum dots can be devised that is similar to the approaches used in double quantum dots. Since hyperfine induced dephasing is caused by low-frequency noise, one can apply refocusing protocols which have already enhanced the coherence properties in double dot systems.\cite{bluhm2011} Another possibility is to consider materials containing fewer nuclear spins. Working in silicon samples is a reasonable approach, as experiments are catching up to the state of the art in GaAs.\cite{maune2012} One advantage of both the subspace and the subsystem implementation is the full controllability of the qubit through the exchange interaction.\cite{divincenzo2000} One does not rely on polarized hyperfine fields\cite{foletti2009,gullans2010} or micromagnets\cite{brunner2011} as for the full controllability for GaAs singlet-triplet qubits.

Overall it is a very interesting task to test the local nature of the error models. Especially for the influence of nuclear spins, which behave on short time scales like classical fluctuating magnetic fields, the local influence of the qubit dynamics is worth testing.  Such an experiment would require the control of the randomness of the hyperfine fields at the positions of the different quantum dots. If it is possible to reduce the randomness at two of the three quantum dots, so that the hyperfine interaction noise acts dominantly on one of the three quantum dots, one can try to test the different scaling behavior of the error rates with the bias parameter $\epsilon$.
Furthermore, our analysis method in the DM should be helpful for the description of other coded qubits implemented in more complex Hilbert spaces. We show in detail here, for the triple-dot qubit, how the interaction with complicated baths can be reduced to just an effective evolution on the coded qubit itself. Such analysis could be extended to other coding strategies when the need arises.

\section{Acknowledgments}
We are grateful for support from the Alexander von Humboldt foundation.
 
\appendix

\section{\label{sec:Simplification}
Simplification of the Analysis}

\subsection{\label{ssec:SimpRot}
Rotating Frame}

When analyzing the master equation, we are interested in the deviation of the qubit evolution from the free evolution $\mathcal{L}_0$ due to the dissipative Lindblad term $\mathcal{L}_D$ (cf. noise description in \autoref{ssec:Noise}). It is therefore meaningful to go for the analysis to a rotating frame with respect to the free Hamiltonian:
\begin{equation}
	\rho\left(t\right)\rightarrow\rho^{rot}\left(t\right)=\mathcal{U}_{rot}\left(t\right)\rho\left(t\right)\mathcal{U}_{rot}^\dagger\left(t\right),
\end{equation}
with $\mathcal{U}_{rot}\left(t\right)=e^{i\mathcal{H}t}$. This leads automatically to a redefinition of the Lindbladian:
\begin{equation}
	\mathcal{L}_0+\mathcal{L}_D\rightarrow\mathcal{L}^{rot}.
\end{equation}
Due to the general conditions of the DM \eqref{eq:DaviesModel}, the Lindbladian in the rotating frame equals the original dissipative Lindbladian: $\mathcal{L}^{rot}=\mathcal{L}_D$.  $\mathcal{L}_D$ consists of a sum of terms, in each appear the coupling operators $\mathcal{A}_{\omega}$ twice, once as a Hermitian conjugate. When the coupling operators $\mathcal{A}_{\omega}^{\left(\dagger\right)}$ are written in the eigenenergy representation of the free Hamiltonian, to each entry a complex argument $e^{i \omega t}$ is added. The phase $\omega$ represents the energy difference of the states that the coupling operators $\mathcal{A}_{\omega}$ connect. Since all Lindblad operators are grouped to couple only equidistant energy levels, these complex factors cancel out.
 
\subsection{\label{ssec:SimpPhase}
Symmetry of Phase Noise}

We want to point out a key symmetry for phase noise, which simplifies our considerations. The action of phase noise through the coupling operators $\sigma_z^i$ ($i=1,2,3$) has an equal effect on the $s_z=\frac{1}{2}$ and the $s_z=-\frac{1}{2}$ subspace (involving also the quadruplet levels). It mixes within these subspaces but never couples subspaces of different $s_z$-quantum number. Furthermore, the corresponding matrix elements in the $s_z=\frac{1}{2}$ and the $s_z=-\frac{1}{2}$ subspace are, up to a sign, identical. This can be understood by the symmetry operation which flips the spins on all dots $\mathcal{U}_{flip}$. It transforms a state from the $s_z=+\frac{1}{2}$ subspace to the corresponding $s_z=-\frac{1}{2}$ subspace and vice versa. It also adds a sign to $\sigma_z^i$. This proves that:
\begin{equation}
\Dirac{W_{1/2}}{\sigma_z^i}{V_{1/2}}=-\Dirac{W_{-1/2}}{\sigma_z^i}{V_{-1/2}}
\end{equation}
for $W,V\in\left\{\Delta,\Delta^\prime,Q\right\}$. Since in every dissipative term these matrix elements appear twice, the factor ``$-1$'' drops out. This symmetry was also identified in the paper by Ladd, which however, did not connect it to the underlying symmetry operator.\cite{ladd2012}

\subsection{\label{ssec:SimpRegimes}
High Symmetry Regimes}

We point out high symmetry regimes of the qubit Hamiltonian, which help us to understand limits of the error rates in \autoref{sec:Errors}.

First of all, without bias we will have effectively a spin zero or spin one particle from the electrons of the outer two dots coupled to a spin-1/2 particle on the middle dot. This can be seen easily when noticing that without bias $J_{12}=J_{23}=J$. The exchange interaction part simplifies to
\begin{equation}
J_{12}\bm{\sigma}_1\cdot\bm{\sigma}_2
+J_{23}\bm{\sigma}_2\cdot\bm{\sigma}_3
=J\ \bm{\sigma}_2\cdot\left(\bm{\sigma}_1+\bm{\sigma}_3\right).
\end{equation}
We can construct eigenstates of the triple-dot Hamiltonian \eqref{eq:Hamiltonian} from spin-1/2 eigenstates on the middle dot and singlet-triplet levels on the outer two dots.

\begin{figure}
\includegraphics[width=0.49\textwidth]{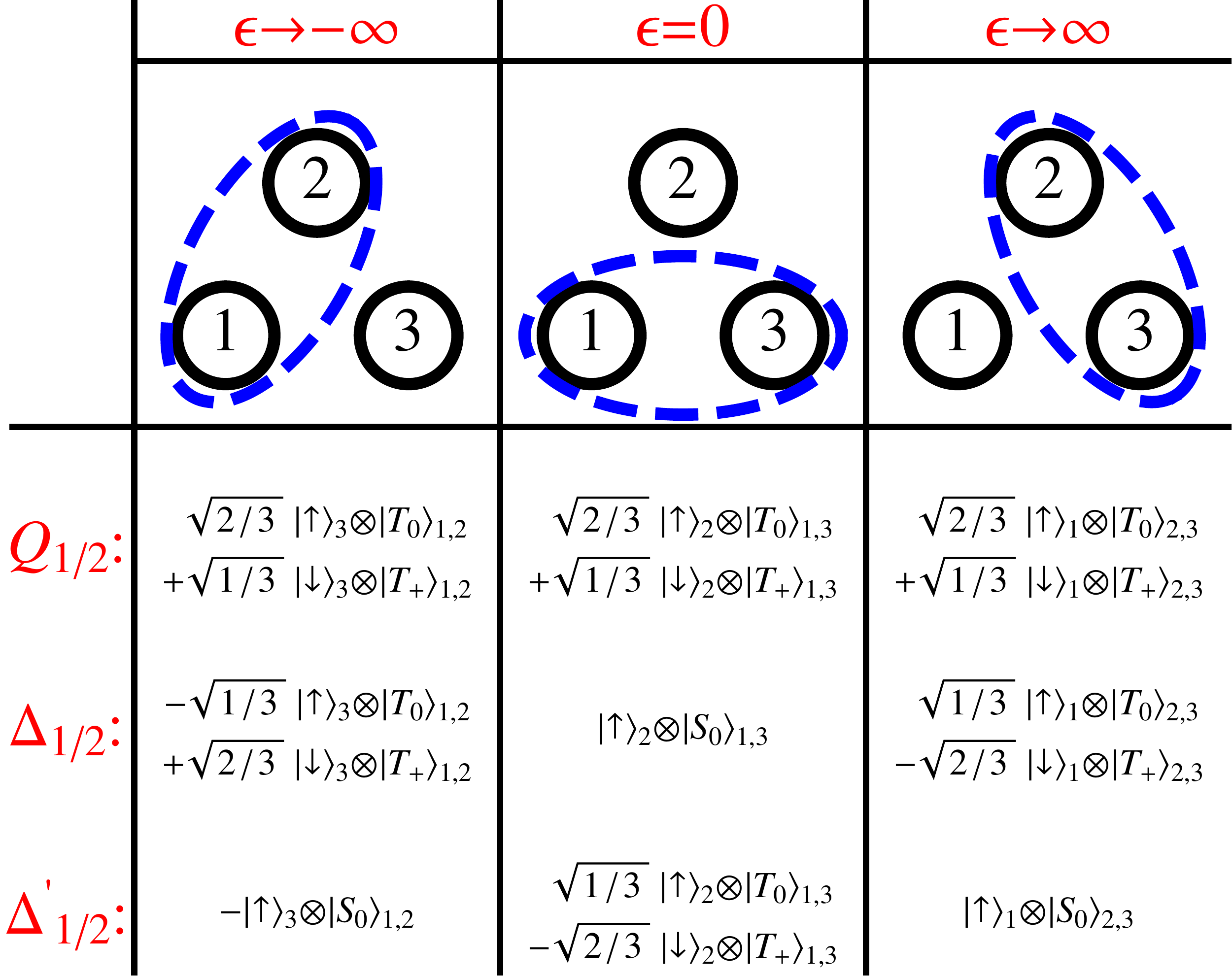}
\caption{\label{fig:HighSymmetryRegimes}
High-symmetry regimes of the exchange interaction Hamiltonian \eqref{eq:Hamiltonian} in the limits of high bias ($\epsilon=\pm\infty$) and no bias. The $s_z=\frac{1}{2}$ eigenstates always describe composite systems of two spin-1/2 levels coupled to one spin-$\frac{1}{2}$ level.}
\end{figure}

Second, in the case of large positive (negative) detuning the exchange interaction parameter $J_{23}$ ($J_{12}$) is dominant. We can ignore the coupling of one dot. Hence, the model describes a strongly coupled double dot and an uncoupled spin-1/2 level. The eigenstates for the exchange interaction Hamiltonian \eqref{eq:Hamiltonian} can again be constructed from the singlet-triplet eigenstates of the double dot and the single electron eigenstates of the uncoupled single dot. We summarize all eigenstates in \figureshortname~\ref{fig:HighSymmetryRegimes}. The corresponding $s_z=-\frac{1}{2}$ states can be obtained by flipping all spins. These eigenstates are agreeing with the limits of \equationshortname~\eqref{eq:state2}-\eqref{eq:state4}.

Additionally, we can understand the action of local noise more easily, when additional symmetries are present. First of all, noise on dot 1 is equivalent to noise on dot 3 when changing the sign of $\epsilon$. This property is true only because in our analysis the tunnel coupling between the dot pairs $1$ and $2$ is identical to the coupling between dots $2$ and $3$. Additionally, we use $\left|\epsilon_+\right|=\left|\epsilon_-\right|$. We therefore never analyze noise on dot 3 individually. For the same reason local noise on dot 2 is equivalent for positive and negative bias. Second, for large negative detuning it does not matter if the noise is acting on the first or the second dot. This result is just a consequence of the situation described earlier. For large negative detuning we couple dots 1 and 2 strongly, while the third quantum dot is effectively decoupled.

\section{\label{sec:Init}
Descriptions of Initial Time Evolution}

In qubit experiments one is usually interested in the time evolution of the qubit on short time scales. With the Nakajima-Zwanzig approach one can construct an effective master equation for the initial time evolution of the qubit.\cite{fick1983,fick1990} The ``common''  master equation describes the time evolution of the full system (with its multiqubit Hilbert space) in a first-order differential equation. Using the Nakajima-Zwanzig approach, one can reduce this equation to the relevant part of the Hilbert space describing just the qubit. In general, the problem of solving this lower dimensional equation is not simpler than solving for the dynamics of the full system. However, with a few additional assumptions we can simplify this lower dimensional equation.

As a first step, one identifies a relevant part of the Hilbert space $\mathbb{H}^{rel}\subset\mathbb{H}$, which is used to define the qubit. One defines a linear map $\mathcal{P}$, which constructs from the full density matrix only the relevant part:
\begin{equation}
\rho^{rel}\left(t\right)=\mathcal{P}\rho\left(t\right).
\end{equation}
We need only two properties for the map $\mathcal{P}$ to be physically meaningful. First of all the map should act on the relevant part of the density matrix like the identity operation. One disposes the condition
\begin{equation}
\mathcal{P}^2=\mathcal{P}.
\label{eq:LinMap1}
\end{equation}
Secondly an observable $\mathcal{F}$ on the relevant part of the Hilbert space should be described in the the same way by the reduced density matrix $\mathcal{P}\rho\left(t\right)$ and the full density matrix $\rho\left(t\right)$. We obtain this physical property by requiring
\begin{equation}
Tr\left(\mathcal{F}\mathcal{P}\bullet \right)=Tr\left(\mathcal{F}\bullet \right),
\label{eq:LinMap2}
\end{equation}
$\bullet$ represents an arbitrary element of Liouville space. Finally for our later purpose we also add a third characteristic. Initially, the qubit is decoupled from the surroundings, which gives $\mathcal{P}\rho\left(0\right)=\rho\left(0\right)$. This requirement is equivalent to the criterion to initialize the qubit into a controlled state.

With these three assumptions we will rewrite our Lindblad master \equationshortname~\eqref{eq:LindblMasEq}:
\begin{equation}
	\dot \rho\left(t\right) = \mathcal{L}\rho\left(t\right).
	\label{eq:DefinitionLindblMasterequ}
\end{equation}
For the upcoming analysis $\mathcal{L}$ can consist of a coherent time evolution $\mathcal{L}_0\left(\bullet\right)=-i\left[\mathcal{H},\bullet\right]$ and it may also include a dissipative Lindblad term $\mathcal{L}_D\left(\bullet\right)=\sum_{\mathcal{A}}\Upsilon_{\mathcal{A}}\mathcal{D}\left[\mathcal{A}\right]\left(\bullet\right)$. One can exactly rewrite the master \equationshortname~\eqref{eq:DefinitionLindblMasterequ} for the relevant part $\rho^{rel}\left(t\right)$ with a time-retarded equation\cite{fick1983,fick1990}:
\begin{align}\label{eq:NakajFinal}
\dot \rho^{rel}\left(t\right)=&\mathcal{P}\mathcal{L}\mathcal{P}\rho^{rel}\left(t\right)\\&
+\int_{0}^{t}dt^\prime\mathcal{P}\mathcal{L}\mathcal{Q}
e^{\mathcal{Q}\mathcal{L}\mathcal{Q}\left(t-t^\prime\right)}\mathcal{Q}\mathcal{L}\mathcal{P}\rho^{rel}\left(t^\prime\right).\nonumber
\end{align}
\equationshortname~\eqref{eq:NakajFinal} is called the Nakajima-Zwanzig equation. We have introduced the projector $\mathcal{Q}\equiv 1-\mathcal{P}$.

To describe the initial time evolution, we divide the full Hilbert space into a relevant part $A$ and an irrelevant  part $B$. $\mathcal{L}_0=\mathcal{L}_A+\mathcal{L}_B$ describes the time evolution of $A$ and $B$ individually. $\mathcal{L}_1$ connects $A$ and $B$. The time evolution should be dominated by $\mathcal{L}_{0}$, while $\mathcal{L}_1$ describes only a ``small'' term. In a second-order Born approximation, we keep terms containing $\mathcal{L}_{1}$ up to second order. The Nakajima-Zwanzig equation in second-order Born approximation reads

\begin{align}\label{eq:NakajBorn}
\dot \rho^{P}\left(t\right)=&\left(\mathcal{P}\mathcal{L}_A\mathcal{P}+\mathcal{P}\mathcal{L}_{AB}\mathcal{P}\right)\rho^{P}\left(t\right)\\&+\int_{0}^{t}dt^\prime\mathcal{P}\mathcal{L}_{AB}\mathcal{Q}
e^{\mathcal{Q}\left(\mathcal{L}_A+\mathcal{L}_B\right)\mathcal{Q}\left(t-t^\prime\right)}\mathcal{Q}\mathcal{L}_{AB}\mathcal{P}\rho^{P}\left(t^\prime\right).\nonumber
\end{align}

\subsection{\label{ssec:InitSubspace}
Subspace Qubit}

To define the subspace qubit one needs to project out parts of the Hilbert space, i.e. one uses a map $\mathcal{P}_P$ made up of projectors. The relevant and irrelevant parts, called $A$ and $B$ in \equationshortname~\eqref{eq:NakajBorn}, are subspaces of the full Hilbert space. We call them $P$ and $Q$, respectively. $\mathcal{P}_P$ constructs from $\rho\left(t\right)\in\mathbb{H}\simeq \mathbb{C}^d$ the relevant density matrix on subspace $P$. $\rho^{P}\left(t\right)$ is only nonzero in $\mathbb{H}^{P}\simeq\mathcal{C}^2$ ($2<d$). The linear map $\mathcal{P}_P$ can be constructed to keep from full density matrix only the relevant components:

\begin{equation}
\mathcal{P}_P:\rho=
\left(
\begin{array}{cc}
\rho^{P} & \rho^{+}\\
\rho^{-} & \rho^{Q}
\end{array}
\right)
\rightarrow\left(
\begin{array}{cc}
\rho^{P} & 0\\
0 & 0
\end{array}
\right).
\label{eq:ProjMap}
\end{equation}

$\mathcal{Q}_P$ is implicitly defined as $1-\mathcal{P}_P$. To rewrite \equationshortname~\eqref{eq:NakajBorn} for $\mathcal{P}=\mathcal{P}_P$ and $\mathcal{Q}=\mathcal{Q}_P$, a bra-ket notation of superoperators turns out to be very useful (for an introduction, see the book of Blum\cite{blum1996}). We use round brackets for superstates in Liouville space. The superprojectors $\sop{i}{i}$ project onto the corresponding part of the density matrix. They also divide Liouville space into four subspaces, which we label by $i$, $i\in\left\{P,Q,+,-\right\}$. We can rewrite all superoperators in this notation and identify projected superoperators. They describe transitions between two of these Liouville subspaces. The superoperator $\mathcal{L}_{P}$ has only components connecting superstates from $P$ and $P$:

\begin{equation}
	\mathcal{L}_P=\sket{P}L_{PP}\sbra{P}.
\end{equation}
$\mathcal{L}_{Q}$ never acts on the relevant subspace:
\begin{equation}
	\mathcal{L}_Q=\sket{Q}L_{QQ}\sbra{Q}.
\end{equation}

The remaining superoperator $\mathcal{L}_{PQ}$ does not just couple the subspaces $P$ and $Q$. It also has contributions to the off-diagonal terms of the density matrix:

\begin{equation}
\mathcal{L}_{PQ}=\sum_{\substack{A,B\in\left\{ P,Q,+,-\right\}\\ AB \notin\left\{PP ,QQ\right\}}} \sket{A}L_{AB}\sbra{B}.
\end{equation}

Using this notation one can rewrite \equationshortname~\eqref{eq:NakajBorn} for the linear map $\mathcal{P}_P$ from \equationshortname~\eqref{eq:ProjMap}. We arrive at a master equation on the relevant subspace $P$:

\begin{align}\nonumber
\dot \rho^{P}\left(t\right)=&L_{PP}\rho^{P}\left(t\right)+\int_{0}^{t}dt^\prime
L_{PQ}e^{L_{QQ}\left(t-t^\prime\right)}L_{QP}
\rho^{P}\left(t^\prime\right)\\
&+\left(L_{P+}L_{+P}+L_{P-}L_{-P}\right)
\int_{0}^{t}dt^\prime\rho^{P}\left(t^\prime\right).
\label{eq:NZBornSubspace}
\end{align}

\subsection{\label{ssec:InitSubsystem}
Subsystem Qubit}

In general, one is interested not only in dividing the Hilbert space into two subsystems, but also in defining a subsystem inside a subspace of the full Hilbert space:

\begin{equation}
	\mathbb{H}=\left(\underbrace{\mathbb{H}_S\otimes\mathbb{H}_B}_{\mathbb{H}_P}\right)\oplus\mathbb{H}_Q.
\end{equation}

We need this approach for the definition of the subsystem qubit (cf. \autoref{ssec:SsubAndSsys}). Here we first project on a four-dimensional subspace $P\equiv span\left\{\Delta_{\frac{1}{2}},\Delta^\prime_{\frac{1}{2}},\Delta_{-\frac{1}{2}},\Delta^\prime_{-\frac{1}{2}}\right\}$. Inside the subspace $P$, we identify a two-dimensional subsystem $S$ to define the qubit. For the subsystem qubit, the subsystem $S$ is specified by the formal quantum number $l$. The irrelevant subsystem $B$ is characterized by the $s_z$-quantum number (cf. \autoref{ssec:SsubAndSsys}).

The projection of the master equation on the $P$ subspace works in the same way as described in Appendix \ref{ssec:InitSubspace}. We only need to use a projector $\mathcal{P}=\mathcal{P}_P$ on a four-dimensional subspace. We now study the modification of the effective master equation due to the introduction of the subsystem $S$ in the $P$ subspace. We start with the master equation $\dot{\rho}\left(t\right)=\mathcal{L}\left(t\right)\rho\left(t\right)$ with a time-dependent superoperator defined in \equationshortname~\eqref{eq:NZBornSubspace}:
\begin{equation}
\mathcal{L}\left(t\right)=L_{PP}+\mathcal{T}\left(t\right),
\label{eq:DefinestartSubsys}
\end{equation}
with
\begin{widetext}
\begin{align}
\mathcal{T}\left(t\right)\rho^P\left(t\right)=\int_{0}^tdt^\prime\left(\underbrace{L_{PQ}e^{L_{QQ}\left(t-t^\prime\right)}L_{QP}+L_{P+}L_{+P}+L_{P-}L_{-P}}_{\bm{\Sigma}\left(t-t^\prime\right)}\right)\rho^P\left(t^\prime\right).
\end{align}
\end{widetext}

$\mathcal{T}\left(t\right)$ integrates the density matrix over all past times. To describe the evolution on the subsystem, one uses a linear map $\mathcal{P}=\mathcal{P}_S$ consisting of a partial trace:

\begin{equation}
\mathcal{P}_S:\rho^{P}\left(t\right)\rightarrow
\rho^{B}_{0}\underbrace{Tr_B\left(\rho^{P}\left(t\right)\right)}_{\rho^{S}\left(t\right)}.
\label{eq:MapSubsystem}
\end{equation}

The linear map $\mathcal{P}_S$ fulfills especially the properties \eqref{eq:LinMap1} and \eqref{eq:LinMap2}. It extracts from the density matrix of the subspace $\rho^{P}\left(t\right)$, the density matrix of the subsystem $\rho^{S}\left(t\right)$. It should be emphasized that we exclude entanglement between the systems $S$ and $B$ through the choice of the map in \equationshortname~\eqref{eq:MapSubsystem}. We fix the subsystem $B$ to a static value $\rho^B_0$. The effective master equation for the subsystem $S$ can be rewritten for the time dependent superoperator \eqref{eq:DefinestartSubsys}, as shown by Fick and Sauermann\cite{fick1983,fick1990}:

\begin{align}\label{eq:NZTime1}
\frac{d}{dt}\rho^{S}\left(t\right)=&\mathcal{P}_S\mathcal{L}\left(t\right)\mathcal{P}_S\rho^{S}\left(t\right)\\&+\int_0^tdt^\prime\mathcal{P}_S\mathcal{L}\left(t\right)\mathcal{Q}_S\mathcal{V}\left(t,t^\prime\right)\mathcal{Q}_S\mathcal{L}\left(t^\prime\right)\mathcal{P}_S\rho^S\left(t^\prime\right),\nonumber
\end{align}
with
\begin{align}
\frac{d}{dt}\mathcal{V}\left(t,t^\prime\right)=\mathcal{Q}_S\mathcal{L}\left(t\right)\mathcal{V}\left(t,t^\prime\right).
\label{eq:NZTime2}
\end{align}

In the analysis of triple quantum dots we try to extract errors for the initial time evolution. For this purpose, we can rewrite the effective master equations \eqref{eq:NZTime1} and \eqref{eq:NZTime2} for the description of short times. We divide the Lindblad operator $L_{PP}$ from \equationshortname~\eqref{eq:DefinestartSubsys} into a part which acts just on the qubit subsystem $S$ ($\mathcal{L}_S$) or the irrelevant subsystem $B$ ($\mathcal{L}_B$) individually. The remaining dissipative term is identified by the operator $\mathcal{L}_{SB}$. $\mathcal{L}_{SB}$ should be small compared to the $\mathcal{L}_S$ and $\mathcal{L}_B$. In second-order Born approximation we get the effective master equation

\begin{align}\label{eq:NZBornSubsys}
\frac{d}{dt}\rho^{S}\left(t\right)=&\mathcal{P}_S\left(\mathcal{L}_{S}+\mathcal{L}_{SB}+\mathcal{T}\left(t\right)\right)\mathcal{P}_S\rho^{S}\left(t\right)\\&+\int_{0}^t dt^\prime\mathcal{P}_S\mathcal{L}_{SB}e^{\left(\mathcal{L}_S+\mathcal{L}_B\right)\left(t-t^\prime\right)}\mathcal{Q}_S\mathcal{L}_{SB}\mathcal{P}_S\rho^{S}\left(t^\prime\right).\nonumber
\end{align}

\section{\label{sec:LongTime}
Long Time Limit of Time Evolution}

Since in the DM the system equilibrates to thermal equilibrium, we can calculate the long time behavior of the models of \autoref{sec:FullTime} analytically. One needs to pay attention that only subspaces that are connected by internal transitions equilibrate.

For the subspace qubit under the influence of phase noise (see analysis in \autoref{sssec:FTR1}), we can restrict ourselves to the subspace $\left\{Q_{\frac{1}{2}}, \Delta_{\frac{1}{2}}, \Delta^\prime_{\frac{1}{2}}\right\}$. In the long time limit, the density matrix will show partial equilibration:
\begin{widetext}
\begin{align}
	&\rho_\infty^{\left\{Q_{\frac{1}{2}}, \Delta_{\frac{1}{2}}, \Delta^\prime_{\frac{1}{2}}\right\}}=
\left(\begin{array}{ccc}
e^{-\frac{E_{Q_{1/2}}}{T_K}}&0&0\\0&e^{-\frac{E_{\Delta_{1/2}}}{T_K}}&0\\0&0&e^{-\frac{E_{\Delta^\prime_{1/2}}}{T_K}}
\end{array}\right)/\sum_{i\in \left\{Q_{\frac{1}{2}},\Delta_{\frac{1}{2}},\Delta^\prime_{\frac{1}{2}}\right\}}\left(e^{-\frac{E_{i}}{T_K}}\right).
\end{align}
\end{widetext}
The long time limit for the population of the subspace qubit can be obtained from the total leakage to the quadruplet state:
\begin{equation}
	O_\infty=1-\frac{e^{-\frac{E_{Q_{1/2}}}{T_K}}}{e^{-\frac{E_{Q_{1/2}}}{T_K}}+e^{-\frac{E_{\Delta_{1/2}}}{T_K}}+e^{-\frac{E_{\Delta^\prime_{1/2}}}{T_K}}}.
	\label{eq:LTL1}
\end{equation}
Since the off-diagonal elements of the density matrix vanish, it is clear that $X_\infty=0$. The long time limit of the qubit's polarization can be calculated from the difference in the population of the $\Delta_{\frac{1}{2}}$ and $\Delta^\prime_{\frac{1}{2}}$ states:
\begin{equation}
	Z_\infty=\frac{e^{-\frac{E_{\Delta_{1/2}}}{T_K}}-e^{-\frac{E_{\Delta^\prime_{1/2}}}{T_K}}}{e^{-\frac{E_{Q_{1/2}}}{T_K}}+e^{-\frac{E_{\Delta_{1/2}}}{T_K}}+e^{-\frac{E_{\Delta^\prime_{1/2}}}{T_K}}}.
	\label{eq:LTL2}
\end{equation}
\equationshortname~\eqref{eq:LTL1} and \equationshortname~\eqref{eq:LTL2} are used to calculate the long time limit for the qubit evolution in \autoref{sssec:FTR1} (see especially insets of \figureshortname~\ref{fig:FullTime1}).

The subsystem qubit with local spin relaxations has a description which is slightly more complicated. The simulation of \autoref{sssec:FTR3} analyzes the specific situation of phase noise near the crossing points of energy levels [see the orange line in energy diagram \figureshortname~\ref{fig:EnergyDiag2}]. We take into account only transitions that occur on the time scale of microseconds. This limits us to transitions in two subspaces $ssp1$ and $ssp2$:
\begin{align}
ssp1&=\left\{Q_{\frac{1}{2}},\Delta_{\frac{1}{2}},\Delta^\prime_{-\frac{1}{2}}\right\},\\
ssp2&=\left\{Q_{\frac{3}{2}},\Delta^\prime_{\frac{1}{2}}\right\}.
\end{align}
In both subspaces thermal equilibrium is reached. Transition rates between these two subspaces and to the remaining states are very small. Only the $l=0$ and $l=1$ states are occupied at $t=0$ for a subsystem qubit (see qubit definition in \autoref{ssec:SsubAndSsys}). The final values for the qubit evolution are dependent on the initial density matrix [cf. \equationshortname~\eqref{eq:subsysbath}],
\begin{equation}
	\rho\left(0\right)=\left(\begin{array}{cc}
	P_{11}&P_{10}\\
	P_{01}&P_{00}
	\end{array}
	\right)\otimes\rho_{0}^{s_z},
	\label{eq:INITDENS}
\end{equation}
where $\rho\left(0\right)$ determines the part of the density matrix, which is initially part of $ssp1$ $\left(O^{ssp1}\right)$, of $ssp2$ $\left(O^{ssp2}\right)$ or remains unchanged $\left(O_u=1-O^{ssp1}-O^{ssp2}\right)$. The initial population of subspace $ssp1$ depends on the occupation of the states $\Delta_{\frac{1}{2}}$ and $\Delta^\prime_{-\frac{1}{2}}$. It can be described by the entries $P_{11}$ and $P_{00}$ of $\rho\left(0\right)$ from \equationshortname~\eqref{eq:INITDENS}, which is itself related to the initial polarization $P_z\left(0\right)$:
\begin{align}	\label{eq:LTL5}
	O^{ssp1}=&\frac{1+P_z\left(0\right)}{2}\frac{1}{1+e^{-\frac{E_z}{T_K}}}\\\nonumber
	&+\frac{1-P_z\left(0\right)}{2}\frac{e^{-\frac{E_z}{T_K}}}{1+e^{-\frac{E_z}{T_K}}}.
\end{align}
For subspace $ssp2$ only the initial occupation of the state $\Delta^\prime_{\frac{1}{2}}$ plays a role, which leads to
\begin{align}
	O^{ssp2}&=\frac{1-P_z\left(0\right)}{2}\frac{1}{1+e^{-\frac{E_z}{T_K}}}.\label{eq:LTL6}
\end{align}
The final population of the subsystem qubit is determined by all transition rates to the quadruplet states:
\begin{align}\label{eq:LTL3}
	O_\infty=&1-O_{Q_{\frac{1}{2}}}-O_{Q_{\frac{3}{2}}}\\\nonumber
	=&1-O^{ssp1}
	\frac{e^{-\frac{E_{Q_{1/2}}}{T_K}}}{e^{-\frac{E_{Q_{1/2}}}{T_K}}+e^{-\frac{E_{\Delta_{1/2}}}{T_K}}+e^{-\frac{E_{\Delta^\prime_{-1/2}}}{T_K}}}\\\nonumber&-
	O^{ssp2}\frac{e^{-\frac{E_{Q_{3/2}}}{T_K}}}{e^{-\frac{E_{Q_{3/2}}}{T_K}}+e^{-\frac{E_{\Delta^\prime_{1/2}}}{T_K}}}.
\end{align}
All superpositions vanish in the long time limit ($X_\infty=0$) and the final polarization can be calculated from the difference in population of the $l=0$ states and the $l=1$ states:
\begin{align}
Z_\infty=&O_{l=0}-O_{l=1}\label{eq:LTL4}\\\nonumber=&
O_u+
O^{ssp1}\frac{e^{-\frac{E_{\Delta_{1/2}}}{T_K}}-e^{-\frac{E_{\Delta^\prime_{-1/2}}}{T_K}}}{e^{-\frac{E_{Q_{1/2}}}{T_K}}+e^{-\frac{E_{\Delta_{1/2}}}{T_K}}+e^{-\frac{E_{\Delta^\prime_{-1/2}}}{T_K}}}
\\\nonumber&-O^{ssp2}\frac{e^{-\frac{E_{\Delta^\prime_{1/2}}}{T_K}}}{e^{-\frac{E_{Q_{3/2}}}{T_K}}+e^{-\frac{E_{\Delta^\prime_{1/2}}}{T_K}}}.
\end{align}
\equationshortname~\eqref{eq:LTL3} and \equationshortname~\eqref{eq:LTL4}, together with \equationshortname~\eqref{eq:LTL5}-\eqref{eq:LTL6}, can be used to describe the long time limit for the subsystem qubit in \autoref{sssec:FTR3} (see especially insets of \figureshortname~\ref{fig:FullTime3}).

\section{\label{sec:EffRates}
Error Description of the Single Qubit Time Evolution}

In this section we connect the description for the single qubit time evolution from the spin-based quantum computation community to the common one in quantum information theory. In the first one, the qubit evolution is described by the evolution on the Bloch sphere (compare, e.g., the recent review of Kloeffel and Loss.\cite{kloeffel2012}) One commonly uses maps on density matrices in an information theoretical approach.\cite{nielsen2000}

\subsection{\label{ssec:ERSolid}
Solid State Approach}

In a solid-state approach, one commonly uses two specific time scales to describe the evolution on the Bloch sphere, which originally came up in the literature of NMR.\cite{slichter1990} First, the longitudinal relaxation time $T_1$ describes the evolution from the excited qubit state $\ket{1}$ to the ground state $\ket{0}$. We call this time scale ``relaxation time'' in the following. Second, the transverse relaxation time $T_2$ (which we call ``dephasing time'') describes the relaxation of a quantum mechanical superposition $\left(\ket{1}+\ket{0}\right)/\sqrt{2}$ to a mixed state.

We describe a complex time evolution in our analysis, including leakage from the computational subspace to the embedding Hilbert space. We characterize this evolution by the introduction of a third time scale, which we call ``leakage time'' $T_0$. Even though all parameters are originally meant to describe the inverse rates of exponential time evolutions, we are fitting our results of more complex dynamics to these parameters. We analyze the initial time evolution from points $\mathbf{P}\left(0\right)=Tr\left(\bm{\sigma}\rho\left(0\right)\right)$ on the Bloch sphere, with $\bm{\sigma}=\left(\sigma_x,\sigma_y,\sigma_z\right)$, and extract the leakage rate $\Gamma_{0}^{\mathbf{P}\left(0\right)}$, the relaxation rate $\Gamma_{1}^{\mathbf{P}\left(0\right)}$ and the dephasing rate $\Gamma_{2}^{\mathbf{P}\left(0\right)}$ of the initial time evolution. We correct all rates by a factor linear in the time argument to help account for the non exponential behavior:
\begin{align}
	\Gamma_{i}^{\mathbf{P}\left(0\right)}\equiv \gamma_{i}^{\mathbf{P}\left(0\right)}+\varphi_{i}^{\mathbf{P}\left(0\right)}\delta t.
	\label{eq:ConventionTimeArg}
\end{align}
In \equationshortname~\eqref{eq:ConventionTimeArg} $\left(\gamma_{i}^{\mathbf{P}\left(0\right)},\varphi_{i}^{\mathbf{P}\left(0\right)}\right)\in\mathbb{R}$. The leakage time $T_{0}^{\mathbf{P}\left(0\right)}$ is described by the corresponding leakage rate $\Gamma_{0}^{\mathbf{P}\left(0\right)}=\left(T_{0}^{\mathbf{P}\left(0\right)}\right)^{-1}$ of the trace evolution of the relevant part of the density matrix (see description in Appendix \ref{sec:Init}):

\begin{equation}
	O_{\mathbf{P}\left(0\right)}\left(\delta t\right)\approx \left[1-\Gamma_0^{\mathbf{P}\left(0\right)}\delta t+\left(\Gamma_0^{\mathbf{P}\left(0\right)}\right)^2\frac{\delta t^2}{2}\right].
		\label{eq:defineleakagerate}
\end{equation}

Since leakage leads to a depopulation of the qubit, we renormalize all Bloch sphere parameters by the trace of the relevant part of the density matrix [$\widehat{P_i}\left(t\right)=P_i\left(t\right)/Tr\left(\rho^{rel}\left(t\right)\right)$]. We assign the relaxation time $T_{1}^{\mathbf{P}\left(0\right)}=\left(\Gamma_{1}^{\mathbf{P}\left(0\right)}\right)^{-1}$ to the $z$ evolution of the qubit from the initial polarization $\widehat{P}_{z}\left(0\right)$ to the final polarization $\widehat{Z}_\infty$:

\begin{align}
\widehat{P}_{z}\left(\delta t\right)\approx& \widehat{P}_{z}\left(0\right)\left[1-\Gamma_1^{\mathbf{P}\left(0\right)}\delta t+\left(\Gamma_1^{\mathbf{P}\left(0\right)}\right)^2\frac{\delta t^2}{2}\right]\\\nonumber&+
\widehat{Z}_\infty\left[\Gamma_1^{\mathbf{P}\left(0\right)}\delta t-\left(\Gamma_1^{\mathbf{P}\left(0\right)}\right)^2\frac{\delta t^2}{2}\right].
\end{align}

Dephasing describes the loss of phase coherence of a qubit. We especially refer to the relaxation of quantum mechanical superpositions to a mixed state. On the Bloch sphere it is connected to the rate at which a point on the surface of the Bloch sphere relaxes to the z-axis. We extract the dephasing time $T_{2}^{\mathbf{P}\left(0\right)}=\left(2\Gamma_{2}^{\mathbf{P}\left(0\right)}\right)^{-1}$\footnote{The factor 2 is used to agree with the definitions of Nielsen and Chuang.\cite{nielsen2000}} from the initial time evolution:

\begin{equation}
	\widehat{P}_{x}\left(\delta t\right)\approx \left[1-\Gamma_2^{\mathbf{P}\left(0\right)}\delta t+\left(\Gamma_2^{\mathbf{P}\left(0\right)}\right)^2\frac{\delta t^2}{2}\right]\widehat{P}_{x}\left(0\right).
	\label{eq:definedephasingrate}
\end{equation}

\begin{figure}
\includegraphics[width=0.4\textwidth]{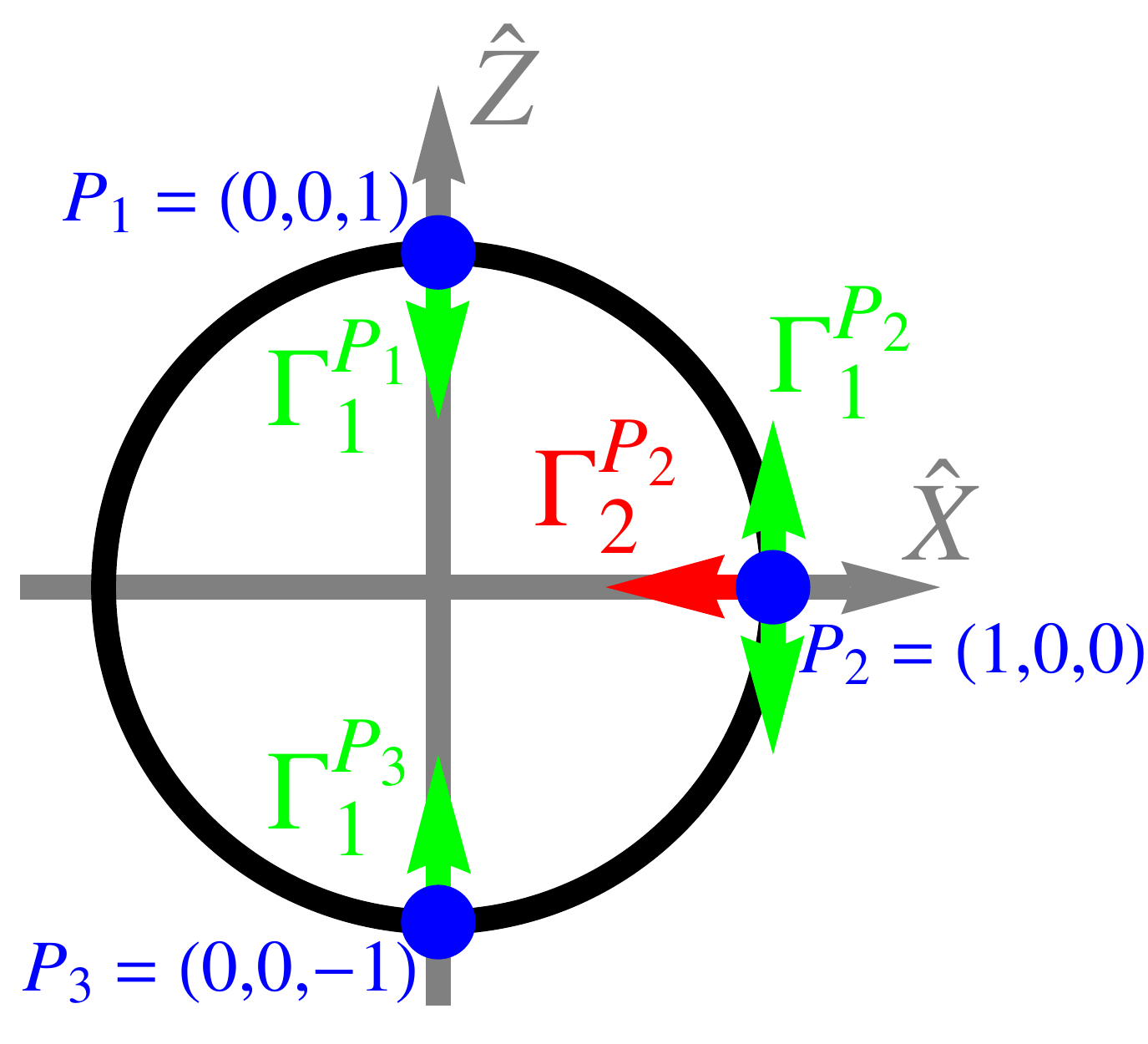}
\caption{\label{fig:EvolutionOnBlochSphere}
Sketch of transition rates, extracted from the initial time evolution of the qubit on the Bloch sphere at three special points. The Bloch sphere parameters are renormalized by the trace evolution of relevant part of the density matrix, which describes the qubit: $\widehat{P}_{x,z}\left(t\right)\equiv P_{x,z}\left(t\right)/Tr\left(\rho^{rel}\right)$.}
\end{figure}

Following the discussion of Appendix \ref{ssec:SimpRate}, we can restrict the analysis to just one plane (e.g. x-z-plane). It is sufficient to extract the parameters $\Gamma_0^{\mathbf{P}\left(0\right)}$, $\Gamma_1^{\mathbf{P}\left(0\right)}$ and $\Gamma_2^{\mathbf{P}\left(0\right)}$ only for three values of $\mathbf{P}\left(0\right)$, i.e. only at three points on the surface of the Bloch sphere, to describe the full time evolution of the qubit (cf. \autoref{ssec:ERInfo}). From the upper and the lower pole $\mathbf{P}\left(0\right)=\mathbf{P}_{1\left(3\right)}=\left(0,0,\left(-\right)1\right)$, we extract the leakage and the relaxation rate to the opposite pole. Because of the properties of the DM, the trajectory exactly follows the z-axis (cf. Appendix \ref{ssec:SimpRate}). From one point of the equator, e.g. $\mathbf{P}\left(0\right)=\mathbf{P}_{2}=\left(1,0,0\right)$, we extract the leakage, relaxation, and dephasing rates. The relaxation rate is extracted from the time evolution to the north or the south pole. Initially just one of the rates, defined in \equationshortname~\eqref{eq:definedephasingrate}, is positive. This positive number defines the relaxation rate $\Gamma_1^{\mathbf{P}_2}$. A sketch of all transition rates on the Bloch sphere is shown in \figureshortname~\ref{fig:EvolutionOnBlochSphere}.

\subsection{\label{ssec:ERInfo}
Information Theoretical Approach}

Commonly, one describes the time evolution through the completely positive linear map $\varepsilon_{\delta t}$ in an information theoretical approach. $\varepsilon_{\delta t}$ constructs from the initial density matrix $\rho\left(0\right)$, the density matrix at some later time $\rho\left(\delta t\right)$:

\begin{equation}
\rho\left(\delta t\right)=\varepsilon_{\delta t}\left(\rho\left(0\right)\right)
\end{equation}

In our analysis $\varepsilon_{\delta t}$ is trace decreasing, since we also take into account leakage to the surroundings. Due to the special trajectory generated in our model (cf. Appendix \ref{ssec:SimpRate}), we only need seven free parameters to completely describe the initial time evolution of our system. Since the map $\varepsilon_{\delta t}$ is linear, it is fixed completely by its action on four pure states $\ket{0}$, $\ket{1}$, and $\ket{+ \atop \left(i\right)}\equiv\left(\ket{1} +{1 \atop \left(i\right)}\ket{0}\right)/\sqrt{2}$:

\begin{align}
\epsilon\left(\op{1}{1}\right)=&
a_{1}\op{1}{1}+
a_{2}\op{0}{0},\\
\epsilon\left(\op{0}{0}\right)=&
a_{3}\op{1}{1}+
a_{4}\op{0}{0},\\
\epsilon\left(\op{+ \atop \left(i\right)}{+ \atop \left(i\right)}\right)=&
a_{5}\op{1}{1}+
a_{6}\op{0}{0}\\\nonumber&+
a_{7}\op{+ \atop \left(i\right)}{+ \atop \left(i\right)}.
\end{align}

It is straightforward to relate the parameters $a_1-a_7$ to the evolution rates $\Gamma_{i}^{\mathbf{P}_j}$, that were defined earlier [cf. \equationshortname~\eqref{eq:defineleakagerate}-\eqref{eq:definedephasingrate}]:

\begin{widetext}
\begin{align}
a_1=&1 - \left(\Gamma_0^{\mathbf{P}_1} + \Gamma_1^{\mathbf{P}_1}\right) \delta t +\left(\Gamma_0^{\mathbf{P}_1} + \Gamma_1^{\mathbf{P}_1}\right)^2\frac{\delta t^2}{2}+\mathcal{O}\left(\delta t^3\right),\\
a_2=& \Gamma_1^{\mathbf{P}_1} \delta t -\Gamma_1^{\mathbf{P}_1}\left(\Gamma_0^{\mathbf{P}_1} + \frac{\Gamma_1^{\mathbf{P}_1}}{2}\right)\delta t^2+\mathcal{O}\left(\delta t^3\right),\\
a_3=& \Gamma_1^{\mathbf{P}_3} \delta t -\Gamma_1^{\mathbf{P}_3}\left(\Gamma_0^{\mathbf{P}_3} + \frac{\Gamma_1^{\mathbf{P}_3}}{2}\right)\delta t^2+\mathcal{O}\left(\delta t^3\right),\\
a_4=& 1 - \left(\Gamma_0^{\mathbf{P}_3} + \Gamma_1^{\mathbf{P}_3}\right) \delta t +\left(\Gamma_0^{\mathbf{P}_3} + \Gamma_1^{\mathbf{P}_3}\right)^2\frac{\delta t^2}{2}+\mathcal{O}\left(\delta t^3\right),\\
a_5=&
\frac{\Gamma_1^{\mathbf{P}_2}+\Gamma_2^{\mathbf{P}_1}}{2}\delta t-
\left(\Gamma_0^{\mathbf{P}_2}\left(\Gamma_1^{\mathbf{P}_2}+\Gamma_2^{\mathbf{P}_2}\right)
+\frac{\left(\Gamma_1^{\mathbf{P}_2}\right)^2+\left(\Gamma_2^{\mathbf{P}_1}\right)^2}{2}\right)\frac{\delta t^2}{2}+\mathcal{O}\left(\delta t^3\right),\\
a_6=&
\frac{-\Gamma_1^{\mathbf{P}_2}+\Gamma_2^{\mathbf{P}_1}}{2}\delta t+
\left(\Gamma_1^{\mathbf{P}_2}-\Gamma_2^{\mathbf{P}_2}\right)\left(\Gamma_0^{\mathbf{P}_2}+\frac{\Gamma_1^{\mathbf{P}_2}+\Gamma_2^{\mathbf{P}_1}}{2}\right)\frac{\delta t^2}{2}+\mathcal{O}\left(\delta t^3\right),\\
a_7=&
1 - \left(\Gamma_0^{\mathbf{P}_2} + \Gamma_2^{\mathbf{P}_2}\right) \delta t +\left(\Gamma_0^{\mathbf{P}_2} + \Gamma_2^{\mathbf{P}_2}\right)^2\frac{\delta t^2}{2}+\mathcal{O}\left(\delta t^3\right).
\end{align}
\end{widetext}

Various properties of interest can be calculated from the map $\varepsilon_{\delta t}$. One example is the entanglement fidelity: $F_{e}=Tr\left[\rho^{RG}\left(\mathbf{1}\otimes\varepsilon_{\delta t}\right)\left(\rho^{RQ}\right)\right]$. $\rho^{RQ}$ is the maximally entangled state of the noisy system $Q$ and the reference system $R$: $\rho^{RQ}=\sum_{ij}\op{ii}{jj}/2$. It describes how well the entanglement between two systems is preserved under the action of the noisy quantum channel $\varepsilon_{\delta t}$\cite{marinescu2012}:

\begin{widetext}
\begin{align}\label{eq:entfid}
	F_{e}=&\frac{a_1+a_4}{4}+\frac{a_7}{2}\\\nonumber=&1-\left[
	\left(\Gamma_0^{\mathbf{P}_1}+\Gamma_1^{\mathbf{P}_1}\right)+
	\left(\Gamma_0^{\mathbf{P}_3}+\Gamma_1^{\mathbf{P}_3}\right)
	+2
	\left(\Gamma_0^{\mathbf{P}_2}+\Gamma_2^{\mathbf{P}_2}\right)\right]\frac{\delta t}{4}+
	\left[
	\left(\Gamma_0^{\mathbf{P}_1}+\Gamma_1^{\mathbf{P}_1}\right)^2
	+
	\left(\Gamma_0^{\mathbf{P}_3}+\Gamma_1^{\mathbf{P}_3}\right)^2+2
	\left(\Gamma_0^{\mathbf{P}_2}+\Gamma_2^{\mathbf{P}_2}\right)^2\right]\frac{\delta t^2}{8}\\\nonumber&
	+\mathcal{O}\left(\delta t^3\right).\nonumber
\end{align}
\end{widetext}

\subsection{\label{ssec:SimpRate}
Error Rates in Our Model}

To describe the initial time evolution, we will see that it is sufficient to use a set of just seven parameters. The initial evolution is a trajectory on the Bloch sphere with full rotation symmetry around the z-axis and reflection symmetry to any plane containing the z-axis. Consequently we can restrict all our analysis to one plane (e.g., the x-z-plane). Additionally, the trajectory starting on the north or the south pole of the Bloch sphere is strictly restricted to the z-axis.

These symmetries are very specific to the analysis of the problem in the DM in \equationshortname~\eqref{eq:DaviesModel} and the specific form of the quantum jump terms [see \equationshortname~\eqref{eq:Lindbl1} and \eqref{eq:Lindbl2}]. The dissipative terms of the DM are collected to generate transitions between equidistant energy levels through the superoperators $\mathcal{D}\left[\mathcal{A}_{\omega}\right]$. This picture will prove to be very helpful to explain the symmetry of the trajectory. The Lindblad operator is the generator of time evolution for the density matrix. It maps the initial density matrix to the density matrix at some later time:
\begin{align}
\rho\left(\delta t\right)&=e^{\mathcal{L}\delta t}\rho\left(0\right)\\
&=\left(1+\mathcal{L}\delta t+\mathcal{L}^2\frac{\delta t^2}{2}+\dots\right)\rho\left(0\right).\label{eq:timeevol2}
\end{align}
\equationshortname~\eqref{eq:timeevol2} makes it clear that all possible combinations of superoperators $\mathcal{D}\left[\mathcal{A}_{\omega}\right] \mathcal{D}\left[\mathcal{B}_{\mu}\right] \mathcal{D}\left[\mathcal{C}_{\nu}\right] \dots$ will act on the initial density matrix to generate the density matrix at some later time. For the subspace qubit we start with a density matrix:
\begin{align}\nonumber
 \rho\left(0\right)=&\left(
 \begin{array}{cc}
 \frac{O_0+Z_0}{2}& \frac{X_0-iY_0}{2}\\
 \frac{X_0+iY_0}{2}& \frac{O_0-Z_0}{2}
 \end{array}
 \right)_{S=\frac{1}{2},s_z=\frac{1}{2}}\\&\oplus
 \left(\mathbf{0}_{2}\right)_{S=\frac{1}{2},s_z=-\frac{1}{2}}\oplus
 \left(\mathbf{0}_{4}\right)_{S=\frac{3}{2}}.
 \end{align}

Initially there is no population in the subspace spanned by $\left\{\Delta_{-\frac{1}{2}},\Delta^\prime_{-\frac{1}{2}}\right\}$ and in the quadruplet subspace. It can be proven easily (see below) that the action of any combination of quantum jumps on the density matrix will lead to a density matrix of this form:

\begin{widetext}
\begin{align}\nonumber
	\mathcal{L}^n\left(\rho\left(0\right)\right)=&
	\left( \begin{array}{cc}
 \alpha_1O_0+\alpha_2Z_0& \alpha_3\left(X_0-iY_0\right)\\
 \alpha_3\left(X_0+iY_0\right)& \alpha_4O_0+\alpha_5Z_0
 \end{array}\right)_{S=\frac{1}{2},s_z=\frac{1}{2}}
 \oplus
 	\left( \begin{array}{cc}
 \beta_1O_0+\beta_2Z_0& \beta_3\left(X_0-iY_0\right)\\
 \beta_3\left(X_0+iY_0\right)& \beta_4O_0+\beta_5Z_0
 \end{array}\right)_{S=\frac{1}{2},s_z=-\frac{1}{2}}\\
 &\oplus
 	\left( \begin{array}{cccc}
\gamma_1O_0+\gamma_2Z_0&0&0&0\\
0&\gamma_3O_0+\gamma_4Z_0&0&0\\
0&0&\gamma_5O_0+\gamma_6Z_0&0\\
0&0&0&\gamma_7O_0+\gamma_8Z_0
 \end{array}\right)_{S=\frac{3}{2}}.\label{eq:timeevssp}
\end{align}
\end{widetext}

The coefficients $\left(\alpha_i,\beta_i,\gamma_i\right)$ are real numbers representing the action of quantum jumps between energy levels. Inspecting the density matrix of \equationshortname~\eqref{eq:timeevssp}, it is clear that the projected part on the qubit subspace will have the same ratio between the x- and y-polarization as the initial density matrix. The trace evolution and the z-evolution is however dependent on the initial z-polarization of the qubit. Since this finding is true for all summands of \equationshortname~\eqref{eq:timeevol2}, it is also true for $\rho\left(\delta t\right)$. Given these restrictions on the generated density matrix, the trajectory on the Bloch sphere will have the specific form described earlier.

We point out how to prove \equationshortname~\eqref{eq:timeevssp} with some easy calculations. All quantum jump transitions can be grouped into two sets. First there are transitions involving only the computational subspace. They can represent pure relaxation (model 1 in Appendix \ref{ssec:MS1}) or pure dephasing (model 2 in Appendix \ref{ssec:MS2}). These models generate transitions in the computational subspace via the coupling operators [cf. \equationshortname~\eqref{eq:DaviesSumExpansion}]:
\begin{equation}
\mathcal{A}\in\left\{
\left(
\begin{array}{cc}
\beta_1&0\\
0&\beta_2
\end{array}
\right),\ \left(
\begin{array}{cc}
0&\beta_3\\
0&0
\end{array}
\right),\ \left(
\begin{array}{cc}
0&0\\
\beta_4&0
\end{array}
\right)\right\},
\end{equation}
with real coefficients $\beta_i$. An initial density matrix $\rho_0=\left( \begin{array}{cc}
 \alpha_1O_0+\alpha_2Z_0& \alpha_3\left(X_0-iY_0\right)\\
 \alpha_3\left(X_0+iY_0\right)& \alpha_4O_0+\alpha_5Z_0
 \end{array}\right)$ will have structurally the same form after the action of one dissipative term ($\mathcal{D}\left[\mathcal{A}\right]\left(\rho_0\right)$). Only the constants $\alpha_i$ will be modified.
 
Transitions involving the remaining Hilbert space will have again two distinct features. First there are quantum jump terms involving just transitions between two energy levels. One can calculate the action in a three-dimensional Hilbert space spanned by the two qubit levels and the coupled external level. An easy calculation shows that the structure of an initial density matrix,
\begin{equation}
\rho_0=\left( \begin{array}{ccc}
 \alpha_1O_0+\alpha_2Z_0& \alpha_3\left(X_0-iY_0\right)&0\\
 \alpha_3\left(X_0+iY_0\right)& \alpha_4O_0+\alpha_5Z_0&0\\
 0&0&\alpha_6O_0+\alpha_7Z_0
 \end{array}\right),
\end{equation}
will remain unchanged. Again only the constants $\alpha_i$ will be modified to different real numbers.\\
Secondly there are quantum jumps involving more than three energy levels. They can be made up of all the transitions introduced so far, but acting on separate subspaces. Consequently they also preserve the structure above. Otherwise, they couple the computational subspace to the $\left(S=\frac{1}{2},s_z=-\frac{1}{2}\right)$ subspace involving correlated quantum jumps between the same $l$ eigenstates. These transitions preserve also the off-diagonal elements of the density matrix on the computational subspace. They never mix diagonal and off-diagonal elements.

The same result is obtained for the subsystem qubit. In fact, all arguments will be identical, since the initial density matrix of the subsystem qubit is already in the form of \equationshortname~\eqref{eq:timeevssp}:

\begin{widetext}
\begin{align}\nonumber
\rho\left(0\right)=& \left(\begin{array}{cc}
 \frac{O_0+Z_0}{2}& \frac{X_0-iY_0}{2}\\
 \frac{X_0+iY_0}{2}& \frac{O_0-Z_0}{2}
 \end{array}
 \right)_{l}
 \otimes \left(\begin{array}{cc}
 \frac{1}{1+e^{-\frac{Ez}{T_K}}}& 0\\
 0&  \frac{e^{-\frac{Ez}{T_K}}}{1+e^{-\frac{Ez}{T_K}}}
 \end{array}
 \right)_{s_z}\oplus \left(\mathbf{0}_4\right)_{S=\frac{3}{2}}
 \\\nonumber
 =&  \frac{1}{1+e^{-\frac{Ez}{T_K}}}\left(\begin{array}{cc}
 \frac{O_0+Z_0}{2}& \frac{X_0-iY_0}{2}\\
 \frac{X_0+iY_0}{2}& \frac{O_0-Z_0}{2}
 \end{array}
 \right)_{S=\frac{1}{2},s_z=\frac{1}{2}}\oplus
 \frac{e^{-\frac{Ez}{T_K}}}{1+e^{-\frac{Ez}{T_K}}}\left(\begin{array}{cc}
 \frac{O_0+Z_0}{2}& \frac{X_0-iY_0}{2}\\
 \frac{X_0+iY_0}{2}& \frac{O_0-Z_0}{2}
 \end{array}
 \right)_{S=\frac{1}{2},s_z=-\frac{1}{2}}
 \oplus \left(\mathbf{0}_4\right)_{S=\frac{3}{2}}.
 \end{align}
 \end{widetext}

\section{\label{sec:MS}
Model Systems}

\begin{figure}
\centering
\subfigure[Model 1: Pure Relaxation]{\label{fig:Model1}\includegraphics[width=0.2375\textwidth]{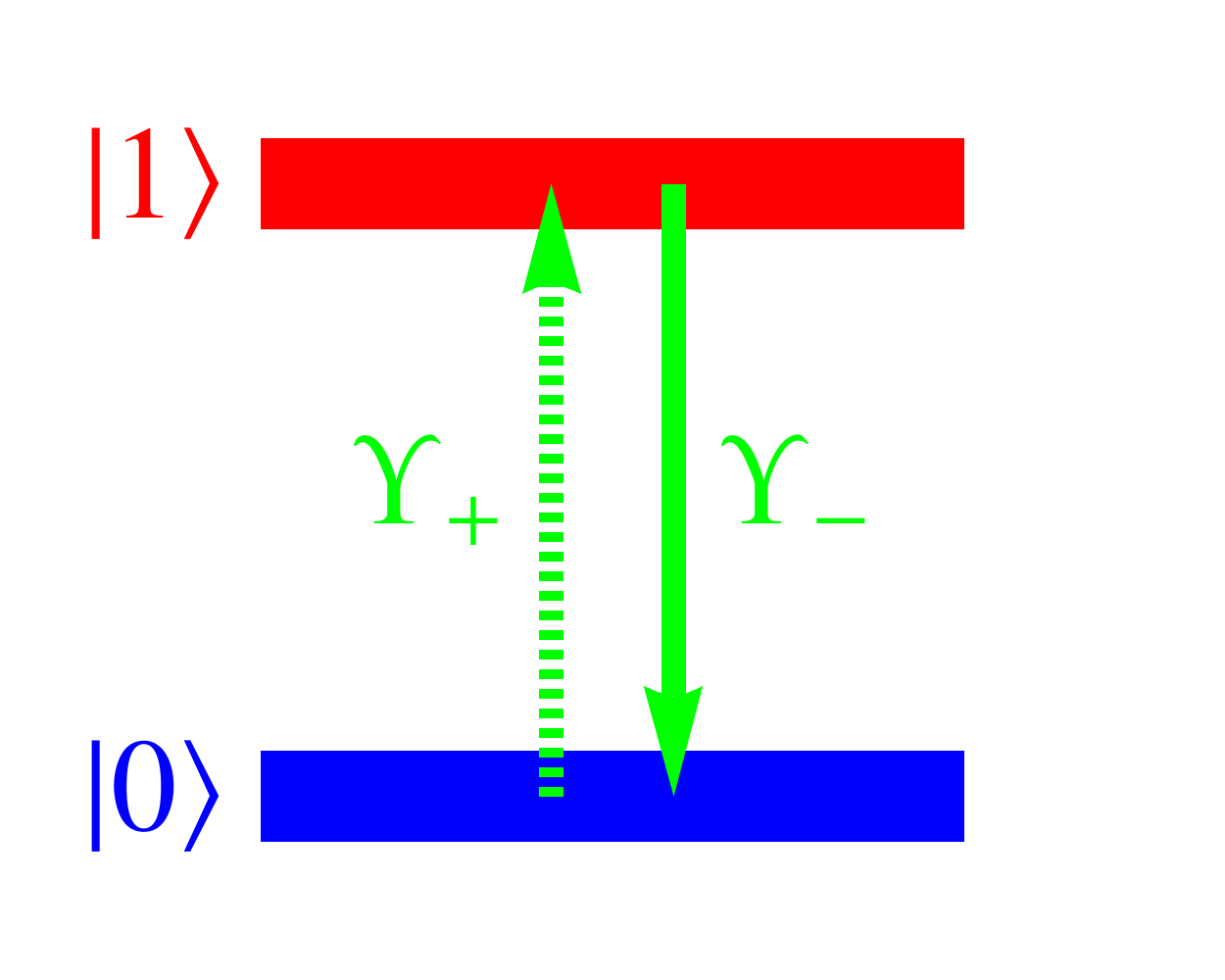}}
\subfigure[Model 2: Pure Dephasing]{\label{fig:Model2}\includegraphics[width=0.2375\textwidth]{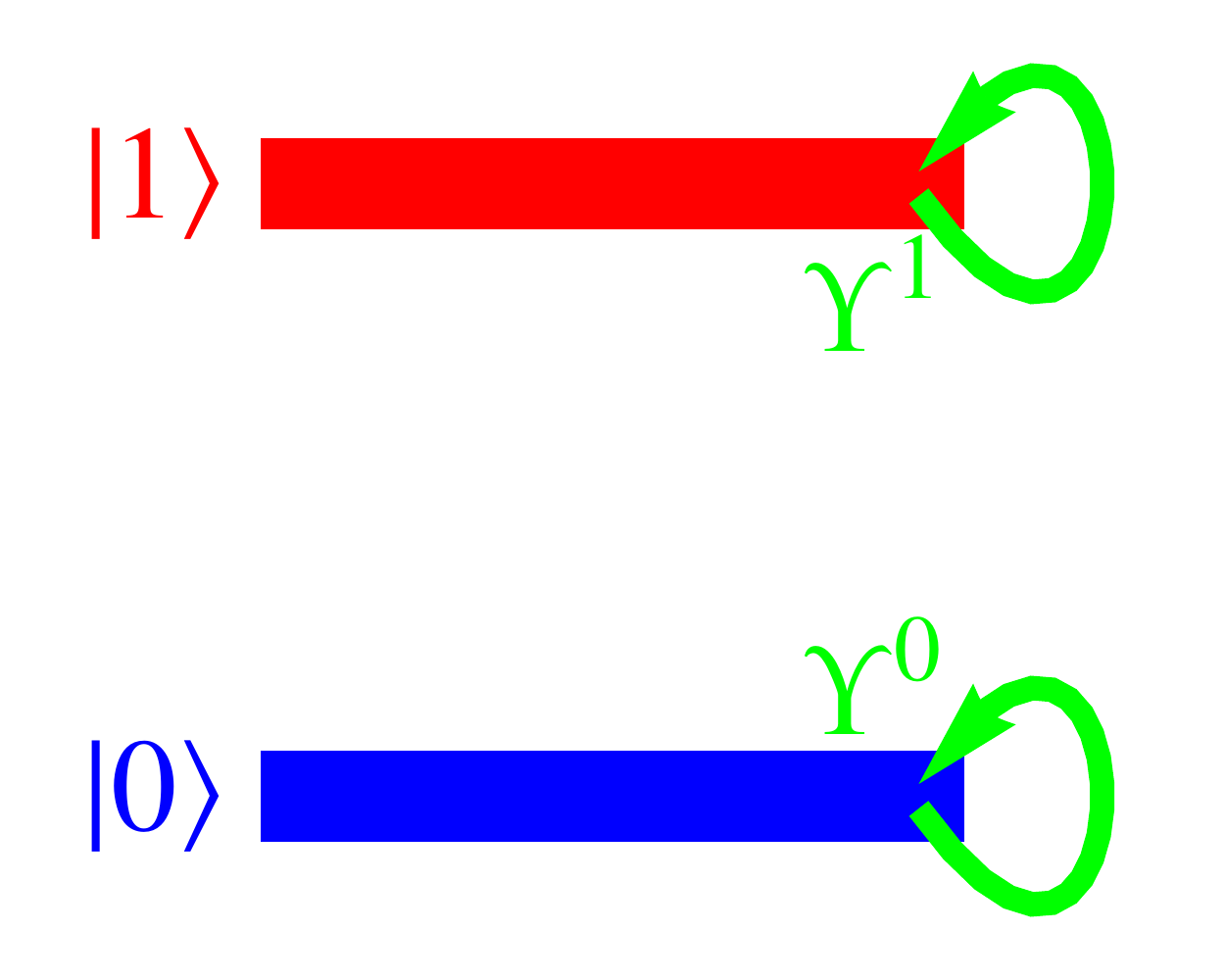}}
\subfigure[Model 3: Two State Leakage]{\label{fig:Model3}\includegraphics[width=0.2375\textwidth]{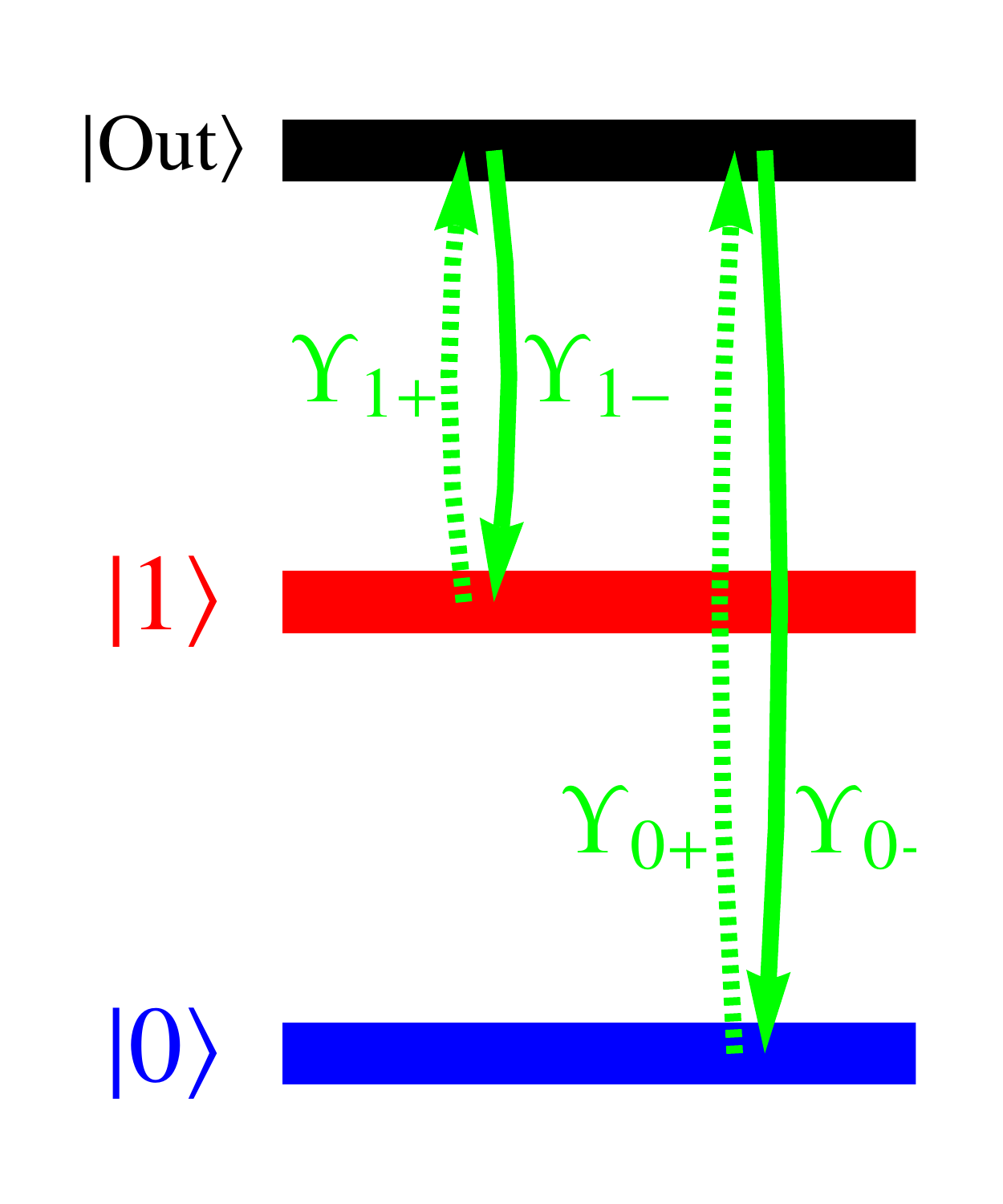}}
\subfigure[Model 4: Internal Transitions of the Subsystem Qubit]{\label{fig:Model4}\includegraphics[width=0.2375\textwidth]{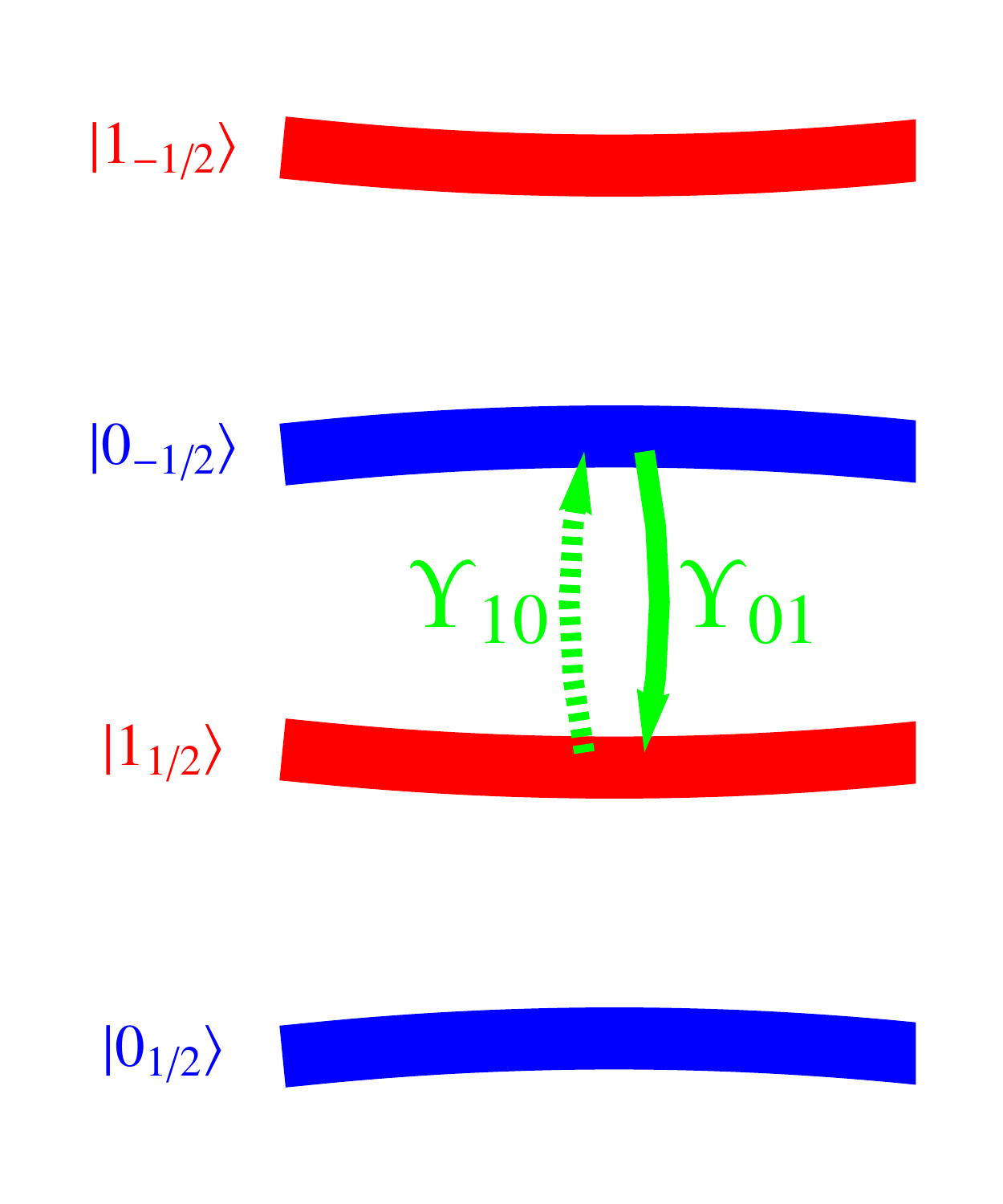}}
\caption{\label{fig:modelsystems}
The time evolution of the qubit in triple quantum dots can be described by four toy models (a)-(d).}
\end{figure}

We present in \figureshortname~\ref{fig:modelsystems} four model systems to describe the effective error rates of the qubit defined in triple quantum dot systems. In the DM energy eigenstates can couple through quantum jumps. Transitions are possible between the qubit levels $\ket{1}$ and $\ket{0}$, but also to other states of the embedding Hilbert space $\ket{\text{Out}}$. The strength of the quantum jumps is specified by the coefficients $\Upsilon_i\in\mathbb{R}$, which are determined by two constants. First of all, there are the transition rates $h\left(\mathcal{A},\omega\right)$, which are extracted from experiments [for the interaction with hyperfine fields, cf. $h\left(\sigma_z^i/\sigma_x^i,\omega\right)$ in \equationshortname~\eqref{eq:ratehyperfine}; for the interaction with phonons, cf. $h\left(\sigma_x^i,\omega\right)$ in \equationshortname~\eqref{eq:ratephonon}]. Secondly, we need also the matrix elements of the transition operator between the energy eigenstates [cf. \equationshortname~\eqref{eq:overlap}]. While $\Upsilon_i$ can be negative, only the positive number $\Upsilon_i^2$ describes a rate. In the following, we describe all four toy model systems individually.

\subsection{\label{ssec:MS1}
Model 1: Pure Relaxation}

The model of pure relaxation describes the transitions in a two-level system through raising and lowering operators: $\Upsilon_{\pm}\sigma_{\pm}$ [cf. \figureshortname~\ref{fig:Model1}]. The full master equation is:
\begin{align}
\dot{\rho}\left(t\right)&=\mathcal{D}\left[\Upsilon_{+}\sigma_{+}\right]\left(\rho\left(t\right)\right)+\mathcal{D}\left[\Upsilon_{-}\sigma_{-}\right]\left(\rho\left(t\right)\right)\\
&=\Upsilon_{+}^2\mathcal{D}\left[\sigma_{+}\right]\left(\rho\left(t\right)\right)+\Upsilon_{-}^2\mathcal{D}\left[\sigma_{-}\right]\left(\rho\left(t\right)\right).
\label{eq:FMEFME}
\end{align}
\equationshortname~\eqref{eq:FMEFME} can easily be solved. The effective error rates, defined in \autoref{sec:EffRates}, can be extracted. Assuming $\Upsilon_-^2>\Upsilon_+^2$, we get
\begin{align}
\Gamma_0^{\mathbf{P}_1,\mathbf{P}_2,\mathbf{P}_3}&=0\\
\Gamma_1^{\mathbf{P}_1}&=\Upsilon_-^2-\frac{1}{2}\Upsilon_+^2\Upsilon_-^2\delta t\\
\Gamma_1^{\mathbf{P}_2}&=\left(\Upsilon_-^2-\Upsilon_+^2\right)-\frac{1}{2}\Upsilon_+^2\left(\Upsilon_-^2-\Upsilon_+^2\right)\delta t\\
\Gamma_3^{\mathbf{P}_1}&=\Upsilon_+^2-\frac{1}{2}\Upsilon_+^2\Upsilon_-^2\delta t\\
\Gamma_2^{\mathbf{P}_2}&=\frac{1}{2}\left(\Upsilon_+^2+\Upsilon_-^2\right).
\end{align}
It is clear that no leakage arises in this model. Relaxation from the $\mathbf{P}_1$ and $\mathbf{P}_3$ is determined by the direct transitions to the opposite pole. The combination of the two transition rates gives a reduction of the overall error. At $\mathbf{P}_2$, we initially see relaxation to the lower pole with a rate determined by the difference of the two transition rates.

\subsection{\label{ssec:MS2}
Model 2: Pure Dephasing}

Pure dephasing [cf. \figureshortname~\ref{fig:Model2}] is described in the DM by the coupling operator:

\begin{equation}
\mathcal{A}=\left(\begin{array}{cc}\Upsilon_1&0\\0&\Upsilon_0\end{array}\right).
\label{eq:coplMod2}
\end{equation}
When solving the master equation $\dot{\rho}\left(t\right)=\mathcal{D}\left[\mathcal{A}\right]\left(\rho\left(t\right)\right)$, we extract the following transition rates:
\begin{align}
\Gamma_0^{\mathbf{P}_1,\mathbf{P}_2,\mathbf{P}_3}&=0,\\
\Gamma_1^{\mathbf{P}_1,\mathbf{P}_2,\mathbf{P}_3}&=0,\\
\Gamma_2^{\mathbf{P}_2}&=\frac{1}{2}\left(\Upsilon_1-\Upsilon_0\right)^2.
\label{eq:cancpuredeph}
\end{align}
The coupling operator \eqref{eq:coplMod2} generates neither relaxation nor leakage. $\mathcal{A}$ describes fluctuating energy levels, which leads to pure phase noise.

\subsection{\label{ssec:MS3}
Model 3: Two State Leakage}

In our calculation, we need to describe leakage of the qubit states to exactly one state of the embedding Hilbert space [cf. \figureshortname~\ref{fig:Model3}]. When solving the master equation 
\begin{align}
\dot{\rho}\left(t\right)=&\mathcal{D}\left[\Upsilon_{1+}\sigma_{+}^{1\rightarrow Out}\right]\left(\rho\left(t\right)\right)+
\mathcal{D}\left[\Upsilon_{1-}\sigma_{-}^{1\rightarrow Out}\right]\left(\rho\left(t\right)\right)\nonumber\\+&
\mathcal{D}\left[\Upsilon_{0+}\sigma_{+}^{0\rightarrow Out}\right]\left(\rho\left(t\right)\right)+
\mathcal{D}\left[\Upsilon_{0-}\sigma_{-}^{0\rightarrow Out}\right]\left(\rho\left(t\right)\right),
\end{align}
we can extract the effective error rates (assuming $\Upsilon_{1+}^2>\Upsilon_{0+}^2$):

\begin{widetext}
\begin{align}
\label{eq:model3-1}
\Gamma_0^{\mathbf{P}_1}=&\Upsilon_{1+}^2- \Upsilon_{1+}^2
\left(\Upsilon_{1-}^2+\Upsilon_{0-}^2\right)\frac{\delta t}{2},\\
\Gamma_0^{\mathbf{P}_2}=&\frac{\Upsilon_{1+}^2+\Upsilon_{0+}^2}{2}
-\frac{
\left(\Upsilon_{1+}^2-\Upsilon_{0+}^2\right)^2+2
\left(\Upsilon_{1+}^2+\Upsilon_{0+}^2\right)\left(\Upsilon_{1-}^2+\Upsilon_{0-}^2\right)}{8}\delta t,\\\label{eq:model3-3}
\Gamma_0^{\mathbf{P}_3}=&\Upsilon_{0+}^2- \Upsilon_{0+}^2
\left(\Upsilon_{1-}^2+\Upsilon_{0-}^2\right)\frac{\delta t}{2},\\
\Gamma_1^{\mathbf{P}_1}=&\Upsilon_{1+}^2 \Upsilon_{0-}^2 \frac{\delta t}{2},\\
\Gamma_1^{\mathbf{P}_2}=&\frac{\Upsilon_{1+}^2-\Upsilon_{0+}^2}{2}
+\frac{
\left(\Upsilon_{1+}^2-\Upsilon_{0+}^2\right)^2-2
\left(\Upsilon_{1+}^2+\Upsilon_{0+}^2\right)\left(\Upsilon_{1-}^2-\Upsilon_{0-}^2\right)}{8}\delta t,\\
\Gamma_1^{\mathbf{P}_3}=&\Upsilon_{0+}^2 \Upsilon_{1-}^2  \frac{\delta t}{2},\\
\Gamma_2^{\mathbf{P}_2}=&\frac{
\left(\Upsilon_{1+}^2-\Upsilon_{0+}^2\right)^2+2
\left(\Upsilon_{1+}^2+\Upsilon_{0+}^2\right)\left(\Upsilon_{1-}^2+\Upsilon_{0-}^2\right)}{8}\delta t.
\end{align}
\end{widetext}

We need in particular two special cases of this model. First of all, for the subspace qubit we analyze the case where only the transition rates from the qubit levels to the surroundings are significant. We can set $\Upsilon_{1+}=\Upsilon_{1}$, $\Upsilon_{0+}=\Upsilon_{2}$ and $\Upsilon_{1-}=\Upsilon_{0-}=0$ and obtain the error rates:
\begin{align}
\Gamma_0^{\mathbf{P}_1}=&\Upsilon_{1}^2\label{eq:model2-b},\\
\Gamma_0^{\mathbf{P}_2}=&\frac{\Upsilon_{1}^2+\Upsilon_{2}^2}{2}-\frac{
\left(\Upsilon_{1}^2-\Upsilon_{2}^2\right)^2}{8}\delta t\label{eq:model2-2},\\
\Gamma_0^{\mathbf{P}_3}=&\Upsilon_{2}^2\label{eq:model2-3},\\
\Gamma_1^{\mathbf{P}_1}=&0\label{eq:model2-4},\\
\Gamma_1^{\mathbf{P}_2}=&\frac{\Upsilon_{1}^2-\Upsilon_{2}^2}{2}+\frac{
\left(\Upsilon_{1}^2+\Upsilon_{2}^2\right)^2}{8}\delta t\label{eq:model2-5},\\
\Gamma_1^{\mathbf{P}_3}=&0\label{eq:model2-6},\\
\Gamma_2^{\mathbf{P}_2}=&\frac{
\left(\Upsilon_{1}^2-\Upsilon_{2}^2\right)^2}{8}\label{eq:model2-e}.
\end{align}
Here, for the north and south pole ($\mathbf{P}_1$ and $\mathbf{P}_3$) no relaxation is generated and only leakage occurs.

Second, for the analysis of the subsystem qubit we need to describe the transition of only one qubit level to the surroundings. For $\Upsilon_{0+}=\Upsilon_+$ and  $\Upsilon_{0-}=\Upsilon_-$, with $\Upsilon_{1+}=\Upsilon_{1-}=0$, we get
\begin{align}
\Gamma_0^{\mathbf{P}_1}=&0,\\
\Gamma_0^{\mathbf{P}_2}=&\frac{\Upsilon_{+}^2}{2}-\frac{\Upsilon_{+}^2}{8}\left(\Upsilon_{+}^2+2\Upsilon_{-}^2\right)\delta t,\\
\Gamma_0^{\mathbf{P}_3}=&\Upsilon_{+}^2-\frac{1}{2} \Upsilon_{+}^2
\Upsilon_{-}^2\delta t,\\
\Gamma_1^{\mathbf{P}_1}=&0,\\
\Gamma_1^{\mathbf{P}_2}=&\frac{\Upsilon_{+}^2}{2}+\frac{\Upsilon_{+}^2}{8}\left(\Upsilon_{+}^2-2\Upsilon_{-}^2\right)\delta t,\\
\Gamma_1^{\mathbf{P}_3}=&0,\\
\Gamma_2^{\mathbf{P}_2}=&\frac{\Upsilon_{+}^2}{8}\left(\Upsilon_{+}^2+2\Upsilon_{-}^2\right)\delta t.
\end{align}
On the north pole $\mathbf{P}_1$ of the Bloch sphere neither leakage nor relaxation is seen. On the south pole $\mathbf{P}_3$, we observe pure leakage.

\subsection{\label{ssec:MS4}
Model 4: Internal Transitions of the Subsystem Qubit}

For the analysis of the subsystem qubit, we will need the extract error rates at the crossing of two levels defining the qubit. We can solve the Davies master equation $\dot{\rho}\left(t\right)=\mathcal{D}\left[\mathcal{A}\right]\left(\rho\left(t\right)\right)$, describing the toy model of \figureshortname~\ref{fig:Model4}. For the case $\Upsilon_{10}^2>e^{-\frac{Ez}{T_K}}\Upsilon_{01}^2$, we get the error rates:

\begin{widetext}
\begin{align}
\Gamma_0^{\mathbf{P}_1,\mathbf{P}_2,\mathbf{P}_3}=&0,\\
\Gamma_1^{\mathbf{P}_1}=&\frac{\Upsilon_{10}^2}{1+e^{-\frac{Ez}{T_K}}}
-\frac{\Upsilon_{10}^2}{1+e^{-\frac{Ez}{T_K}}}
\left(\frac{e^{-\frac{Ez}{T_K}}}{1+e^{-\frac{Ez}{T_K}}}\Upsilon_{10}^2+\Upsilon_{01}^2\right)
\frac{\delta t}{2},\\
\Gamma_1^{\mathbf{P}_2}=&\frac{\Upsilon_{10}^2-e^{-\frac{Ez}{T_K}}\Upsilon_{01}^2}{1+e^{-\frac{Ez}{T_K}}}+\left(\frac{\Upsilon_{10}^2-e^{-\frac{Ez}{T_K}}\Upsilon_{01}^2}{1+e^{-\frac{Ez}{T_K}}}\right)
\left(\frac{e^{-\frac{Ez}{T_K}}\left(\Upsilon_{10}^2+\Upsilon_{01}^2\right)}{1+e^{-\frac{Ez}{T_K}}}+\Upsilon_{01}^2\right)
\frac{\delta t}{2},\\
\Gamma_1^{\mathbf{P}_3}=&\frac{e^{-\frac{Ez}{T_K}}}{1+e^{-\frac{Ez}{T_K}}}\Upsilon_{01}^2
-\frac{e^{-\frac{Ez}{T_K}}}{1+e^{-\frac{Ez}{T_K}}}\Upsilon_{01}^2
\left(\Upsilon_{10}^2+\frac{e^{-\frac{Ez}{T_K}}}{1+e^{-\frac{Ez}{T_K}}}\Upsilon_{01}^2\right)
\frac{\delta t}{2},\\
\Gamma_2^{\mathbf{P}_2}=&\frac{\Upsilon_{10}^2+e^{-\frac{Ez}{T_K}}\Upsilon_{01}^2}{2\left(1+e^{-\frac{Ez}{T_K}}\right)}
-\frac{e^{-\frac{Ez}{T_K}}}{\left(1+e^{-\frac{Ez}{T_K}}\right)^2}
\left(\Upsilon_{10}^2-\Upsilon_{01}^2\right)^2
\frac{\delta t}{8}.
\end{align}
\end{widetext}

\bibliography{library}

\begin{thebibliography}{49}%
\makeatletter
\providecommand \@ifxundefined [1]{%
 \@ifx{#1\undefined}
}%
\providecommand \@ifnum [1]{%
 \ifnum #1\expandafter \@firstoftwo
 \else \expandafter \@secondoftwo
 \fi
}%
\providecommand \@ifx [1]{%
 \ifx #1\expandafter \@firstoftwo
 \else \expandafter \@secondoftwo
 \fi
}%
\providecommand \natexlab [1]{#1}%
\providecommand \enquote  [1]{``#1''}%
\providecommand \bibnamefont  [1]{#1}%
\providecommand \bibfnamefont [1]{#1}%
\providecommand \citenamefont [1]{#1}%
\providecommand \href@noop [0]{\@secondoftwo}%
\providecommand \href [0]{\begingroup \@sanitize@url \@href}%
\providecommand \@href[1]{\@@startlink{#1}\@@href}%
\providecommand \@@href[1]{\endgroup#1\@@endlink}%
\providecommand \@sanitize@url [0]{\catcode `\\12\catcode `\$12\catcode
  `\&12\catcode `\#12\catcode `\^12\catcode `\_12\catcode `\%12\relax}%
\providecommand \@@startlink[1]{}%
\providecommand \@@endlink[0]{}%
\providecommand \url  [0]{\begingroup\@sanitize@url \@url }%
\providecommand \@url [1]{\endgroup\@href {#1}{\urlprefix }}%
\providecommand \urlprefix  [0]{URL }%
\providecommand \Eprint [0]{\href }%
\providecommand \doibase [0]{http://dx.doi.org/}%
\providecommand \selectlanguage [0]{\@gobble}%
\providecommand \bibinfo  [0]{\@secondoftwo}%
\providecommand \bibfield  [0]{\@secondoftwo}%
\providecommand \translation [1]{[#1]}%
\providecommand \BibitemOpen [0]{}%
\providecommand \bibitemStop [0]{}%
\providecommand \bibitemNoStop [0]{.\EOS\space}%
\providecommand \EOS [0]{\spacefactor3000\relax}%
\providecommand \BibitemShut  [1]{\csname bibitem#1\endcsname}%
\let\auto@bib@innerbib\@empty
\bibitem [{\citenamefont {Hanson}\ \emph {et~al.}(2007)\citenamefont {Hanson},
  \citenamefont {Kouwenhoven}, \citenamefont {Petta}, \citenamefont {Tarucha},\
  and\ \citenamefont {Vandersypen}}]{hanson2007}%
  \BibitemOpen
  \bibfield  {author} {\bibinfo {author} {\bibfnamefont {R.}~\bibnamefont
  {Hanson}}, \bibinfo {author} {\bibfnamefont {L.~P.}\ \bibnamefont
  {Kouwenhoven}}, \bibinfo {author} {\bibfnamefont {J.~R.}\ \bibnamefont
  {Petta}}, \bibinfo {author} {\bibfnamefont {S.}~\bibnamefont {Tarucha}}, \
  and\ \bibinfo {author} {\bibfnamefont {L.~M.~K.}\ \bibnamefont
  {Vandersypen}},\ }\href {\doibase 10.1103/RevModPhys.79.1217} {\bibfield
  {journal} {\bibinfo  {journal} {Reviews of Modern Physics}\ }\textbf
  {\bibinfo {volume} {79}},\ \bibinfo {pages} {1217} (\bibinfo {year}
  {2007})}\BibitemShut {NoStop}%
\bibitem [{\citenamefont {Loss}\ and\ \citenamefont
  {DiVincenzo}(1998)}]{loss1998}%
  \BibitemOpen
  \bibfield  {author} {\bibinfo {author} {\bibfnamefont {D.}~\bibnamefont
  {Loss}}\ and\ \bibinfo {author} {\bibfnamefont {D.~P.}\ \bibnamefont
  {DiVincenzo}},\ }\href {\doibase 10.1103/PhysRevA.57.120} {\bibfield
  {journal} {\bibinfo  {journal} {Physical Review A}\ }\textbf {\bibinfo
  {volume} {57}},\ \bibinfo {pages} {120} (\bibinfo {year} {1998})}\BibitemShut
  {NoStop}%
\bibitem [{\citenamefont {Levy}(2002)}]{levy2002}%
  \BibitemOpen
  \bibfield  {author} {\bibinfo {author} {\bibfnamefont {J.}~\bibnamefont
  {Levy}},\ }\href {\doibase 10.1103/PhysRevLett.89.147902} {\bibfield
  {journal} {\bibinfo  {journal} {Physical Review Letters}\ }\textbf {\bibinfo
  {volume} {89}},\ \bibinfo {pages} {147902} (\bibinfo {year}
  {2002})}\BibitemShut {NoStop}%
\bibitem [{\citenamefont {Taylor}\ \emph {et~al.}(2005)\citenamefont {Taylor},
  \citenamefont {Engel}, \citenamefont {D{\"u}r}, \citenamefont {Yacoby},
  \citenamefont {Marcus}, \citenamefont {Zoller},\ and\ \citenamefont
  {Lukin}}]{taylor2005}%
  \BibitemOpen
  \bibfield  {author} {\bibinfo {author} {\bibfnamefont {J.~M.}\ \bibnamefont
  {Taylor}}, \bibinfo {author} {\bibfnamefont {H.-A.}\ \bibnamefont {Engel}},
  \bibinfo {author} {\bibfnamefont {W.}~\bibnamefont {D{\"u}r}}, \bibinfo
  {author} {\bibfnamefont {A.}~\bibnamefont {Yacoby}}, \bibinfo {author}
  {\bibfnamefont {C.~M.}\ \bibnamefont {Marcus}}, \bibinfo {author}
  {\bibfnamefont {P.}~\bibnamefont {Zoller}}, \ and\ \bibinfo {author}
  {\bibfnamefont {M.~D.}\ \bibnamefont {Lukin}},\ }\href {\doibase
  10.1038/nphys174} {\bibfield  {journal} {\bibinfo  {journal} {Nature
  Physics}\ }\textbf {\bibinfo {volume} {1}},\ \bibinfo {pages} {177} (\bibinfo
  {year} {2005})}\BibitemShut {NoStop}%
\bibitem [{\citenamefont {Petta}\ \emph {et~al.}(2005)\citenamefont {Petta},
  \citenamefont {Johnson}, \citenamefont {Taylor}, \citenamefont {Laird},
  \citenamefont {Yacoby}, \citenamefont {Lukin}, \citenamefont {Marcus},
  \citenamefont {Hanson},\ and\ \citenamefont {Gossard}}]{petta2005}%
  \BibitemOpen
  \bibfield  {author} {\bibinfo {author} {\bibfnamefont {J.~R.}\ \bibnamefont
  {Petta}}, \bibinfo {author} {\bibfnamefont {A.~C.}\ \bibnamefont {Johnson}},
  \bibinfo {author} {\bibfnamefont {J.~M.}\ \bibnamefont {Taylor}}, \bibinfo
  {author} {\bibfnamefont {E.~A.}\ \bibnamefont {Laird}}, \bibinfo {author}
  {\bibfnamefont {A.}~\bibnamefont {Yacoby}}, \bibinfo {author} {\bibfnamefont
  {M.~D.}\ \bibnamefont {Lukin}}, \bibinfo {author} {\bibfnamefont {C.~M.}\
  \bibnamefont {Marcus}}, \bibinfo {author} {\bibfnamefont {M.~P.}\
  \bibnamefont {Hanson}}, \ and\ \bibinfo {author} {\bibfnamefont {A.~C.}\
  \bibnamefont {Gossard}},\ }\href {\doibase 10.1126/science.1116955}
  {\bibfield  {journal} {\bibinfo  {journal} {Science}\ }\textbf {\bibinfo
  {volume} {309}},\ \bibinfo {pages} {2180} (\bibinfo {year}
  {2005})}\BibitemShut {NoStop}%
\bibitem [{\citenamefont {Johnson}\ \emph {et~al.}(2005)\citenamefont
  {Johnson}, \citenamefont {Petta}, \citenamefont {Taylor}, \citenamefont
  {Yacoby}, \citenamefont {Lukin}, \citenamefont {Marcus}, \citenamefont
  {Hanson},\ and\ \citenamefont {Gossard}}]{johnson2005}%
  \BibitemOpen
  \bibfield  {author} {\bibinfo {author} {\bibfnamefont {A.~C.}\ \bibnamefont
  {Johnson}}, \bibinfo {author} {\bibfnamefont {J.~R.}\ \bibnamefont {Petta}},
  \bibinfo {author} {\bibfnamefont {J.~M.}\ \bibnamefont {Taylor}}, \bibinfo
  {author} {\bibfnamefont {A.}~\bibnamefont {Yacoby}}, \bibinfo {author}
  {\bibfnamefont {M.~D.}\ \bibnamefont {Lukin}}, \bibinfo {author}
  {\bibfnamefont {C.~M.}\ \bibnamefont {Marcus}}, \bibinfo {author}
  {\bibfnamefont {M.~P.}\ \bibnamefont {Hanson}}, \ and\ \bibinfo {author}
  {\bibfnamefont {A.~C.}\ \bibnamefont {Gossard}},\ }\href {\doibase
  10.1038/nature03815} {\bibfield  {journal} {\bibinfo  {journal} {Nature}\
  }\textbf {\bibinfo {volume} {435}},\ \bibinfo {pages} {925} (\bibinfo {year}
  {2005})}\BibitemShut {NoStop}%
\bibitem [{\citenamefont {DiVincenzo}\ \emph {et~al.}(2000)\citenamefont
  {DiVincenzo}, \citenamefont {Bacon}, \citenamefont {Kempe}, \citenamefont
  {Burkard},\ and\ \citenamefont {Whaley}}]{divincenzo2000}%
  \BibitemOpen
  \bibfield  {author} {\bibinfo {author} {\bibfnamefont {D.~P.}\ \bibnamefont
  {DiVincenzo}}, \bibinfo {author} {\bibfnamefont {D.}~\bibnamefont {Bacon}},
  \bibinfo {author} {\bibfnamefont {J.}~\bibnamefont {Kempe}}, \bibinfo
  {author} {\bibfnamefont {G.}~\bibnamefont {Burkard}}, \ and\ \bibinfo
  {author} {\bibfnamefont {K.~B.}\ \bibnamefont {Whaley}},\ }\href {\doibase
  10.1038/35042541} {\bibfield  {journal} {\bibinfo  {journal} {Nature}\
  }\textbf {\bibinfo {volume} {408}},\ \bibinfo {pages} {339} (\bibinfo {year}
  {2000})}\BibitemShut {NoStop}%
\bibitem [{\citenamefont {Laird}\ \emph {et~al.}(2010)\citenamefont {Laird},
  \citenamefont {Taylor}, \citenamefont {DiVincenzo}, \citenamefont {Marcus},
  \citenamefont {Hanson},\ and\ \citenamefont {Gossard}}]{laird2010}%
  \BibitemOpen
  \bibfield  {author} {\bibinfo {author} {\bibfnamefont {E.~A.}\ \bibnamefont
  {Laird}}, \bibinfo {author} {\bibfnamefont {J.~M.}\ \bibnamefont {Taylor}},
  \bibinfo {author} {\bibfnamefont {D.~P.}\ \bibnamefont {DiVincenzo}},
  \bibinfo {author} {\bibfnamefont {C.~M.}\ \bibnamefont {Marcus}}, \bibinfo
  {author} {\bibfnamefont {M.~P.}\ \bibnamefont {Hanson}}, \ and\ \bibinfo
  {author} {\bibfnamefont {A.~C.}\ \bibnamefont {Gossard}},\ }\href {\doibase
  10.1103/PhysRevB.82.075403} {\bibfield  {journal} {\bibinfo  {journal}
  {Physical Review B}\ }\textbf {\bibinfo {volume} {82}},\ \bibinfo {pages}
  {075403} (\bibinfo {year} {2010})}\BibitemShut {NoStop}%
\bibitem [{\citenamefont {Gaudreau}\ \emph {et~al.}(2012)\citenamefont
  {Gaudreau}, \citenamefont {Granger}, \citenamefont {Kam}, \citenamefont
  {Aers}, \citenamefont {Studenikin}, \citenamefont {Zawadzki}, \citenamefont
  {Pioro-Ladriere}, \citenamefont {Wasilewski},\ and\ \citenamefont
  {Sachrajda}}]{gaudreau2012}%
  \BibitemOpen
  \bibfield  {author} {\bibinfo {author} {\bibfnamefont {L.}~\bibnamefont
  {Gaudreau}}, \bibinfo {author} {\bibfnamefont {G.}~\bibnamefont {Granger}},
  \bibinfo {author} {\bibfnamefont {A.}~\bibnamefont {Kam}}, \bibinfo {author}
  {\bibfnamefont {G.~C.}\ \bibnamefont {Aers}}, \bibinfo {author}
  {\bibfnamefont {S.~A.}\ \bibnamefont {Studenikin}}, \bibinfo {author}
  {\bibfnamefont {P.}~\bibnamefont {Zawadzki}}, \bibinfo {author}
  {\bibfnamefont {M.}~\bibnamefont {Pioro-Ladriere}}, \bibinfo {author}
  {\bibfnamefont {Z.~R.}\ \bibnamefont {Wasilewski}}, \ and\ \bibinfo {author}
  {\bibfnamefont {A.~S.}\ \bibnamefont {Sachrajda}},\ }\href {\doibase
  10.1038/nphys2149} {\bibfield  {journal} {\bibinfo  {journal} {Nature
  Physics}\ }\textbf {\bibinfo {volume} {8}},\ \bibinfo {pages} {54} (\bibinfo
  {year} {2012})}\BibitemShut {NoStop}%
\bibitem [{\citenamefont {Viola}\ \emph {et~al.}(2001)\citenamefont {Viola},
  \citenamefont {Knill},\ and\ \citenamefont {Laflamme}}]{viola2001}%
  \BibitemOpen
  \bibfield  {author} {\bibinfo {author} {\bibfnamefont {L.}~\bibnamefont
  {Viola}}, \bibinfo {author} {\bibfnamefont {E.}~\bibnamefont {Knill}}, \ and\
  \bibinfo {author} {\bibfnamefont {R.}~\bibnamefont {Laflamme}},\ }\href
  {\doibase 10.1088/0305-4470/34/35/331} {\bibfield  {journal} {\bibinfo
  {journal} {Journal of Physics A}\ }\textbf {\bibinfo {volume} {34}},\
  \bibinfo {pages} {7067} (\bibinfo {year} {2001})}\BibitemShut {NoStop}%
\bibitem [{\citenamefont {Fong}\ and\ \citenamefont
  {Wandzura}(2011)}]{fong2011}%
  \BibitemOpen
  \bibfield  {author} {\bibinfo {author} {\bibfnamefont {B.~H.}\ \bibnamefont
  {Fong}}\ and\ \bibinfo {author} {\bibfnamefont {S.~M.}\ \bibnamefont
  {Wandzura}},\ }\href@noop {} {\bibfield  {journal} {\bibinfo  {journal}
  {Quantum Info. Comput.}\ }\textbf {\bibinfo {volume} {11}},\ \bibinfo {pages}
  {1003} (\bibinfo {year} {2011})}\BibitemShut {NoStop}%
\bibitem [{\citenamefont {West}\ and\ \citenamefont {Fong}(2012)}]{west2012}%
  \BibitemOpen
  \bibfield  {author} {\bibinfo {author} {\bibfnamefont {J.~R.}\ \bibnamefont
  {West}}\ and\ \bibinfo {author} {\bibfnamefont {B.~H.}\ \bibnamefont
  {Fong}},\ }\href {\doibase 10.1088/1367-2630/14/8/083002} {\bibfield
  {journal} {\bibinfo  {journal} {New Journal of Physics}\ }\textbf {\bibinfo
  {volume} {14}},\ \bibinfo {pages} {083002} (\bibinfo {year}
  {2012})}\BibitemShut {NoStop}%
\bibitem [{\citenamefont {Kempe}\ \emph {et~al.}(2001)\citenamefont {Kempe},
  \citenamefont {Bacon}, \citenamefont {Lidar},\ and\ \citenamefont
  {Whaley}}]{kempe2001}%
  \BibitemOpen
  \bibfield  {author} {\bibinfo {author} {\bibfnamefont {J.}~\bibnamefont
  {Kempe}}, \bibinfo {author} {\bibfnamefont {D.}~\bibnamefont {Bacon}},
  \bibinfo {author} {\bibfnamefont {D.~A.}\ \bibnamefont {Lidar}}, \ and\
  \bibinfo {author} {\bibfnamefont {K.~B.}\ \bibnamefont {Whaley}},\ }\href
  {\doibase 10.1103/PhysRevA.63.042307} {\bibfield  {journal} {\bibinfo
  {journal} {Physical Review A}\ }\textbf {\bibinfo {volume} {63}},\ \bibinfo
  {pages} {042307} (\bibinfo {year} {2001})}\BibitemShut {NoStop}%
\bibitem [{\citenamefont {Lidar}\ and\ \citenamefont
  {Whaley}(2003)}]{lidar2003}%
  \BibitemOpen
  \bibfield  {author} {\bibinfo {author} {\bibfnamefont {D.~A.}\ \bibnamefont
  {Lidar}}\ and\ \bibinfo {author} {\bibfnamefont {K.~B.}\ \bibnamefont
  {Whaley}},\ }\href@noop {} {\bibfield  {journal} {\bibinfo  {journal}
  {Lecture Notes in Physics}\ }\textbf {\bibinfo {volume} {622}},\ \bibinfo
  {pages} {83} (\bibinfo {year} {2003})}\BibitemShut {NoStop}%
\bibitem [{\citenamefont {Davies}(1974)}]{davies1974}%
  \BibitemOpen
  \bibfield  {author} {\bibinfo {author} {\bibfnamefont {E.~B.}\ \bibnamefont
  {Davies}},\ }\href {\doibase 10.1007/BF01608389} {\bibfield  {journal}
  {\bibinfo  {journal} {Communications in Mathematical Physics}\ }\textbf
  {\bibinfo {volume} {39}},\ \bibinfo {pages} {91} (\bibinfo {year}
  {1974})}\BibitemShut {NoStop}%
\bibitem [{\citenamefont {Fick}\ and\ \citenamefont
  {Sauermann}(1983)}]{fick1983}%
  \BibitemOpen
  \bibfield  {author} {\bibinfo {author} {\bibfnamefont {E.}~\bibnamefont
  {Fick}}\ and\ \bibinfo {author} {\bibfnamefont {G.}~\bibnamefont
  {Sauermann}},\ }\href@noop {} {\emph {\bibinfo {title} {Quantenstatistik
  dynamischer Prozesse}}}\ (\bibinfo  {publisher} {Harri Deutsch},\ \bibinfo
  {year} {1983})\BibitemShut {NoStop}%
\bibitem [{\citenamefont {Fick}\ and\ \citenamefont
  {Sauermann}(1990)}]{fick1990}%
  \BibitemOpen
  \bibfield  {author} {\bibinfo {author} {\bibfnamefont {E.}~\bibnamefont
  {Fick}}\ and\ \bibinfo {author} {\bibfnamefont {G.}~\bibnamefont
  {Sauermann}},\ }\href@noop {} {\emph {\bibinfo {title} {The Quantum
  Statistics of Dynamic Processes}}}\ (\bibinfo  {publisher} {Springer},\
  \bibinfo {year} {1990})\BibitemShut {NoStop}%
\bibitem [{\citenamefont {Coish}\ and\ \citenamefont {Loss}(2005)}]{coish2005}%
  \BibitemOpen
  \bibfield  {author} {\bibinfo {author} {\bibfnamefont {W.~A.}\ \bibnamefont
  {Coish}}\ and\ \bibinfo {author} {\bibfnamefont {D.}~\bibnamefont {Loss}},\
  }\href {\doibase 10.1103/PhysRevB.72.125337} {\bibfield  {journal} {\bibinfo
  {journal} {Physical Review B}\ }\textbf {\bibinfo {volume} {72}},\ \bibinfo
  {pages} {125337} (\bibinfo {year} {2005})}\BibitemShut {NoStop}%
\bibitem [{\citenamefont {Taylor}\ \emph {et~al.}(2007)\citenamefont {Taylor},
  \citenamefont {Petta}, \citenamefont {Johnson}, \citenamefont {Yacoby},
  \citenamefont {Marcus},\ and\ \citenamefont {Lukin}}]{taylor2007}%
  \BibitemOpen
  \bibfield  {author} {\bibinfo {author} {\bibfnamefont {J.~M.}\ \bibnamefont
  {Taylor}}, \bibinfo {author} {\bibfnamefont {J.~R.}\ \bibnamefont {Petta}},
  \bibinfo {author} {\bibfnamefont {A.~C.}\ \bibnamefont {Johnson}}, \bibinfo
  {author} {\bibfnamefont {A.}~\bibnamefont {Yacoby}}, \bibinfo {author}
  {\bibfnamefont {C.~M.}\ \bibnamefont {Marcus}}, \ and\ \bibinfo {author}
  {\bibfnamefont {M.~D.}\ \bibnamefont {Lukin}},\ }\href {\doibase
  10.1103/PhysRevB.76.035315} {\bibfield  {journal} {\bibinfo  {journal}
  {Physical Review B}\ }\textbf {\bibinfo {volume} {76}},\ \bibinfo {pages}
  {035315} (\bibinfo {year} {2007})}\BibitemShut {NoStop}%
\bibitem [{Note1()}]{Note1}%
  \BibitemOpen
  \bibinfo {note} {These parameters are related to the ones from the Hubbard
  Hamiltonian in Eq.~\protect \textup {\hbox {\mathsurround \z@ \protect
  \normalfont (\ignorespaces \ref {eq:Hubbard}\unskip \@@italiccorr )}} by
  $\epsilon _{-}\equiv -\left (\epsilon _{1}-\epsilon _{2}+U_{1}\right )$ and
  $\epsilon _{+}\equiv \epsilon _{3}-\epsilon _{2}+U_{3}$}\BibitemShut
  {NoStop}%
\bibitem [{\citenamefont {Celio}\ and\ \citenamefont {Loss}(1989)}]{celio1989}%
  \BibitemOpen
  \bibfield  {author} {\bibinfo {author} {\bibfnamefont {M.}~\bibnamefont
  {Celio}}\ and\ \bibinfo {author} {\bibfnamefont {D.}~\bibnamefont {Loss}},\
  }\href {\doibase 10.1016/0378-4371(89)90490-1} {\bibfield  {journal}
  {\bibinfo  {journal} {Physica A}\ }\textbf {\bibinfo {volume} {158}},\
  \bibinfo {pages} {769} (\bibinfo {year} {1989})}\BibitemShut {NoStop}%
\bibitem [{\citenamefont {Bravyi}\ and\ \citenamefont
  {Haah}(2011)}]{bravyi2011}%
  \BibitemOpen
  \bibfield  {author} {\bibinfo {author} {\bibfnamefont {S.}~\bibnamefont
  {Bravyi}}\ and\ \bibinfo {author} {\bibfnamefont {J.}~\bibnamefont {Haah}},\
  }\href {http://arxiv.org/abs/1112.3252v1} {\bibfield  {journal} {\bibinfo
  {journal} {arXiv:1112.3252 [quant-ph]}\ } (\bibinfo {year}
  {2011})}\BibitemShut {NoStop}%
\bibitem [{\citenamefont {Spohn}(1977)}]{spohn1977}%
  \BibitemOpen
  \bibfield  {author} {\bibinfo {author} {\bibfnamefont {H.}~\bibnamefont
  {Spohn}},\ }\href {\doibase 10.1007/BF00420668} {\bibfield  {journal}
  {\bibinfo  {journal} {Letters in Mathematical Physics}\ }\textbf {\bibinfo
  {volume} {2}},\ \bibinfo {pages} {33} (\bibinfo {year} {1977})}\BibitemShut
  {NoStop}%
\bibitem [{Note2()}]{Note2}%
  \BibitemOpen
  \bibinfo {note} {We refer especially to typical qubit manipulation times. In
  the mentioned publication by Gaudreau et al. pulse times well below $10$ ns
  are used\cite {gaudreau2012}}\BibitemShut {NoStop}%
\bibitem [{Note3()}]{Note3}%
  \BibitemOpen
  \bibinfo {note} {In the paper by Gaudreau et al. the tunnel couplings $T$ are
  defined differently than the parameters $t$ in Eq.~\protect \textup {\hbox
  {\mathsurround \z@ \protect \normalfont (\ignorespaces \ref
  {eq:ExInt1}\unskip \@@italiccorr )}} and \protect \textup {\hbox
  {\mathsurround \z@ \protect \normalfont (\ignorespaces \ref
  {eq:ExInt2}\unskip \@@italiccorr )}}. The two constants are however connected
  by $T=t/\protect \sqrt {2}$}\BibitemShut {NoStop}%
\bibitem [{\citenamefont {Coish}\ and\ \citenamefont {Loss}(2004)}]{coish2004}%
  \BibitemOpen
  \bibfield  {author} {\bibinfo {author} {\bibfnamefont {W.~A.}\ \bibnamefont
  {Coish}}\ and\ \bibinfo {author} {\bibfnamefont {D.}~\bibnamefont {Loss}},\
  }\href {\doibase 10.1103/PhysRevB.70.195340} {\bibfield  {journal} {\bibinfo
  {journal} {Physical Review B}\ }\textbf {\bibinfo {volume} {70}},\ \bibinfo
  {pages} {195340} (\bibinfo {year} {2004})}\BibitemShut {NoStop}%
\bibitem [{\citenamefont {Coish}\ \emph {et~al.}(2010)\citenamefont {Coish},
  \citenamefont {Fischer},\ and\ \citenamefont {Loss}}]{coish2010}%
  \BibitemOpen
  \bibfield  {author} {\bibinfo {author} {\bibfnamefont {W.~A.}\ \bibnamefont
  {Coish}}, \bibinfo {author} {\bibfnamefont {J.}~\bibnamefont {Fischer}}, \
  and\ \bibinfo {author} {\bibfnamefont {D.}~\bibnamefont {Loss}},\ }\href
  {\doibase 10.1103/PhysRevB.81.165315} {\bibfield  {journal} {\bibinfo
  {journal} {Physical Review B}\ }\textbf {\bibinfo {volume} {81}},\ \bibinfo
  {pages} {165315} (\bibinfo {year} {2010})}\BibitemShut {NoStop}%
\bibitem [{\citenamefont {Cywinski}\ \emph
  {et~al.}(2009{\natexlab{a}})\citenamefont {Cywinski}, \citenamefont
  {Witzel},\ and\ \citenamefont {DasSarma}}]{cywinski2009}%
  \BibitemOpen
  \bibfield  {author} {\bibinfo {author} {\bibfnamefont {L.}~\bibnamefont
  {Cywinski}}, \bibinfo {author} {\bibfnamefont {W.~M.}\ \bibnamefont
  {Witzel}}, \ and\ \bibinfo {author} {\bibfnamefont {S.}~\bibnamefont
  {DasSarma}},\ }\href {\doibase 10.1103/PhysRevLett.102.057601} {\bibfield
  {journal} {\bibinfo  {journal} {Physical Review Letters}\ }\textbf {\bibinfo
  {volume} {102}},\ \bibinfo {pages} {057601} (\bibinfo {year}
  {2009}{\natexlab{a}})}\BibitemShut {NoStop}%
\bibitem [{\citenamefont {Cywinski}\ \emph
  {et~al.}(2009{\natexlab{b}})\citenamefont {Cywinski}, \citenamefont
  {Witzel},\ and\ \citenamefont {DasSarma}}]{cywinski2009-2}%
  \BibitemOpen
  \bibfield  {author} {\bibinfo {author} {\bibfnamefont {L.}~\bibnamefont
  {Cywinski}}, \bibinfo {author} {\bibfnamefont {W.~M.}\ \bibnamefont
  {Witzel}}, \ and\ \bibinfo {author} {\bibfnamefont {S.}~\bibnamefont
  {DasSarma}},\ }\href {\doibase 10.1103/PhysRevB.79.245314} {\bibfield
  {journal} {\bibinfo  {journal} {Physical Review B}\ }\textbf {\bibinfo
  {volume} {79}},\ \bibinfo {pages} {245314} (\bibinfo {year}
  {2009}{\natexlab{b}})}\BibitemShut {NoStop}%
\bibitem [{\citenamefont {Ladd}(2012)}]{ladd2012}%
  \BibitemOpen
  \bibfield  {author} {\bibinfo {author} {\bibfnamefont {T.~D.}\ \bibnamefont
  {Ladd}},\ }\href {\doibase 10.1103/PhysRevB.86.125408} {\bibfield  {journal}
  {\bibinfo  {journal} {Physical Review B}\ }\textbf {\bibinfo {volume} {86}},\
  \bibinfo {pages} {125408} (\bibinfo {year} {2012})}\BibitemShut {NoStop}%
\bibitem [{\citenamefont {Merkulov}\ \emph {et~al.}(2002)\citenamefont
  {Merkulov}, \citenamefont {Efros},\ and\ \citenamefont
  {Rosen}}]{merkulov2002}%
  \BibitemOpen
  \bibfield  {author} {\bibinfo {author} {\bibfnamefont {I.~A.}\ \bibnamefont
  {Merkulov}}, \bibinfo {author} {\bibfnamefont {A.~L.}\ \bibnamefont {Efros}},
  \ and\ \bibinfo {author} {\bibfnamefont {M.}~\bibnamefont {Rosen}},\ }\href
  {\doibase 10.1103/PhysRevB.65.205309} {\bibfield  {journal} {\bibinfo
  {journal} {Physical Review B}\ }\textbf {\bibinfo {volume} {65}},\ \bibinfo
  {pages} {205309} (\bibinfo {year} {2002})}\BibitemShut {NoStop}%
\bibitem [{Note4()}]{Note4}%
  \BibitemOpen
  \bibinfo {note} {In single spin experiments fluctuating in-plane magnetic
  fields, both in x- and y-direction, cause spin flips. The generated dynamics
  is very similar. We just consider the coupling operators $\sigma _x^i$, since
  they directly relate to single spin flips $\sigma _{\pm }$}\BibitemShut
  {NoStop}%
\bibitem [{\citenamefont {Amasha}\ \emph {et~al.}(2008)\citenamefont {Amasha},
  \citenamefont {MacLean}, \citenamefont {Radu}, \citenamefont {Zumb\"{u}hl},
  \citenamefont {Kastner}, \citenamefont {Hanson},\ and\ \citenamefont
  {Gossard}}]{amasha2008}%
  \BibitemOpen
  \bibfield  {author} {\bibinfo {author} {\bibfnamefont {S.}~\bibnamefont
  {Amasha}}, \bibinfo {author} {\bibfnamefont {K.}~\bibnamefont {MacLean}},
  \bibinfo {author} {\bibfnamefont {I.~P.}\ \bibnamefont {Radu}}, \bibinfo
  {author} {\bibfnamefont {D.~M.}\ \bibnamefont {Zumb\"{u}hl}}, \bibinfo
  {author} {\bibfnamefont {M.~A.}\ \bibnamefont {Kastner}}, \bibinfo {author}
  {\bibfnamefont {M.~P.}\ \bibnamefont {Hanson}}, \ and\ \bibinfo {author}
  {\bibfnamefont {A.~C.}\ \bibnamefont {Gossard}},\ }\href {\doibase
  10.1103/PhysRevLett.100.046803} {\bibfield  {journal} {\bibinfo  {journal}
  {Physical Review Letters}\ }\textbf {\bibinfo {volume} {100}},\ \bibinfo
  {pages} {046803} (\bibinfo {year} {2008})}\BibitemShut {NoStop}%
\bibitem [{\citenamefont {Nowack}\ \emph {et~al.}(2011)\citenamefont {Nowack},
  \citenamefont {Shafiei}, \citenamefont {Laforest}, \citenamefont
  {Prawiroatmodjo}, \citenamefont {Schreiber}, \citenamefont {Reichl},
  \citenamefont {Wegscheider},\ and\ \citenamefont {Vandersypen}}]{nowack2011}%
  \BibitemOpen
  \bibfield  {author} {\bibinfo {author} {\bibfnamefont {K.~C.}\ \bibnamefont
  {Nowack}}, \bibinfo {author} {\bibfnamefont {M.}~\bibnamefont {Shafiei}},
  \bibinfo {author} {\bibfnamefont {M.}~\bibnamefont {Laforest}}, \bibinfo
  {author} {\bibfnamefont {G.~E. D.~K.}\ \bibnamefont {Prawiroatmodjo}},
  \bibinfo {author} {\bibfnamefont {L.~R.}\ \bibnamefont {Schreiber}}, \bibinfo
  {author} {\bibfnamefont {C.}~\bibnamefont {Reichl}}, \bibinfo {author}
  {\bibfnamefont {W.}~\bibnamefont {Wegscheider}}, \ and\ \bibinfo {author}
  {\bibfnamefont {L.~M.~K.}\ \bibnamefont {Vandersypen}},\ }\href {\doibase
  10.1126/science.1209524} {\bibfield  {journal} {\bibinfo  {journal}
  {Science}\ }\textbf {\bibinfo {volume} {333}},\ \bibinfo {pages} {1269}
  (\bibinfo {year} {2011})}\BibitemShut {NoStop}%
\bibitem [{\citenamefont {Hu}(2011)}]{hu2011}%
  \BibitemOpen
  \bibfield  {author} {\bibinfo {author} {\bibfnamefont {X.}~\bibnamefont
  {Hu}},\ }\href {\doibase 10.1103/PhysRevB.83.165322} {\bibfield  {journal}
  {\bibinfo  {journal} {Physical Review B}\ }\textbf {\bibinfo {volume} {83}},\
  \bibinfo {pages} {165322} (\bibinfo {year} {2011})}\BibitemShut {NoStop}%
\bibitem [{\citenamefont {Gamble}\ \emph {et~al.}(2012)\citenamefont {Gamble},
  \citenamefont {Friesen}, \citenamefont {Coppersmith},\ and\ \citenamefont
  {Hu}}]{gamble2012}%
  \BibitemOpen
  \bibfield  {author} {\bibinfo {author} {\bibfnamefont {J.~K.}\ \bibnamefont
  {Gamble}}, \bibinfo {author} {\bibfnamefont {M.}~\bibnamefont {Friesen}},
  \bibinfo {author} {\bibfnamefont {S.~N.}\ \bibnamefont {Coppersmith}}, \ and\
  \bibinfo {author} {\bibfnamefont {X.}~\bibnamefont {Hu}},\ }\href {\doibase
  10.1103/PhysRevB.86.035302} {\bibfield  {journal} {\bibinfo  {journal}
  {Physical Review B}\ }\textbf {\bibinfo {volume} {86}},\ \bibinfo {pages}
  {035302} (\bibinfo {year} {2012})}\BibitemShut {NoStop}%
\bibitem [{\citenamefont {Hu}\ and\ \citenamefont {DasSarma}(2006)}]{hu2006}%
  \BibitemOpen
  \bibfield  {author} {\bibinfo {author} {\bibfnamefont {X.}~\bibnamefont
  {Hu}}\ and\ \bibinfo {author} {\bibfnamefont {S.}~\bibnamefont {DasSarma}},\
  }\href {\doibase 10.1103/PhysRevLett.96.100501} {\bibfield  {journal}
  {\bibinfo  {journal} {Physical Review Letters}\ }\textbf {\bibinfo {volume}
  {96}},\ \bibinfo {pages} {100501} (\bibinfo {year} {2006})}\BibitemShut
  {NoStop}%
\bibitem [{\citenamefont {Ribeiro}\ \emph {et~al.}(2010)\citenamefont
  {Ribeiro}, \citenamefont {Petta},\ and\ \citenamefont
  {Burkard}}]{ribeiro2010}%
  \BibitemOpen
  \bibfield  {author} {\bibinfo {author} {\bibfnamefont {H.}~\bibnamefont
  {Ribeiro}}, \bibinfo {author} {\bibfnamefont {J.~R.}\ \bibnamefont {Petta}},
  \ and\ \bibinfo {author} {\bibfnamefont {G.}~\bibnamefont {Burkard}},\ }\href
  {\doibase 10.1103/PhysRevB.82.115445} {\bibfield  {journal} {\bibinfo
  {journal} {Physical Review B}\ }\textbf {\bibinfo {volume} {82}},\ \bibinfo
  {pages} {115445} (\bibinfo {year} {2010})}\BibitemShut {NoStop}%
\bibitem [{\citenamefont {Bluhm}\ \emph {et~al.}(2011)\citenamefont {Bluhm},
  \citenamefont {Foletti}, \citenamefont {Neder}, \citenamefont {Rudner},
  \citenamefont {Mahalu}, \citenamefont {Umansky},\ and\ \citenamefont
  {Yacoby}}]{bluhm2011}%
  \BibitemOpen
  \bibfield  {author} {\bibinfo {author} {\bibfnamefont {H.}~\bibnamefont
  {Bluhm}}, \bibinfo {author} {\bibfnamefont {S.}~\bibnamefont {Foletti}},
  \bibinfo {author} {\bibfnamefont {I.}~\bibnamefont {Neder}}, \bibinfo
  {author} {\bibfnamefont {M.}~\bibnamefont {Rudner}}, \bibinfo {author}
  {\bibfnamefont {D.}~\bibnamefont {Mahalu}}, \bibinfo {author} {\bibfnamefont
  {V.}~\bibnamefont {Umansky}}, \ and\ \bibinfo {author} {\bibfnamefont
  {A.}~\bibnamefont {Yacoby}},\ }\href {\doibase doi:10.1038/nphys1856}
  {\bibfield  {journal} {\bibinfo  {journal} {Nature Physics}\ }\textbf
  {\bibinfo {volume} {7}},\ \bibinfo {pages} {109} (\bibinfo {year}
  {2011})}\BibitemShut {NoStop}%
\bibitem [{\citenamefont {Maune}\ \emph {et~al.}(2012)\citenamefont {Maune},
  \citenamefont {Borselli}, \citenamefont {Huang}, \citenamefont {Ladd},
  \citenamefont {Deelman}, \citenamefont {Holabird}, \citenamefont {Kiselev},
  \citenamefont {Alvarado-Rodriguez}, \citenamefont {Ross}, \citenamefont
  {Schmitz}, \citenamefont {Sokolich}, \citenamefont {Watson}, \citenamefont
  {Gyure},\ and\ \citenamefont {Hunter}}]{maune2012}%
  \BibitemOpen
  \bibfield  {author} {\bibinfo {author} {\bibfnamefont {B.~M.}\ \bibnamefont
  {Maune}}, \bibinfo {author} {\bibfnamefont {M.~G.}\ \bibnamefont {Borselli}},
  \bibinfo {author} {\bibfnamefont {B.}~\bibnamefont {Huang}}, \bibinfo
  {author} {\bibfnamefont {T.~D.}\ \bibnamefont {Ladd}}, \bibinfo {author}
  {\bibfnamefont {P.~W.}\ \bibnamefont {Deelman}}, \bibinfo {author}
  {\bibfnamefont {K.~S.}\ \bibnamefont {Holabird}}, \bibinfo {author}
  {\bibfnamefont {A.~A.}\ \bibnamefont {Kiselev}}, \bibinfo {author}
  {\bibfnamefont {I.}~\bibnamefont {Alvarado-Rodriguez}}, \bibinfo {author}
  {\bibfnamefont {R.~S.}\ \bibnamefont {Ross}}, \bibinfo {author}
  {\bibfnamefont {A.~E.}\ \bibnamefont {Schmitz}}, \bibinfo {author}
  {\bibfnamefont {M.}~\bibnamefont {Sokolich}}, \bibinfo {author}
  {\bibfnamefont {C.~A.}\ \bibnamefont {Watson}}, \bibinfo {author}
  {\bibfnamefont {M.~F.}\ \bibnamefont {Gyure}}, \ and\ \bibinfo {author}
  {\bibfnamefont {A.~T.}\ \bibnamefont {Hunter}},\ }\href {\doibase
  10.1038/nature10707} {\bibfield  {journal} {\bibinfo  {journal} {Nature}\
  }\textbf {\bibinfo {volume} {481}},\ \bibinfo {pages} {344} (\bibinfo {year}
  {2012})}\BibitemShut {NoStop}%
\bibitem [{\citenamefont {Foletti}\ \emph {et~al.}(2009)\citenamefont
  {Foletti}, \citenamefont {Bluhm}, \citenamefont {Mahalu}, \citenamefont
  {Umansky},\ and\ \citenamefont {Yacoby}}]{foletti2009}%
  \BibitemOpen
  \bibfield  {author} {\bibinfo {author} {\bibfnamefont {S.}~\bibnamefont
  {Foletti}}, \bibinfo {author} {\bibfnamefont {H.}~\bibnamefont {Bluhm}},
  \bibinfo {author} {\bibfnamefont {D.}~\bibnamefont {Mahalu}}, \bibinfo
  {author} {\bibfnamefont {V.}~\bibnamefont {Umansky}}, \ and\ \bibinfo
  {author} {\bibfnamefont {A.}~\bibnamefont {Yacoby}},\ }\href {\doibase
  10.1038/nphys1424} {\bibfield  {journal} {\bibinfo  {journal} {Nature
  Physics}\ }\textbf {\bibinfo {volume} {5}},\ \bibinfo {pages} {903} (\bibinfo
  {year} {2009})}\BibitemShut {NoStop}%
\bibitem [{\citenamefont {Gullans}\ \emph {et~al.}(2010)\citenamefont
  {Gullans}, \citenamefont {Krich}, \citenamefont {Taylor}, \citenamefont
  {Bluhm}, \citenamefont {Halperin}, \citenamefont {Marcus}, \citenamefont
  {Stopa}, \citenamefont {Yacoby},\ and\ \citenamefont {Lukin}}]{gullans2010}%
  \BibitemOpen
  \bibfield  {author} {\bibinfo {author} {\bibfnamefont {M.}~\bibnamefont
  {Gullans}}, \bibinfo {author} {\bibfnamefont {J.~J.}\ \bibnamefont {Krich}},
  \bibinfo {author} {\bibfnamefont {J.~M.}\ \bibnamefont {Taylor}}, \bibinfo
  {author} {\bibfnamefont {H.}~\bibnamefont {Bluhm}}, \bibinfo {author}
  {\bibfnamefont {B.~I.}\ \bibnamefont {Halperin}}, \bibinfo {author}
  {\bibfnamefont {C.~M.}\ \bibnamefont {Marcus}}, \bibinfo {author}
  {\bibfnamefont {M.}~\bibnamefont {Stopa}}, \bibinfo {author} {\bibfnamefont
  {A.}~\bibnamefont {Yacoby}}, \ and\ \bibinfo {author} {\bibfnamefont {M.~D.}\
  \bibnamefont {Lukin}},\ }\href {\doibase 10.1103/PhysRevLett.104.226807}
  {\bibfield  {journal} {\bibinfo  {journal} {Physical Review Letters}\
  }\textbf {\bibinfo {volume} {104}},\ \bibinfo {pages} {226807} (\bibinfo
  {year} {2010})}\BibitemShut {NoStop}%
\bibitem [{\citenamefont {Brunner}\ \emph {et~al.}(2011)\citenamefont
  {Brunner}, \citenamefont {Shin}, \citenamefont {Obata}, \citenamefont
  {Pioro-Ladri{\`e}re}, \citenamefont {Kubo}, \citenamefont {Yoshida},
  \citenamefont {Taniyama}, \citenamefont {Tokura},\ and\ \citenamefont
  {Tarucha}}]{brunner2011}%
  \BibitemOpen
  \bibfield  {author} {\bibinfo {author} {\bibfnamefont {R.}~\bibnamefont
  {Brunner}}, \bibinfo {author} {\bibfnamefont {Y.-S.}\ \bibnamefont {Shin}},
  \bibinfo {author} {\bibfnamefont {T.}~\bibnamefont {Obata}}, \bibinfo
  {author} {\bibfnamefont {M.}~\bibnamefont {Pioro-Ladri{\`e}re}}, \bibinfo
  {author} {\bibfnamefont {T.}~\bibnamefont {Kubo}}, \bibinfo {author}
  {\bibfnamefont {K.}~\bibnamefont {Yoshida}}, \bibinfo {author} {\bibfnamefont
  {T.}~\bibnamefont {Taniyama}}, \bibinfo {author} {\bibfnamefont
  {Y.}~\bibnamefont {Tokura}}, \ and\ \bibinfo {author} {\bibfnamefont
  {S.}~\bibnamefont {Tarucha}},\ }\href {\doibase
  10.1103/PhysRevLett.107.146801} {\bibfield  {journal} {\bibinfo  {journal}
  {Physical Review Letters}\ }\textbf {\bibinfo {volume} {107}},\ \bibinfo
  {pages} {146801} (\bibinfo {year} {2011})}\BibitemShut {NoStop}%
\bibitem [{\citenamefont {Blum}(1996)}]{blum1996}%
  \BibitemOpen
  \bibfield  {author} {\bibinfo {author} {\bibfnamefont {K.}~\bibnamefont
  {Blum}},\ }\href@noop {} {\emph {\bibinfo {title} {Density matrix theory and
  applications}}}\ (\bibinfo  {publisher} {Plenum},\ \bibinfo {address} {New
  York},\ \bibinfo {year} {1996})\BibitemShut {NoStop}%
\bibitem [{\citenamefont {Kloeffel}\ and\ \citenamefont
  {Loss}(2013)}]{kloeffel2012}%
  \BibitemOpen
  \bibfield  {author} {\bibinfo {author} {\bibfnamefont {C.}~\bibnamefont
  {Kloeffel}}\ and\ \bibinfo {author} {\bibfnamefont {D.}~\bibnamefont
  {Loss}},\ }\href {\doibase 10.1146/annurev-conmatphys-030212-184248}
  {\bibfield  {journal} {\bibinfo  {journal} {Annual Review of Condensed Matter
  Physics}\ }\textbf {\bibinfo {volume} {4}},\ \bibinfo {pages} {51} (\bibinfo
  {year} {2013})}\BibitemShut {NoStop}%
\bibitem [{\citenamefont {Nielsen}\ and\ \citenamefont
  {Chuang}(2000)}]{nielsen2000}%
  \BibitemOpen
  \bibfield  {author} {\bibinfo {author} {\bibfnamefont {M.~A.}\ \bibnamefont
  {Nielsen}}\ and\ \bibinfo {author} {\bibfnamefont {I.~L.}\ \bibnamefont
  {Chuang}},\ }\href@noop {} {\emph {\bibinfo {title} {Quantum computation and
  quantum information}}}\ (\bibinfo  {publisher} {Cambridge University Press},\
  \bibinfo {address} {Cambridge},\ \bibinfo {year} {2000})\BibitemShut
  {NoStop}%
\bibitem [{\citenamefont {Slichter}(1990)}]{slichter1990}%
  \BibitemOpen
  \bibfield  {author} {\bibinfo {author} {\bibfnamefont {C.~P.}\ \bibnamefont
  {Slichter}},\ }\href@noop {} {\emph {\bibinfo {title} {Principles of magnetic
  resonance}}}\ (\bibinfo  {publisher} {Springer},\ \bibinfo {address}
  {Berlin},\ \bibinfo {year} {1990})\BibitemShut {NoStop}%
\bibitem [{Note5()}]{Note5}%
  \BibitemOpen
  \bibinfo {note} {The factor 2 is used to agree with the definitions of
  Nielsen and Chuang.\cite {nielsen2000}}\BibitemShut {NoStop}%
\bibitem [{\citenamefont {Marinescu}\ and\ \citenamefont
  {Marinescu}(2012)}]{marinescu2012}%
  \BibitemOpen
  \bibfield  {author} {\bibinfo {author} {\bibfnamefont {D.~C.}\ \bibnamefont
  {Marinescu}}\ and\ \bibinfo {author} {\bibfnamefont {G.~M.}\ \bibnamefont
  {Marinescu}},\ }\href@noop {} {\emph {\bibinfo {title} {Classical and quantum
  information}}}\ (\bibinfo  {publisher} {Elsevier Academic Press},\ \bibinfo
  {address} {Amsterdam},\ \bibinfo {year} {2012})\BibitemShut {NoStop}%
\end{thebibliography}%
\end{document}